 \documentclass[12pt,preprint]{aastex}

\newcommand{\lya}          {Ly$\alpha$}

\newcommand{\cdens}         {\mbox{cm$^{-2}$}}
\newcommand{\hi}          {\mbox{\rm \ion{H}{1}}}

\newcommand{\kms}         {km~s$^{-1}$}

\newcommand{\lymana}      {\mbox{Lyman-$\alpha$}}

\newcommand{\mstar}       {\mbox{$M_{*}$}}
\newcommand{\msun}        {\mbox{$M_{\odot}$}}

\def\spose#1{\hbox to 0pt{#1\hss}}
\def\lta{\mathrel{\spose{\lower 3pt\hbox{$\mathchar"218$}}
     \raise 2.0pt\hbox{$\mathchar"13C$}}}
\def\gta{\mathrel{\spose{\lower 3pt\hbox{$\mathchar"218$}}
     \raise 2.0pt\hbox{$\mathchar"13E$}}}

\shorttitle{}
\shortauthors{Penprase et al.}
\slugcomment{Accepted for Publication in Ap. J. June 16, 2010 }
\usepackage{graphicx}

\begin{document}

\title{Keck ESI Observations of Metal-Poor Damped Lyman-$\alpha$ Systems\altaffilmark{1}}

\author{Bryan E. Penprase\altaffilmark{2,3}, J. Xavier Prochaska\altaffilmark{4}, Wallace L. W. Sargent\altaffilmark{3}, Irene Toro Martinez\altaffilmark{2}, and Daniel J. Beeler\altaffilmark{2}}

\altaffiltext{1}{Some of the data presented herein were obtained at the
  W. M. Keck Observatory, which is operated as a scientific
  partnership among the California Institute of Technology, the
  University of California, and NASA.  
  The Observatory was made possible by the generous
  financial support of the W. M. Keck Foundation.}

\altaffiltext{2}{Department of Physics and Astronomy,
                 Pomona College, 610 N. College Ave., Claremont, CA  91711;
                 penprase@dci.pomona.edu}

\altaffiltext{3}{Department of Astronomy, California Institute of
  Technology, 1200 E. California Blvd., MS 105-24, Pasadena, CA 91125;
  wws@astro.caltech.edu}

\altaffiltext{4}{Department of Astronomy and Astrophysics, UCO/Lick Observatory, University of California, 1156 High Street, Santa Cruz, CA 95064; xavier@ucolick.org}

\begin{abstract}

We present the first results from a survey of SDSS quasars selected for 
strong \hi\ damped \lymana\ (DLA) absorption with corresponding
low equivalent width absorption from strong low-ion transitions
(e.g.\ \ion{C}{2}~$\lambda$1334 and \ion{Si}{2}~$\lambda$1260).
These metal-poor DLA candidates were selected from 
the SDSS DR5 quasar spectroscopic database, 
and comprise a large new sample for probing low metallicity galaxies.
Medium-resolution echellette spectra from the Keck ESI spectrograph for 
an initial sample of 35 systems were obtained to explore the metal-poor
tail of the DLA distribution and to investigate the nucleosynthetic patterns
at these metallicities. We have estimated saturation corrections for the moderately under-resolved spectra, and systems with very narrow Doppler parameter ( b  $\le$ 5 \kms) will likely have underestimated abundances. For those systems with Doppler parameters b $>$ 5 \kms, we have measured low metallicity DLA gas with 
[X/H] $<$ -2.4 for at least one of C, O, Si, or Fe.  Assuming non-saturated components, we estimate that several DLA systems have [X/H] $< -2.8$,  including five DLA systems with both low equivalent widths and  low metallicity in transitions of both \ion{C}{2} and \ion{O}{1}. All of the measured DLA metallicities, however, exceed or are consistent with a metallicity of at least
1/1000 of solar, regardless of the effects of saturation in our spectra.  Our results indicate that the metal-poor tail of galaxies 
at $z \sim 3$ drops exponentially at [X/H]~$\lesssim -3$. If the distribution of metallicity is Gaussian, the
probability of identifying ISM gas with lower abundance is extremely small, and our results suggest that DLA systems with [X/H] $<$ -4.0 are extremely rare, and could comprise only 8$\times10^{-7}$ of  DLA systems. The relative abundances of species within these low-metallicity DLA 
systems is compared with stellar nucleosynthesis models, and are consistent with
stars having masses of 30 \msun\ $<$ \mstar\ $<$ 100 \msun. The observed ratio of [C/O] for values of [O/H] $< -2.5$ exceeds values seen in moderate metallicity DLA systems, and also exceeds theoretical nucleosynthesis predictions for higher mass Population III stars. We also have observed a correlation between the column density N(\ion{C}{4}) 
with [Si/H] metallicity, suggestive of a trend between mass of the DLA system and its metallicity.  

\end{abstract}


\section{INTRODUCTION}
\label{intro}

A key probe of nucleosynthesis in the early universe has been low metallicity stars
discovered in the Galaxy and its nearest neighbors.
New samples of stars developed for this purpose have detected iron abundances of [Fe/H] $<$ -5.0, and similarly low abundances of other elements. 
(Frebel et al, 2006, 2007; Cohen et al, 2007).
The abundance patterns of these metal poor stars have been compared with stellar nucleosynthesis models, and have provided useful constraints on the initial mass function and other properties of the Pop III stars  (Matteucci, 2003; Meynet and Maeder, 2002; Woosley and Weaver, 1995). Despite the success in finding many low metallicity stars, the abundances derived from 
these stars have uncertainty related to 
the unknown mixing  and convection in the star, the possible 
of separation of dust and gas in star formation, and enrichment 
of the stellar atmosphere from nearby companion stars.

The Damped \lymana\ (DLA) systems have been a useful complementary probe of the 
early universe, and of metallicity in the intergalactic medium. 
The \ion{H}{1} threshold of N(HI) $= 2\times 10^{20} \rm cm^{-2}$ 
defines the DLAs.  Above this threshold the gas may safely be assumed 
to be predominantly neutral due to shielding 
by Hydrogen, allowing the determination of metallicity from coexisting species such as \ion{C}{2},\ion{O}{1},\ion{Si}{2},\ion{Al}{2}, and \ion{Fe}{2} compared to N(HI) with minimal corrections for ionization.
Since early surveys of DLA systems 
(Wolfe et al.\ 1995; Sargent et al 1989),
the number of DLA systems has continued to rise into the hundreds, 
allowing a statistically significant sample (Prochaska et al 2005). 
The larger surveys of DLAs 
show increasing metallicity as a function of cosmic time (Prochaska et al 2003),
but with significant 
scatter, and some evidence of dust depletion in the medium
(e.g.\ Pettini et al.\ 1994).  
Using the largest samples of DLA systems, it is possible to probe the
low metallicity tail of DLAs.
Previous works have reported a ``floor'' to the metallicity distribution
at [X/H] $\approx -2.7$ (Prochaska et al. 2003), with only a 
small fraction of only 10 percent or less of DLA systems showing 
[X/H] $< -2.5$. 

Recent work has focused on the abundance patterns of these very low
metallicity systems.
Pettini et al.\ (2008) provide high resolution spectra 
of a sample of three low metallicity DLA systems at redshifts ranging from 2.07 $< $ z(DLA) $<$ 2.81, and detect values of [O/H] in the range of -2.04 $<$ [O/H] $<$ -2.42, with lower metallicities in [Fe/H], which ranges from -2.55 $<$ [Fe/H] $<$ -2.80. The UVES spectra also provide useful information on the linewidths of these low-metallicity DLAs and indicate that 
metal-poor DLAs tend to have quiescient kinematics with $b$ values in the range of 3.7 \kms $<$ b $<$ 25.1 \kms, 
with an average/median $b$-values of 9.24 \kms\ and 6.5 \kms\ respectively.

In order to study the lowest metallicity DLA systems, we began 
a survey to examine a very large sample of quasars with DLA 
systems of low metallicity using the Sloan Digitial Sky Survey (SDSS) 
fifth release (DR5), which contains 77,229 newly detected quasars 
(Schneider, et al, 2007).  The DLA sample was assembled from fits to 
the SDSS spectra in which the estimated \ion{H}{1}
column density satisfied log(N(HI)) $\ge $ 20.3 (Prochaska et al.\ 2005).
From the sample of 968 DLA systems, a subset was chosen that were toward 
relatively bright ($r<20$) quasars, and which had DLAs with unusually
weak absorption from generally strong low-ion transitions 
(e.g.\ \ion{C}{2}~$\lambda$1334, \ion{Si}{2}~$\lambda$1260, \ion{O}{1}~1302) 
based on equivalent width analysis of the SDSS spectra.  
Note that follow-up observations and re-analysis of the \lymana\ transition
may reveal new estimates for N(HI) that fall below the DLA threshold.
In this manuscript, we will consider the metallicities and abundance
patterns of these systems even if they are now formally referred to
as sub-DLAs (or super Lyman limit systems; SLLS).

A Keck observing program with the ESI echellete spectrograph was then conducted to provide more accurate measurements of column densities of \ion{C}{2}, \ion{O}{1}, \ion{Si}{2}, and other species, to determine the actual metallicities of the DLA systems. By targeting the best newly discovered DLA systems from such a large sample of quasars,  we aim to efficiently identify the lowest metallicity DLA systems and, in turn, constrain the low metallicity distribution function of high $z$ galaxies
and enable detailed comparison with nucleosynthetic models. Our sample probes a wide redshift range, 2.3 $< z({\rm DLA}) <$ 4.12, addressing 
the metallicity distribution at higher redshifts than previous DLA surveys. 

Models of Pop III stars have gained in sophistication and predictive ability. The current generation of models (Woosley and Heger, 2007) can provide exact nucleosynthetic yields as a function of stellar mass, to enable comparison with stellar or DLA observations. 
In most of the Pop III stars, a strong "odd-even effect" is seen in which C,O,Si and other even atomic number elements are as much as 1 dex more abundant than odd atomic number elements such as N, Al, and Mn.  The odd-even effect is well known as the result of rapid nucleosynthesis in C and O core burning stages through the triple alpha process. 
The end point of Pop III stars differs from other stars and is thought to involve either Type II or pair-pair annihilation supernovae, depending on the mass of the progenitor stars.  
The yields of high mass stars have been modeled by a number of groups, 
including the effects of rotation, mass loss through winds, and turbulent mixing within the stars (\cite{chie02},\cite{heg02},\cite{ume02}).

The latter effects, combined with an unknown IMF, lead
to a large range of predictions for the integrated yields of Pop~III stars.
New observations of our low metallicity DLA sample can complement the work of theoretical models, and stellar studies to help better determine the nature of both the high mass Pop III stars and star formation in the early universe.

Despite the importance of low-metallicity systems to constraining nucleosynthesis within the early universe, very few DLA systems of low metallicity have been discovered to date.
Previous work has compared DLA derived metallicities with nucleosynthesis yields. Akerman et al (2004) adopted yields from Meynet and Maeder (2002) for stars of mass 8 \msun $<$ \mstar\ $<$ 80  \msun\ , and for lower mass stars with 0.8 \msun $<$ \mstar\ $<$ 8 \msun\ those of van den Hoek and Groenewegen (1997).  Models of Heger and Woosley (2002), Chieffi and Limongi (2002), and Umeda and Nomoto (2002) agree in their predictions of yields of some elements, but provide divergent estimates of the ratios of [C/O] for high mass nucleosynthesis. 

In this paper, we present a large set of Keck ESI observations 
that demonstrate the efficiency of pre-selection, and provide estimates of metallicity for 35 DLA systems.  We present the first results on the low metallicity distribution function
of high $z$ galaxies and compare these results to earlier studies and to nucleosynthesis models.

\section{OBSERVATIONS AND DATA REDUCTION}
\label{observations}

\subsection{Sample Selection}

The damped \lya\ systems studied in this paper were drawn from
a much larger list of metal-poor DLA candidates that was
generated as follows.  We started first with the complete set
of DLA candidates discovered by \cite{pro05} and \cite{pw09} in
the Sloan Digital Sky Survey, Data Releases 1 to 5.  The candidates
were identified using an automated algorithm that focused solely
on \lya\ absorption.  These authors also searched independently 
for metal-line systems in the same quasar spectra, restricting the
search to the spectral region redward of the \lya\ forest
\citep[see][for a description]{shf06}.  All of the metal-poor
DLA candidates had fewer than three significantly $(4\sigma$)
detected metal-line transitions.
Furthermore, when analyzing the \lya\ line, Prochaska and 
collaborators searched again for metal-line absorption using
the estimated centroid of \lya\ as a strong prior.  Those without
any obvious metal-line absorption (corresponding to $\lesssim 0.5$\AA)
were flagged, in part to allow for an additional systematic uncertainty
when estimating the \ion{H}{1} column density.
Altogether, we culled 405 
metal-poor DLA candidates as listed
in Table~\ref{tab:candidates}.  Note that these systems are not
required to satisfy the selection criteria used by \cite{pro05} and
\cite{pw09} in their analysis.  Many have poorer S/N spectra and many
have $N_{\rm HI}$ values estimated to be lower than the standard
DLA threshold of $2 \times 10^{20}$\cdens.   We also emphasize that
the overwhelming majority of DLAs detected in the SDSS survey do
exhibit significant metal-line absorption indicating a metallicity in
excess of $\approx 1/100$ solar.   The sample studied in this paper,
therefore represents the most promising candidates for systems with
very low metallicity.

From Table~\ref{tab:candidates}, we chose a subset for follow-up
obsevations at higher spectral resolution and S/N.  Our principal
goal was to discover the most metal-poor DLAs, not to choose
targets that broadly sample the metal-poor candidate list.  To this end,
we selected targets with the following characteristics:
(i) the background quasar's RA coincided with our Winter/Spring
observing allocation;
(ii) brighter quasars to maximize the efficiency of our observing time;
(iii) candidates with larger $N_{\rm HI}$ values; 
(iv) candidates with absorption redshifts near the quasar emission
redshift such that key transitions (e.g.\ \ion{O}{1}~1302, \ion{C}{2}~1334)
lay outside the \lya\ forest;
(v) quasars with more than one metal-poor DLA candidates.
The targets eventually observed represent a balance between these
various factors.  Note that the combined factors generally imply
that the DLA candidates satisfy the statistical criteria of \cite{pro05}.

\subsection{Data Acquisition and Reduction}

The data for this paper were acquired during two observing runs at the Keck II telescope in March 2007 and May 2008. We describe the observations of both observing runs below.
Observations of a sample of 10 quasars selected for low metallicity were taken at the Keck II telescope for $\sim$ 5.5 hours on 16 March 2007. 
The Echellete Spectrograph and Imager (ESI) (\cite{schein02})
was used in echellette mode, which provides a free spectral 
range of $3900$\AA\ to $10900$~\AA\ at a dispersion ranging from $0.16~$\AA~pixel$^{-1}$ 
to $0.30$~\AA~pixel$^{-1}$, corresponding to a constant velocity dispersion
of 11.5 \kms\ pixel$^{-1}$. 
The slit width remained constant at 0.75\arcsec\ for all of the observations,
providing a FWHM resolution of $\approx 57 $\kms.
The Clear S filter was used for the observations. 
Exposure times on the target quasars ranged from 16 to 42 minutes, and resulted in a median S/N in the observed spectra of S/N $=$ 29 per resolution element, 
with a range of values of 23 $<$ S/N $<$ 64. 
A second sample of 23 quasars was observed in a second observing run 
at the Keck~II telescope over three nights during 7-9 May 2008. The ESI 
spectrograph and instrument parameters were the same as on the first 
observing run, and greatly improved atmospheric conditions enabled a 
very high S/N to be obtained for most of the 23 quasars, resulting 
in a large sample of spectra with an average S/N of approximately S/N = 50.

The data from both observing runs were reduced using the IDL-based 
ESIRedux\footnote{http://www2.keck.hawaii.edu/inst/esi/ESIRedux/index.html}
pipeline (Prochaska et al.\ 2003b), and further processed 
using IDL to fit a high order spline to remove the quasar continuum, 
and extract column densities.  The observed sample of quasars, 
apparent magnitudes, exposure times, the DLA redshifts, and other quantities are presented below in Table \ref{tab:obstab}.

For each of the quasars, the HI profile was fit using XIDL and IDL routines, and the damping wings of the HI profiles were visually compared to the observed continuum in the quasar spectrum. The HI profiles were fit independently by two of the team (JXP and BEP) and our HI columns agreed in all cases within the quoted uncertainty, in most cases to within 0.1 dex in N(HI). 
In a number of cases, our new estimates for the N(HI) values indicate that the absorber does not meet the DLA threshold criterion of $2 \times 10^{20} \rm cm^{-2}$. The effects of these sub-DLA systems and their ionization corrections is discussed in $\S$~\ref{sec:coabund}.

Column densities of a wide range of species were derived using both the weak line limit and apparent optical depth (AOD) technique (Savage and Sembach, 1991).  In some cases a local continuum fit was performed during the AOD column density measurement to improve accuracy.   Where multiple transitions existed, the transitions were combined with a weighted average among selected transitions to created a single optical depth profile for which the column density can be derived. For species with multiple transitions, we rejected strongly blended transitions and transitions in which complete saturation was observed (i.e. pixel values $<$ 0.1) either due to blending or saturation.  A lower limit of column density was adopted for strongly saturated lines (such as \ion{Mg}{2}) based on the corresponding column density from the equivalent width of the transition with the smallest oscillator strength in the weak line limit. The column densities derived from low-ion transitions of the elements C, O, Si, Al, Fe, Mg, and Mn were compiled and converted to logarithmic abundances using the solar abundances of Lodders (2003), and the derived \ion{H}{1} column density. 

Concern about saturation of some species resulted in a separate analysis of the effects of saturation within the ESI spectra using standard COG analysis.  These procedures are detailed in the following section.

\subsection{Saturation Criteria}
\label{sec:saturate}

Since the limited resolution of the ESI spectrograph precludes a definitive measurement of component b values, even when a large number of transitions allow a determination from the curve of growth, we have adopted the conservative assumption that many of our DLA systems will have narrow b values, which would decrease the equivalent width limit
at which saturation effects would become significant.  In the case of such narrow b values, lines which appear to be weak in our ESI spectra could begin to become saturated, implying some of the AOD derived column densities underestimate the actual values.

\begin{figure}[t!]
\epsscale{1.50}
\includegraphics[scale=0.60]{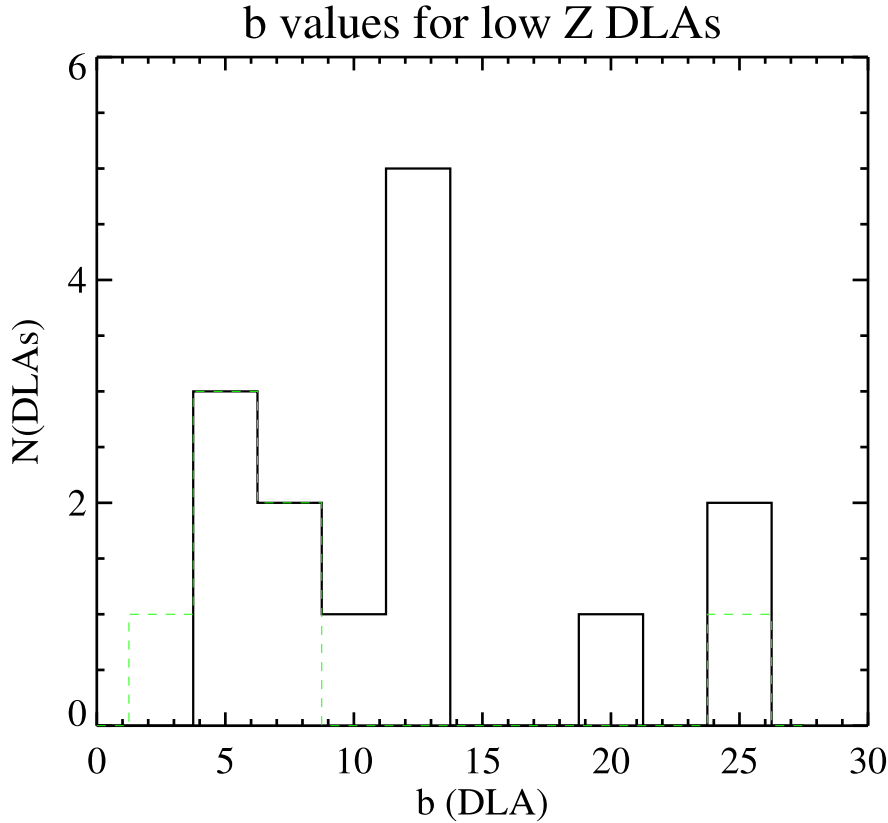}
\caption{
Histogram of best fit b values for our ESI sample (solid black line), based on fitting of a normalized curve of growth for multiple species of varying f-values for those DLAs with non-blended weak lines in several species. The DLA components within the low-metallicity sample of Pettini, et al (2008) is also included (dashed green line). The median b value for our ESI DLA sample is determined to be b=7.5 \kms, and agrees well with the median b value for the Pettini et al sample of b=6.5 \kms.}
\label{coghist}
\end{figure}

Within Pettini et al (2008), three low-metallicity DLA systems were observed with high resolution and included 7 separate velocity components with b values ranging from 3.7 to 25.1 \kms, with a mean b value of 9.24 \kms, and a median b value of 6.5 \kms.  For our ESI sample we have also examined the resulting b values from the best fit to a normalized curve of growth for multiple transitions. When unblended absorption profiles exist over  multiple transitions spanning a range of f-values and equivalent widths, we have been able to determine b values for our ESI spectra, which lie within the range of values reported in Pettini, et al (2008).  Figure \ref{coghist} presents a histogram of measured b values within our ESI sample (solid line), compared to the histogram of component b values within the low metallicity DLAs reported in Pettini et al (2008). Our curve of growth fits showed b values consistent with the range of  reported b values from Pettini, et al (2008), with a mean value of 12.5 \kms, a median value of b=7.5 \kms, and a range of values between 5 \kms $<$ b $<$ 25 \kms.

\begin{figure}[t!]
\epsscale{0.30}
\includegraphics[scale=0.35]{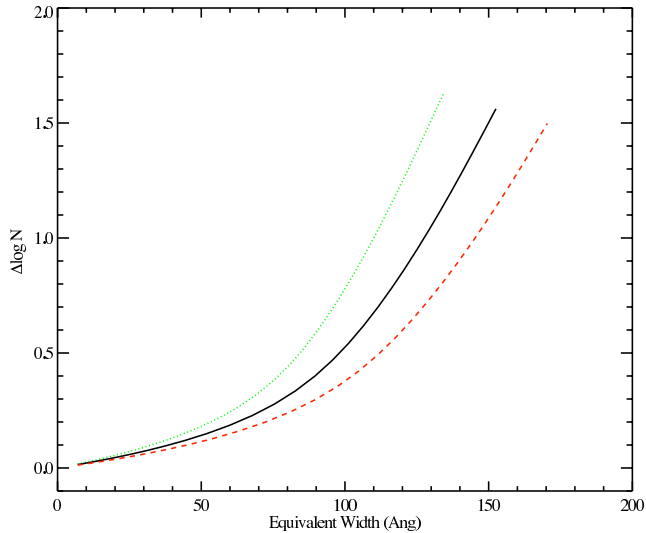}
\caption{
Plot of modeled column density offset $(\Delta \log N \equiv \log N_{tru} 
- \log N_{model}$)
obtained using the AOD technique for b values of 6.5,7.5, and 8.5 \kms (top to bottom), using the ESI instrument parameters from our observing run, with a 0.75 arcsecond slit, on a single OI 1302\AA\ line.  The possibility of low b values of low metallicity DLA systems requires caution in deriving column densities from the moderate resolution spectra of the ESI spectrograph. 
}
\label{satfig}
\end{figure}

Figure \ref{satfig} shows the results of a simulation comparing derived AOD column densities 
for an absorption profile of an OI line at the simulated ESI resolution with 
the "true" column densities, for a DLA system with low b-values consistent with 
some of the low-metallicity DLA systems as measured by Pettini et al (2008). The three lines represent the correction needed to derive the "true" column densities as a function of equivalent widths for b values of 6.5, 7.5, and 8.5 \kms . This plot graphically displays the difficulty in using under-resolved spectra for strong absorption lines, as the actual column densities diverge from the AOD column densities as saturation becomes an issue for lines with equivalent widths greater than 50 m\AA, and by equivalent widths of 130 m\AA, the saturation corrections become too large to allow any reliable determination of column densities (see also Prochaska 2006). 
Fortunately many of the DLA absorption lines in this work are weak,
and have equivalent widths which show that for b values in the range
of 6.5 \kms $<$ b $<$ 8.5 \kms, the lines are either unsaturated or
mildly saturated.  

We do caution, however, that some of the DLAs could have
multiple components with very low Doppler parameter (e.g. b=3 \kms); such systems would require very
large corrections to the reported column densities.  Indeed, the
highly conservative reader may wish to consider our measurements as 
lower limits to the column densities until higher resolution
spectroscopy is obtained.  We wish to emphasize, however, that such
low Doppler parameters have not been observed for the majority of DLA
systems, nor are they expected.  The line-profile fitting of
individual DLAs generally report $b$-values of 5-10 \kms
\citep[e.g.][]{pw96,dz06}.  The only obvious exceptions are the 
subset of DLAs which show molecular (i.e.\ cold) gas
\citep[e.g.][]{nlp+08,jwp+09}.
Second, apparent optical depth analysis have not indicated any
significant `hidden saturation' in DLA profiles \citep{pw96}.
Third, if the gas is a warm neutral medium \citep[$T \approx
7000$K;][]{ksb+09}.
then even ignoring macroscopic motions one predicts b $>$ 3 \kms
for Carbon.  Therefore, we expect few of the systems presented here to
exhibit such low Doppler parameters.

From the results of our simulation shown in Figure \ref{satfig}, we have developed criteria for determining whether our absorption lines should be considered saturated, mildly saturated or non-saturated, and for those transitions which are mildly saturated, we have calculated "saturation corrections" to enable an estimation of the range of possible column densities in the case of narrow lines.  Transitions were considered mildly saturated when their equivalent widths exceed 50 m\AA, but were less than 100 m\AA.  To determine column densities, we have applied corrections to the AOD column density $\delta$N  in the regime from 50-130 m\AA, while in all cases we consider column densities from species  exceeding 130 m\AA\ to be lower limits due to saturation effects.

To correct the column density when the equivalent width is between 50-130 m\AA, we adopt the correction shown in Figure \ref{satfig} for an adopted b value of 7.5  \kms.  Our correction involves adjusting the derived AOD column densities upward to a value midway between the non-adjusted (weak-line limit) column density and the fully corrected column density, with error estimates increased to span the range of possible column densities. By choosing to adjust our adopted column densities upwards (with corresponding increases in our error estimates), we are able to account for the probable range of column densities if the DLA sightline has narrow absorption lines in the range of b=7.5\kms, but we also preserve as a lower limit of the error bar the estimated column density for larger b values.

The result of this procedure is not entirely satisfactory, but represents our best attempt to provide the most information available from our moderate resolution ESI spectra, while recognizing the limitations imposed by the limited resolution. A future study with higher resolution would enable more accurate determination of column densities for the stronger lines, but with this compromise we provide a range of possible values for these column densities. 

\subsection{Sensitivity limit for extremely low-metallicity systems}
\label{sec:lowlimit}

We also computed the lowest metallicities detectable within our program, as it is important to confirm that our search was capable of detecting extremely low-metallicity systems. For a typical value of \ion{H}{1}=20.5, and redshift z=3.0, we can calculate the lowest limiting metallicity for various transitions, given a 4 pixel wide absorption profile close to the instrumental resolution of FWHM~$\approx 57$\kms, 
and a 4 $\sigma$ detection. Table \ref{limitstable} provides an estimate of the lowest possible observable values of metallicity for a number of species, for spectra with $S/N = 10$ and $S/N = 20$. These limits were calculated using parameters for a typical DLA where we adopted values for N(HI) = 20.5, z=3.0, and a metal line profile of 4 pixels, and a 4 $\sigma$ detection. Since our spectra all had values of $S/N$ $\ge$ 20 for most of the free spectral range, the right column of Table \ref{limitstable} describes the limiting sensitivity of our survey to low metallicity systems. 
All of the spectra have \ion{O}{1} and \ion{C}{2} equivalent widths larger than our minimum limiting sensitivity, but in some spectra species such as \ion{Fe}{2} and \ion{Si}{2} approached this limiting minimum value, and therefore for some DLA systems we report upper limits only in these species. The following section describes our results in detail for the complete sample of DLA systems.

\section{Analysis}
\label{results}

We have conducted a uniform analysis of the entire sample of 38 quasars using 
AOD code and apply corrections for line saturation as discussed in $\S$~\ref{sec:saturate}. In cases where the metallicity and elemental abundances are very low, we provide complete details about the DLA system, its detected absorption lines, and the resulting column densities and abundances in individual subsections. For those DLAs not discussed below, the results are included in the summary tables and included in our considerations of the statistics of metal abundances in the later sections of the paper.

\subsection{Results for Individual DLA systems with Low Metallicity}

\subsubsection{SDSS 0955+4116}
\label{sec:firstind}

\begin{figure}[ht]
\begin{minipage}[c]{0.45\linewidth}
\centering

\includegraphics[angle=0,%
width=3.0in]{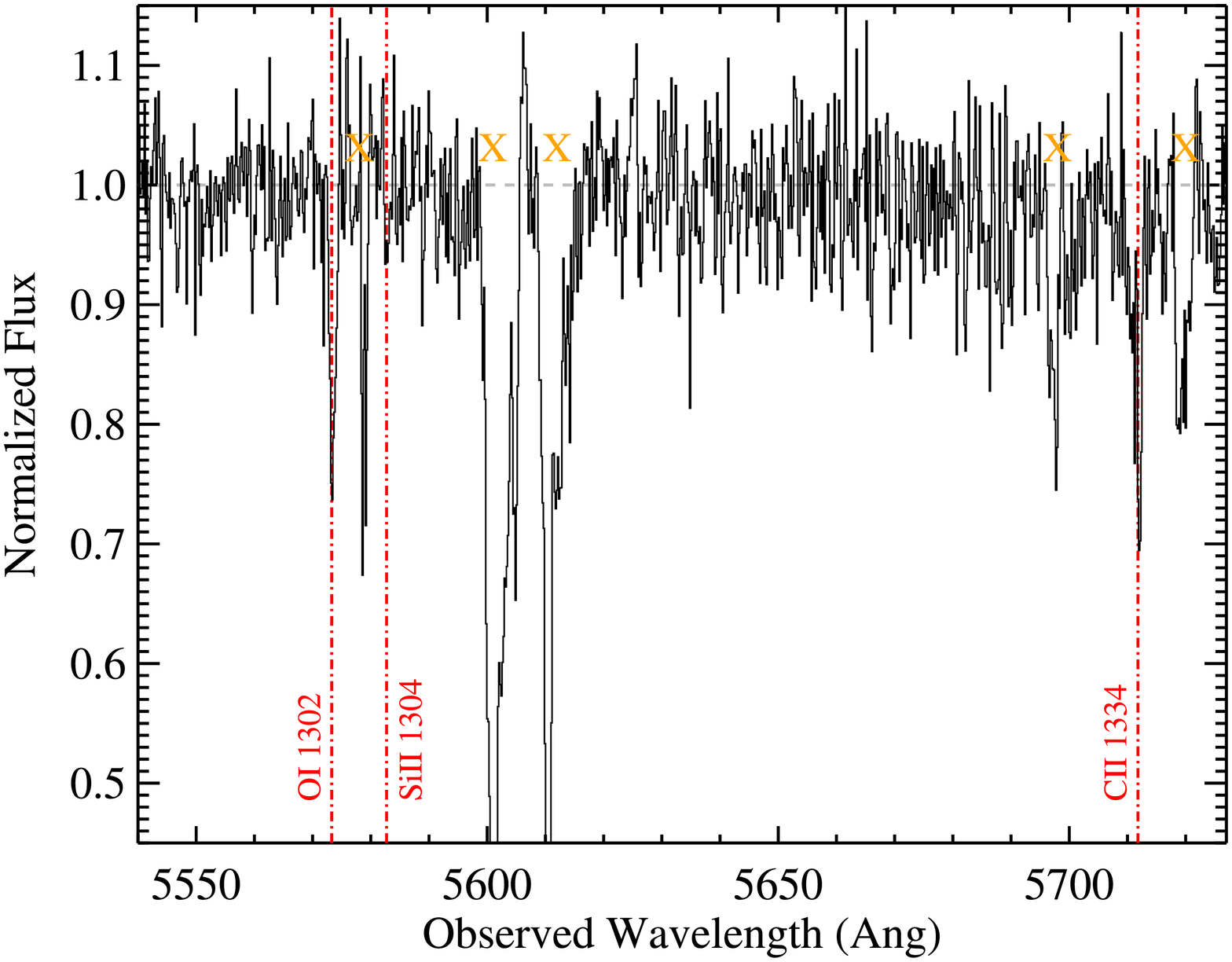}

\includegraphics[width=3.0in,%
angle=0]{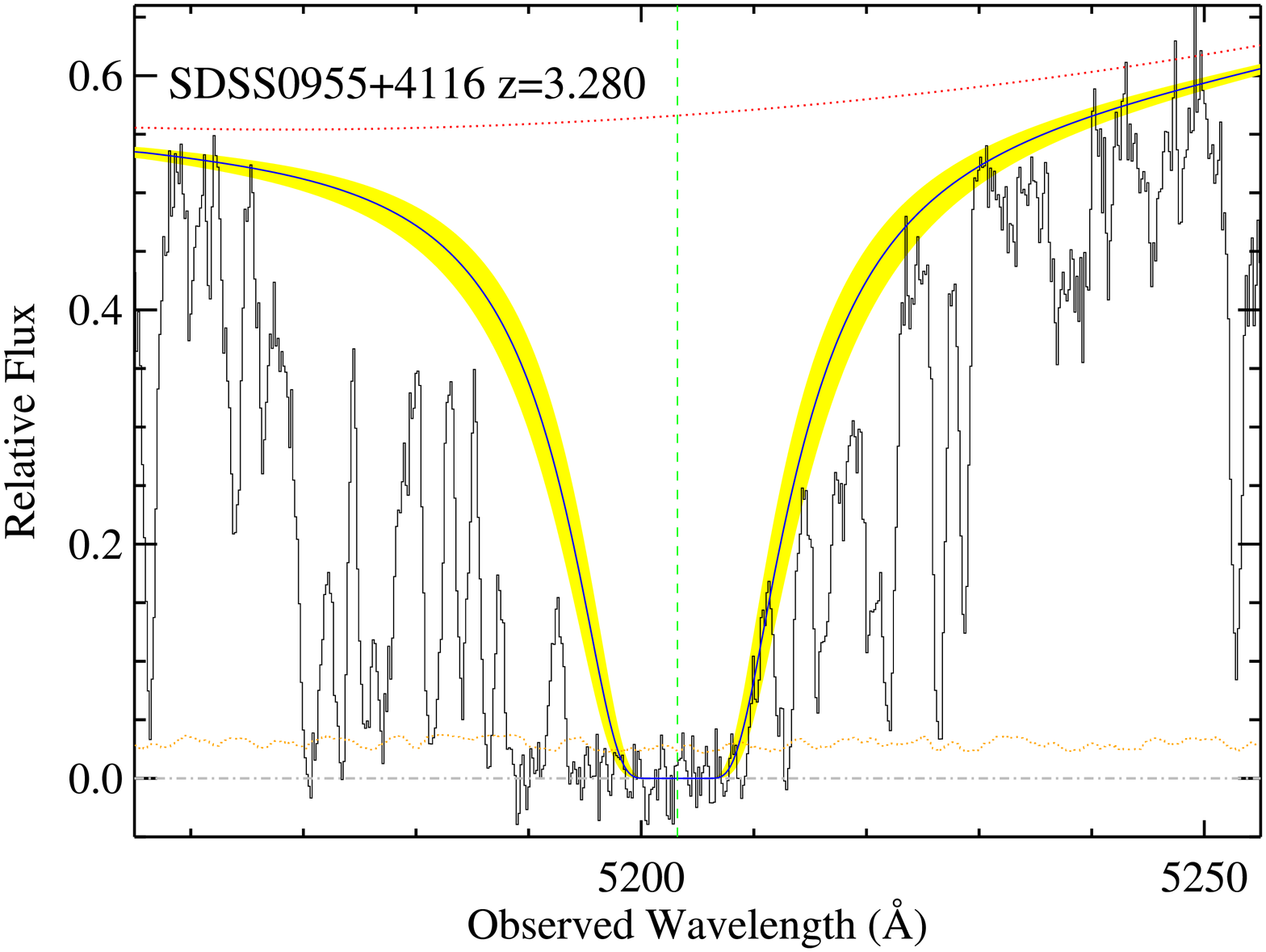}

\end{minipage}
\hspace{0.2cm}
\begin{minipage}[c]{0.45\linewidth}
\centering

\includegraphics[width=2.8in,%
angle=0]{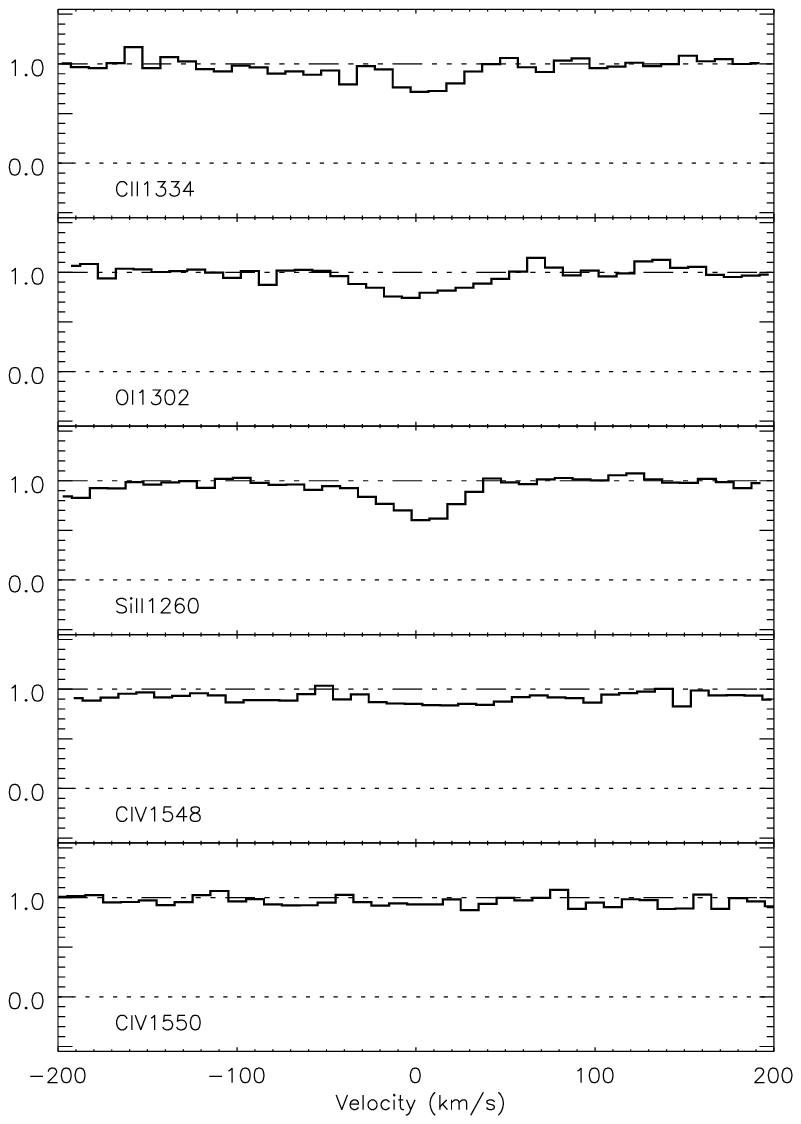}

\end{minipage}

\caption{
Section of our spectrum SDSS0955+4116 showing the lines of \ion{C}{2}~$\lambda$1334, \ion{Si}{2}~$\lambda$1304, and \ion{O}{1}~$\lambda$1302 for the DLA system at z=3.28102 (left), and the \ion{H}{1} damped Lyman alpha profile at the same redshift (right), along with our best fit model (solid line) for log(N(\ion{H}{1})) = 20.10, and one sigma limits to the HI fit (yellow/shaded region).
}
\label{0955+4116fig}
\end{figure}

We present in the left panel of Figure \ref{0955+4116fig} a plot of our Keck/ESI spectra for the DLA system toward SDSS0955+4116,
in the wavelength range which includes important lines of \ion{C}{2}~$\lambda$1334, \ion{Si}{2}~$\lambda$1304, and \ion{O}{1}~$\lambda$1302.  The right panel of Figure \ref{0955+4116fig} shows the \ion{H}{1} Damped \lymana\ profile, and we indicate our best fit model for N(\ion{H}{1})=20.10, along with the one $\sigma$ range of N(\ion{H}{1}) in the shaded region. A curve of growth analysis of the sightline suggests a b-value of 7.5 \kms, although all of the lines are weak, and an accurate measurement of the $b$ value is difficult. 
Since all species have equivalent widths less than 100 m\AA, we have corrected the derived AOD column densities according to the prescription in $\S$~\ref{sec:saturate}. The results of our adopted column densities for the various species are presented in Table \ref{tab:J0955+4116}. 
This DLA also is observed to have relatively weak absorption lines for both \ion{C}{2} and \ion{O}{1}, and is one of five DLA systems from our survey where both equivalent widths of these species are less than 130 m\AA, which forms the weak line subsample used in Figure \ref{coratplot}. 
In Table \ref{tab:abunds} we summarize the abundances for this DLA, for which we derive values of [C/H] = -2.82 $\pm$ 0.14, [O/H]=-2.82 $\pm$ 0.10, [Si/H]=-2.75 $\pm$ 0.18.  The observed value of [C/H] is the second lowest of our sample, while the observed value of [O/H] is the sixth lowest of our sample. Non-detections of the species \ion{Al}{2} and \ion{Fe}{2} give upper limits of [Al/H] $<$ -2.74, and [Fe/H] $<$ -2.30.

\subsubsection{SDSS 1001+0343}

\begin{figure}[ht]
\begin{minipage}[c]{0.45\linewidth}
\centering

\includegraphics[angle=0,%
width=3.0in]{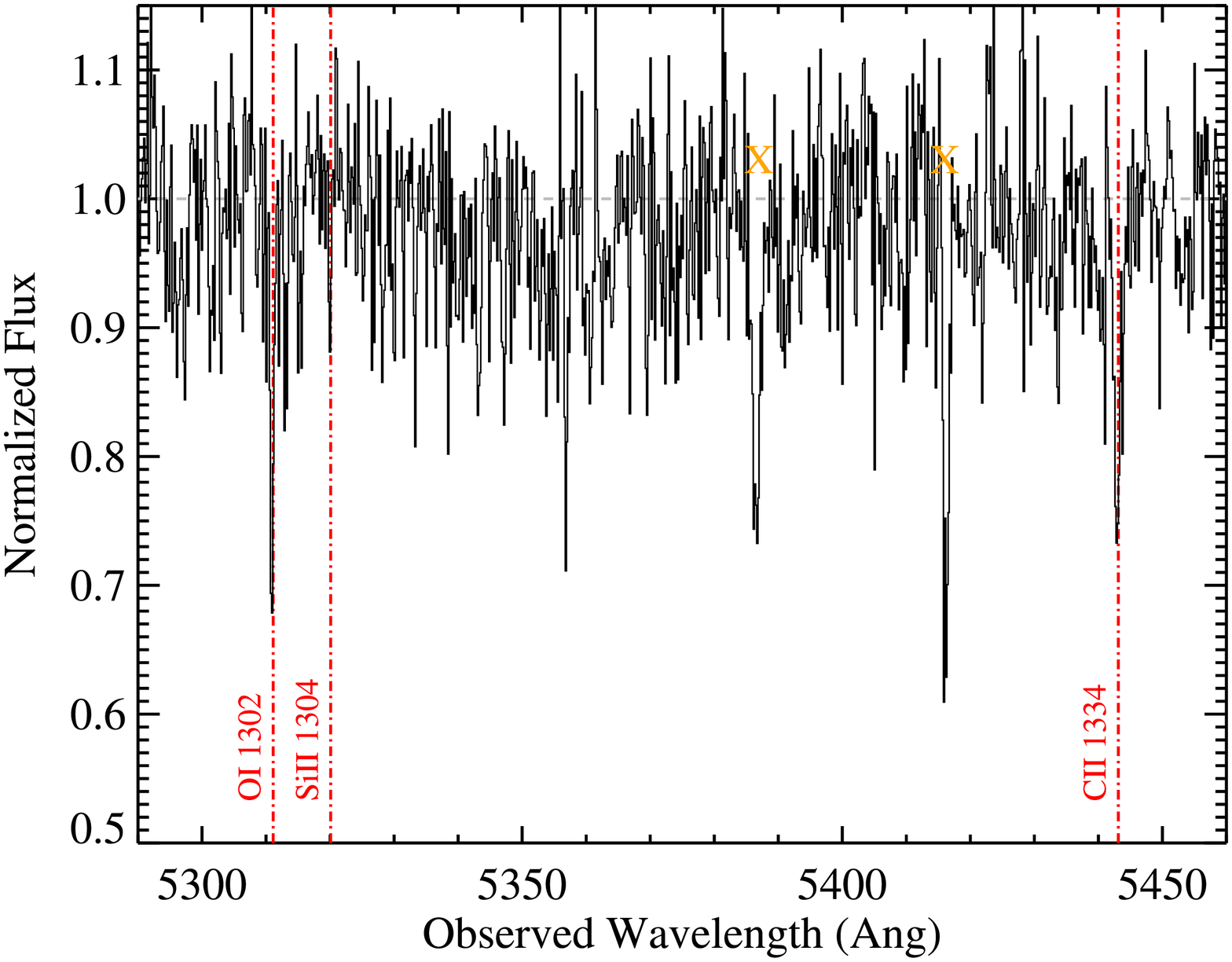}

\includegraphics[width=3.0in,%
angle=0]{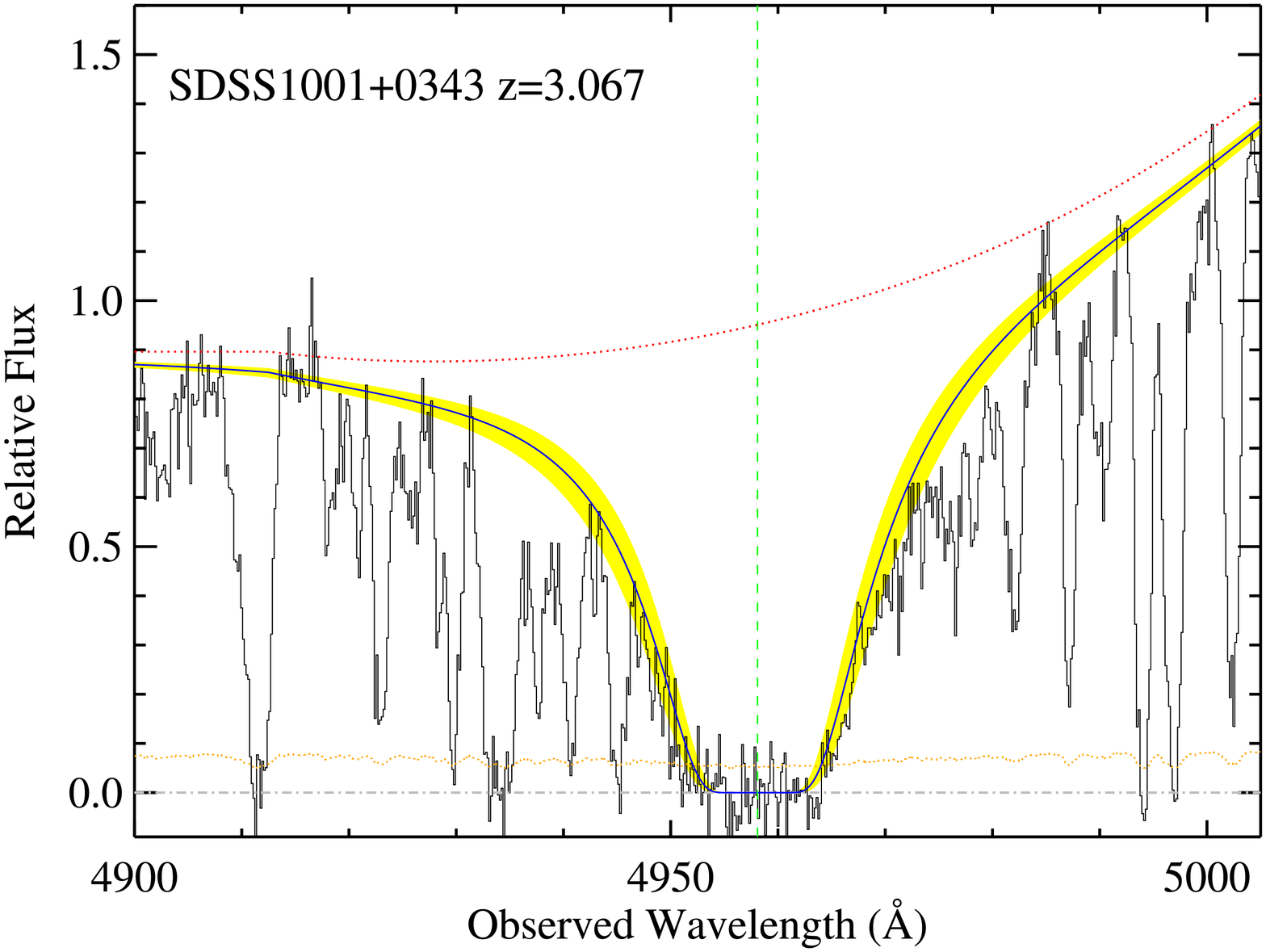}

\end{minipage}
\hspace{0.2cm}
\begin{minipage}[c]{0.45\linewidth}
\centering

\includegraphics[width=2.8in,%
angle=0]{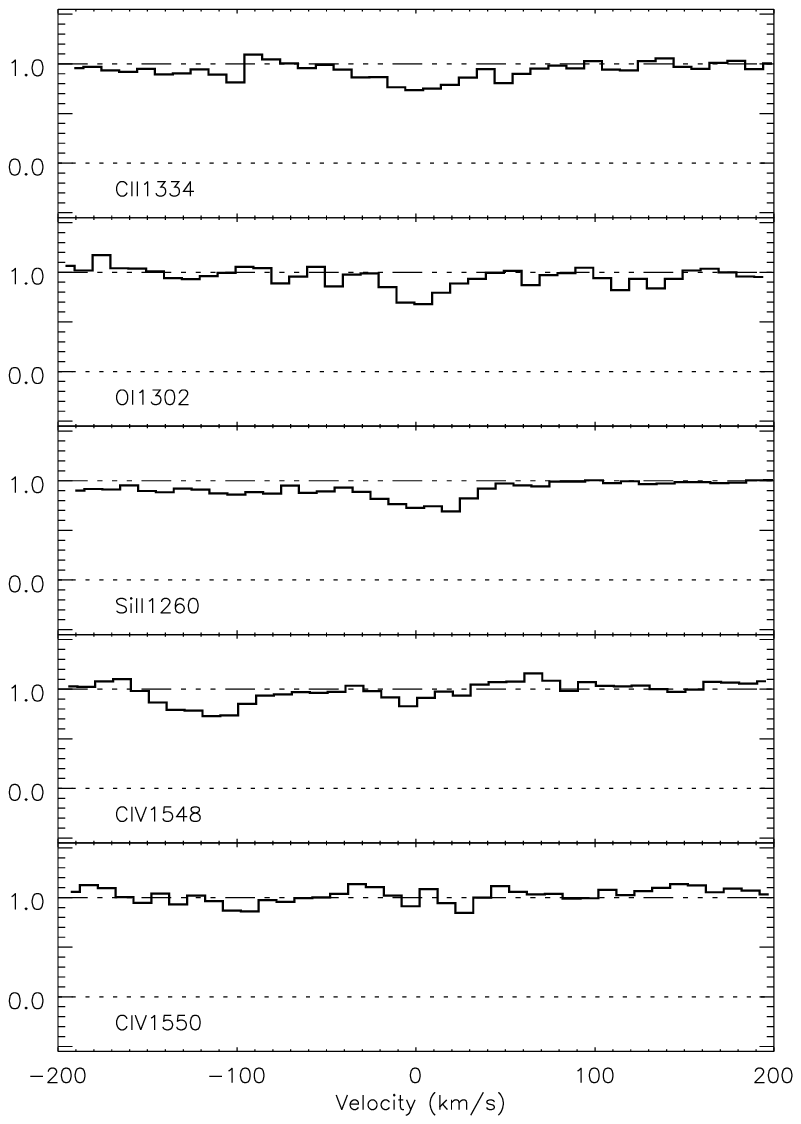}

\end{minipage}
\caption{
Section of our spectrum SDSS1001+0343 showing the lines of \ion{C}{2}~$\lambda$1334, \ion{Si}{2}~$\lambda$1304, and \ion{O}{1}~$\lambda$1302 for the DLA system at z=3.07867 (left), and the \ion{H}{1} damped Lyman alpha profile at the same redshift (right), along with our best fit model (solid line) for log(N(\ion{H}{1})) = 20.15, and one sigma limits to the HI fit (yellow/shaded region).
}
\label{1001+0343fig}
\end{figure}

The DLA system toward SDSS1001+0343 is one of our lowest metallicity systems, and we present in the left panel of Figure \ref{1001+0343fig} a plot of our spectra for the DLA system of SDSS1001+0343, as in $\S$~\ref{sec:firstind}, where the right panel of Figure \ref{1001+0343fig} shows the \ion{H}{1} Damped \lymana\ profile, with the best fit model for N(\ion{H}{1})=20.15. A curve of growth analysis of the available transitions for this sightline suggests a b-value of 7.5 \kms, and all transitions all have equivalent widths less than 80 m\AA, where possible saturation effects are limited. We have adjusted the stronger lines for possible effects of saturation as described above, typically at a level of 0.1 to 0.2 dex. The results of our adopted column densities for the various species are presented in Table \ref{tab:J1001+0343}. In Table \ref{tab:abunds} we summarize the abundances for this DLA, for which we derive values of [C/H] = -2.85 $\pm$ 0.13, [O/H]=-2.93 $\pm$ 0.08,  [Si/H] = -2.94 $\pm$ 0.12, [Al/H] $<$ -2.82 $\pm$ 0.13, and an upper limit of Fe/H] $<$ -2.32 . This DLA also is observed to have very weak absorption lines for both \ion{C}{2} and \ion{O}{1}, and is one of five DLA systems from our survey where both equivalent widths of these species are less than 130 m\AA, which forms the weak line subsample used in Figure \ref{coratplot}. The observed value of [C/H] is the lowest in our sample, while the [O/H] is the third lowest of the sample.

\subsubsection{SDSS 1219+1603}

\begin{figure}[ht]
\begin{minipage}[c]{0.45\linewidth}
\centering

\includegraphics[angle=0,%
width=3.0in]{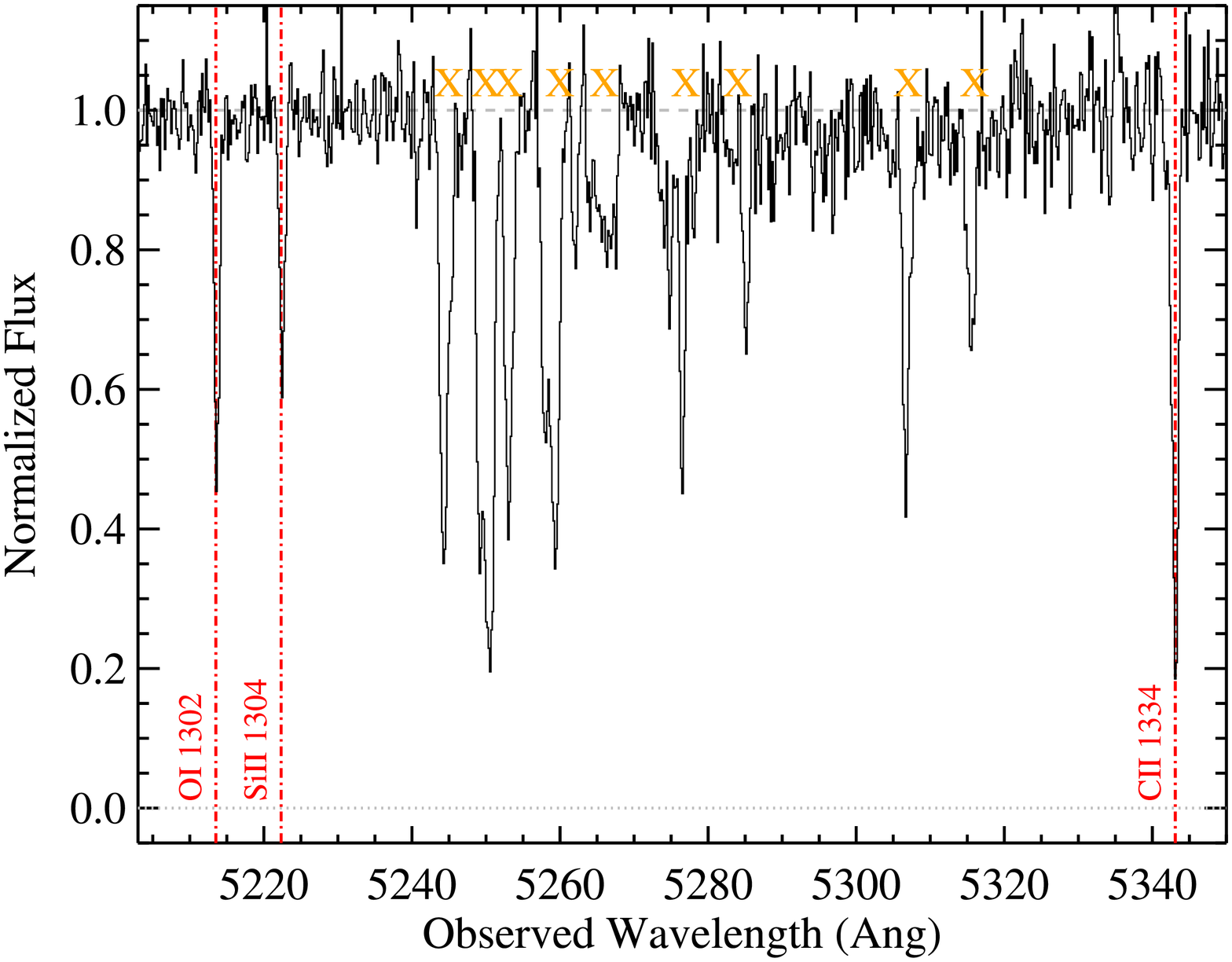}

\includegraphics[width=3.0in,%
angle=0]{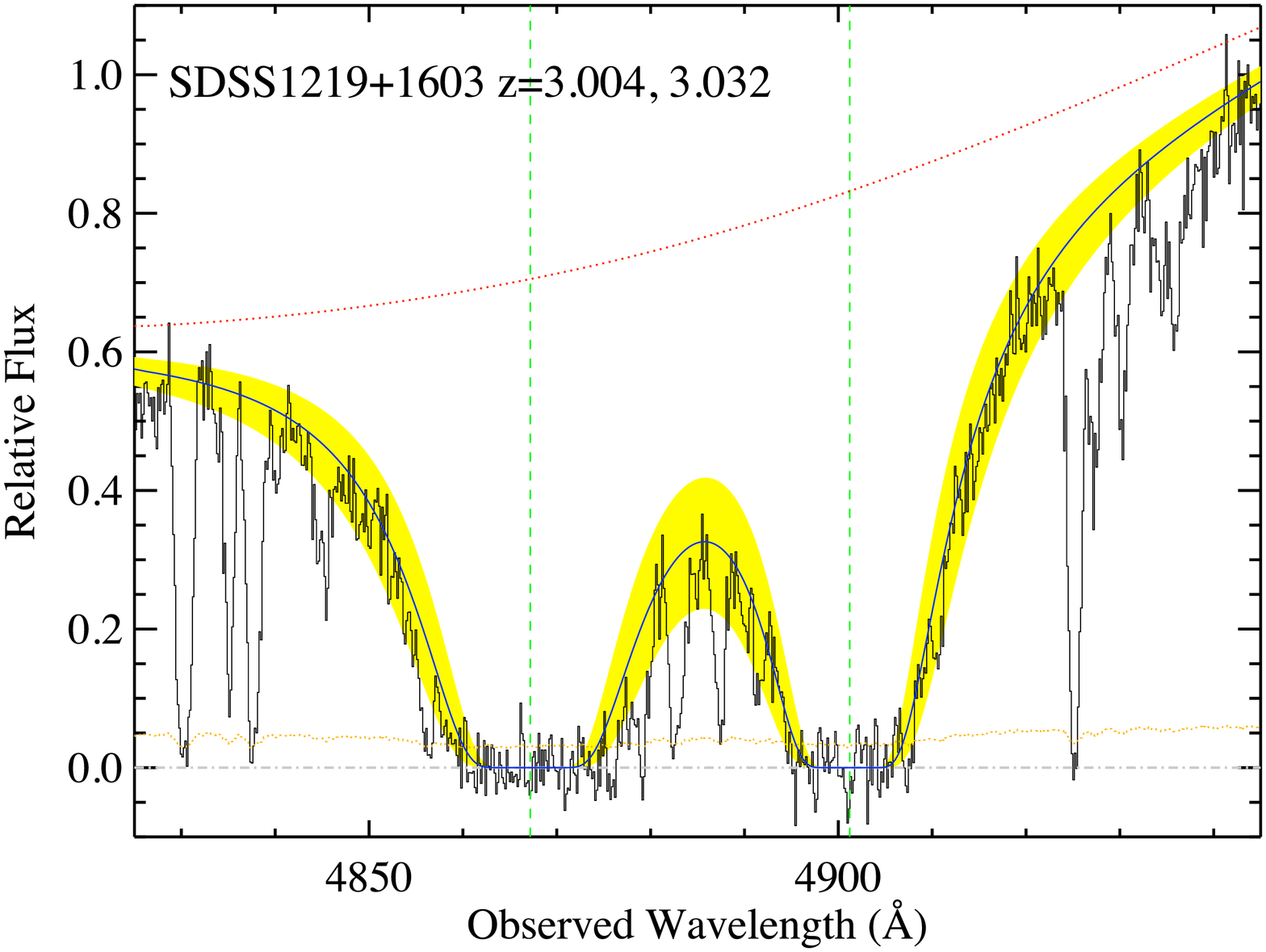}

\end{minipage}
\hspace{0.2cm}
\begin{minipage}[c]{0.45\linewidth}
\centering

\includegraphics[width=2.8in,%
angle=0]{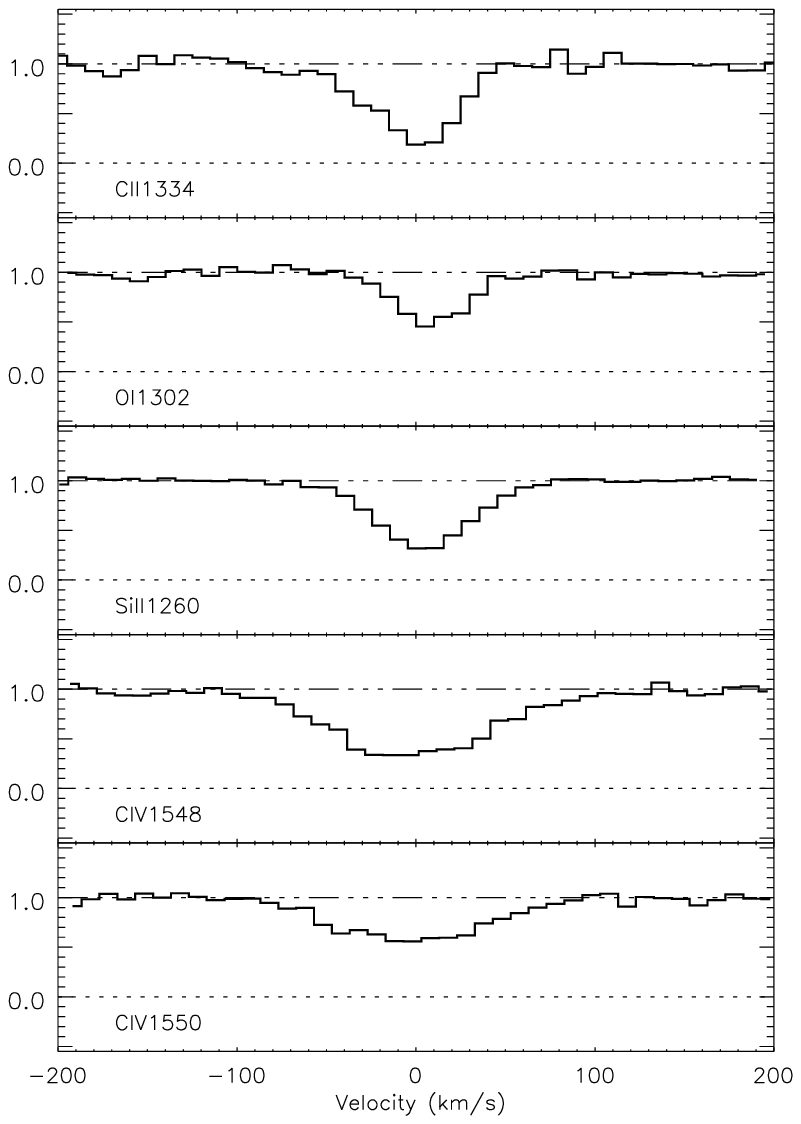}

\end{minipage}
\caption{
Section of our spectrum SDSS1219+1603 showing the lines of \ion{C}{2}~$\lambda$1334, \ion{Si}{2}~$\lambda$1304, and \ion{O}{1}~$\lambda$1302 for the DLA system at z=3.00372 (left), and the \ion{H}{1} damped Lyman alpha profile at the same redshift (right), along with our best fit model (solid line) for log(N(\ion{H}{1})) = 20.35, and one sigma limits to the HI fit (yellow/shaded region). A smaller sub-DLA with log(N(\ion{H}{1})) = 20.15 appears adjacent to this DLA at z=3.032.
}
\label{1219+1603fig}
\end{figure}

Figure \ref{1219+1603fig}  shows details of the spectrum of the DLA system of SDSS1219+1603, with two panels showing the spectrum near key transitions of \ion{C}{2}, \ion{Si}{2}, and \ion{O}{1} (left), and the damped \ion{H}{1} profile (right)
as in $\S$~\ref{sec:firstind}. Our best fit model for N(\ion{H}{1})=20.35, and the SDSS1219+1603 DLA at z=3.00372 appears adjacent to a sub-DLA with N(\ion{H}{1})=20.15 at z=3.032.  The results of our adopted column densities for the various species are presented in Table \ref{tab:J1219+1603}. In Table \ref{tab:abunds} we summarize the abundances for this DLA, for which we derive values of  [O/H]=-2.59 $\pm$ 0.34, [Si/H]=-2.08 $\pm$ 0.15, [Fe/H]=-2.09 $\pm$ 0.10, and we report only lower limits for the metallicities of [C/H] $>$ -2.59, and [Al/H] $>$ -2.29 due to strong absorption of both species with equivalent widths exceeding 130 m\AA. The SDSS1219+1603 DLA shows metal abundances at the high end of our sample, although much less than the mean [X/H] = -1.52 from the survey of Prochaska et al (2003).

\subsubsection{SDSS 1305+2902}

\begin{figure}[ht]
\begin{minipage}[c]{0.45\linewidth}
\centering

\includegraphics[angle=0,%
width=3.0in]{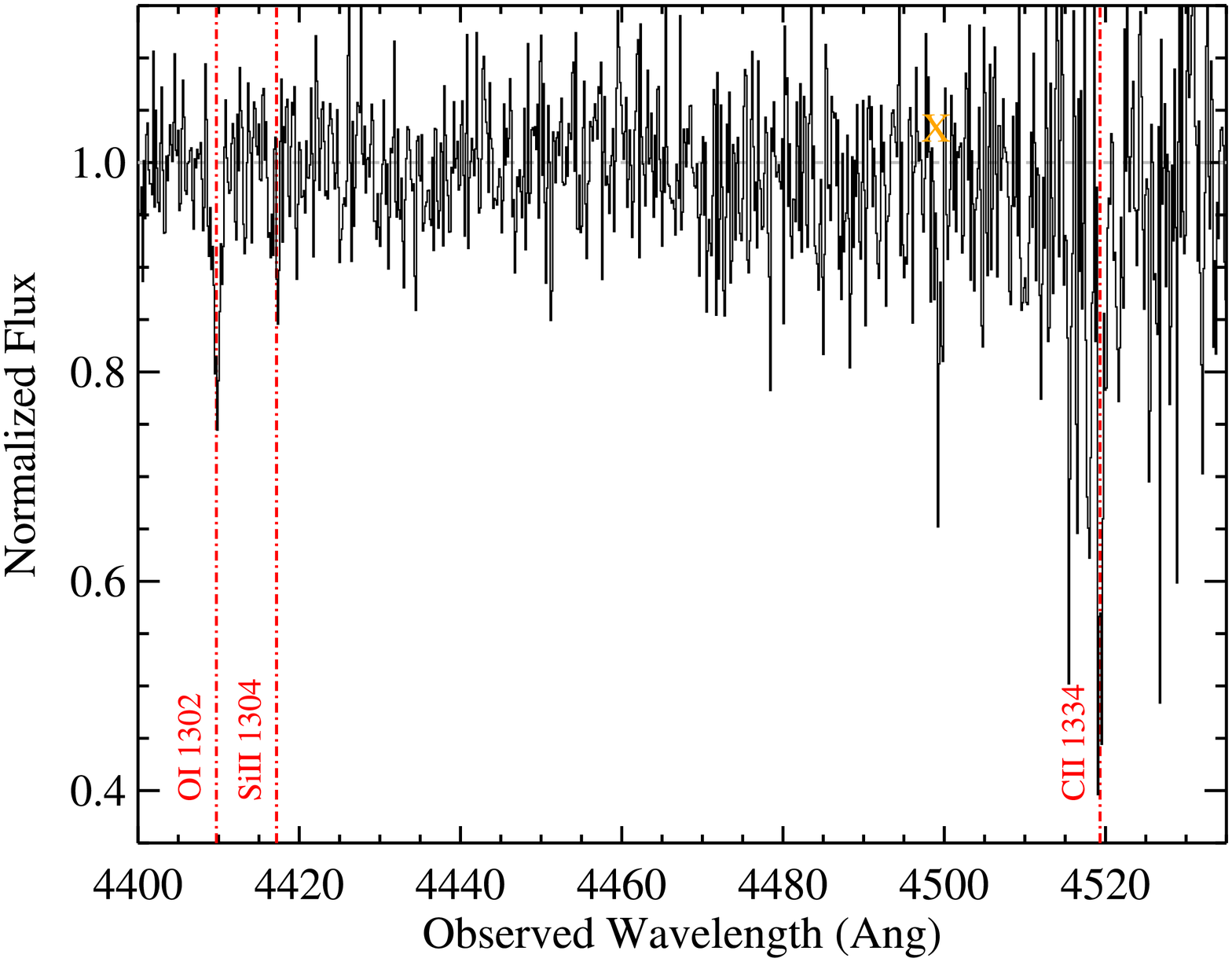}

\includegraphics[width=3.0in,%
angle=0]{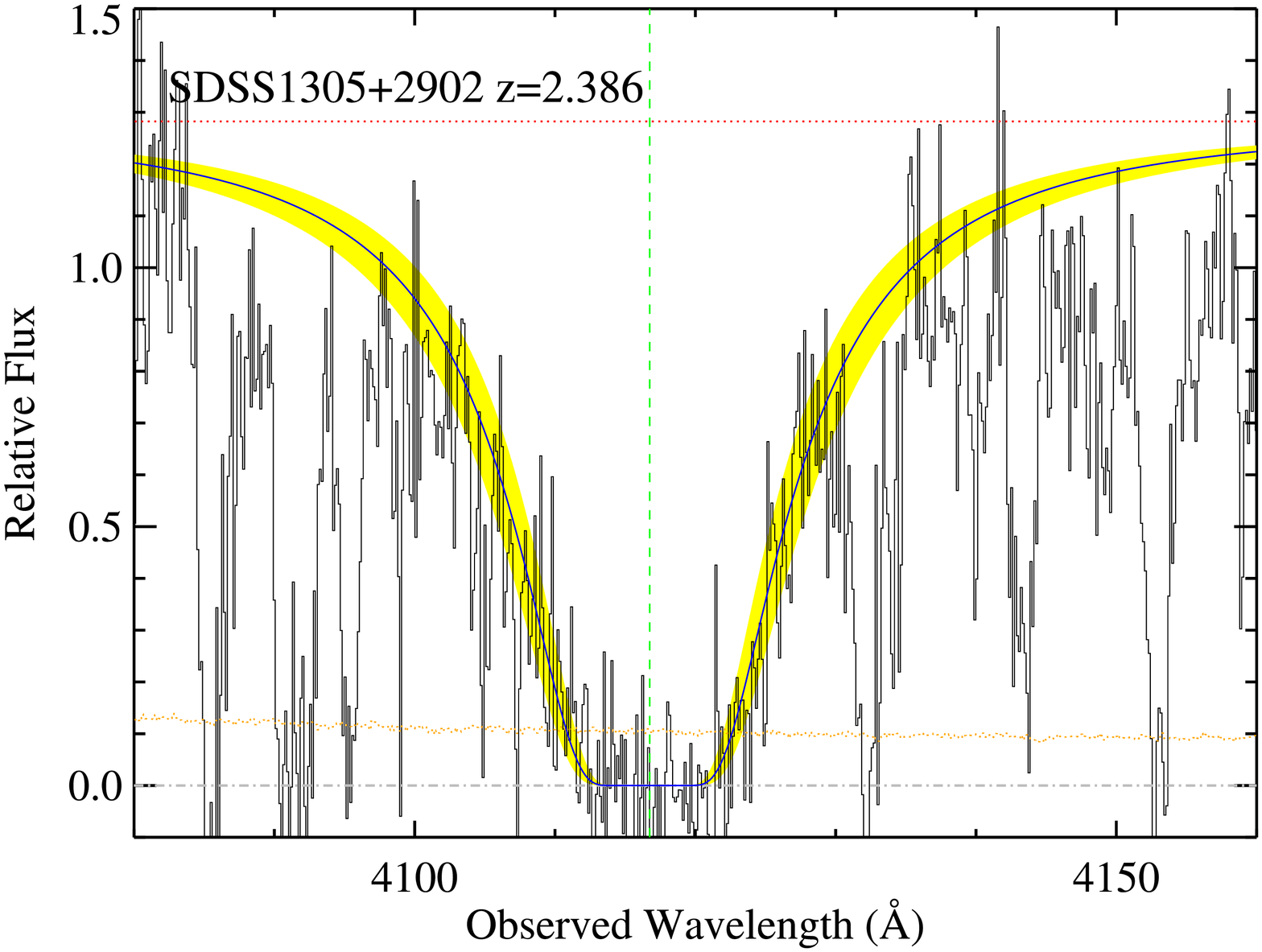}

\end{minipage}
\hspace{0.2cm}
\begin{minipage}[c]{0.45\linewidth}
\centering

\includegraphics[width=2.8in,%
angle=0]{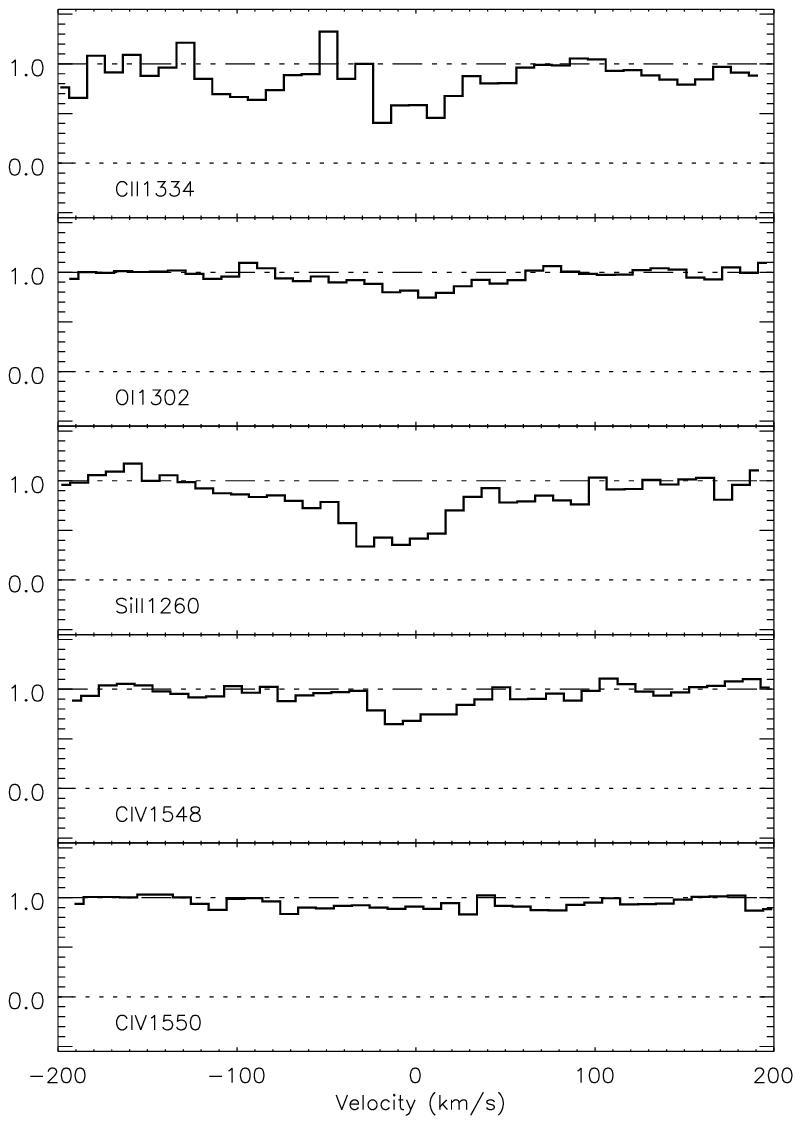}

\end{minipage}
\caption{
Section of our spectrum SDSS1305+2902 showing the lines of \ion{C}{2}~$\lambda$1334, \ion{Si}{2}~$\lambda$1304, and \ion{O}{1}~$\lambda$1302 for the DLA system at z=2.38645 (left), and the \ion{H}{1} damped Lyman alpha profile at the same redshift (right), along with our best fit model (solid line) for log(N(\ion{H}{1})) = 20.25, and one sigma limits to the HI fit (yellow/shaded region).
}
\label{1305+2902fig}
\end{figure}

Figure \ref{1305+2902fig}  shows details of the spectrum of the DLA system of SDSS1305+2902, with two panels showing the spectrum near key transitions of \ion{C}{2}, \ion{Si}{2}, and \ion{O}{1} (left), and the damped \ion{H}{1} profile (right)
as in $\S$~\ref{sec:firstind}. Our best fit model for N(\ion{H}{1})=20.25,  which places this system very close to the lower limit of log(N(\ion{H}{1})) = 20.3 that provides self-shielding in a DLA system. A curve of growth analysis for the SDSS 1305+2902 DLA suggests a b-value of approximately 10.0 \kms, and some saturation corrections were applied for lines with strengths in the 50-130 m\AA\ range. The resulting correction for the \ion{C}{2} column density was quite large, and gives us only a very uncertain estimate of its column to within 0.45 dex, but we include the \ion{C}{2} column density to allow for comparison with the \ion{O}{1}, which had a smaller correction of 0.13 dex corresponding to its weaker absorption of 70 m\AA\ 
equivalent width.  Absorption lines of the species \ion{Si}{2} and \ion{Fe}{2} both included detections of absorption with weak transitions with equivalent width less than 50 m\AA\ which were not saturated, and therefore we have the AOD adopted column densities for these species using only very small corrections for possible saturation. The results of our adopted column densities for the various species are presented in Table \ref{tab:J1305+2902}. In Table \ref{tab:abunds} we summarize the abundances for this DLA, for which we derive values of [C/H] = -2.46 $\pm$ 0.45, [O/H]=-2.90 $\pm$ 0.12, [Si/H]=-2.54, $\pm$ 0.10, [Al/H] = -2.83 $\pm$ 0.10, and [Fe/H] = -2.79 $\pm$ 0.13 . This DLA has reported column densities for both \ion{C}{2} and \ion{O}{1}, and is one of five DLA systems from our survey where both equivalent widths of these species are less than 130 m\AA, which forms the "weak line" subsample used in Figure \ref{coratplot}, and has the fourth lowest measured value of [O/H] in our sample. The SDSS 1305+2902 DLA is also unique in that it has a complete sample of metallicities for the elements C, O, Si, Fe, and Al, which despite the very large uncertainty in the estimated [C/H] metallicity, helps constrain the relative metal abundances in the low metallicity DLA gas.

\subsubsection{SDSS 1358+0349}

\begin{figure}[ht]
\begin{minipage}[c]{0.45\linewidth}
\centering

\includegraphics[angle=0,%
width=3.0in]{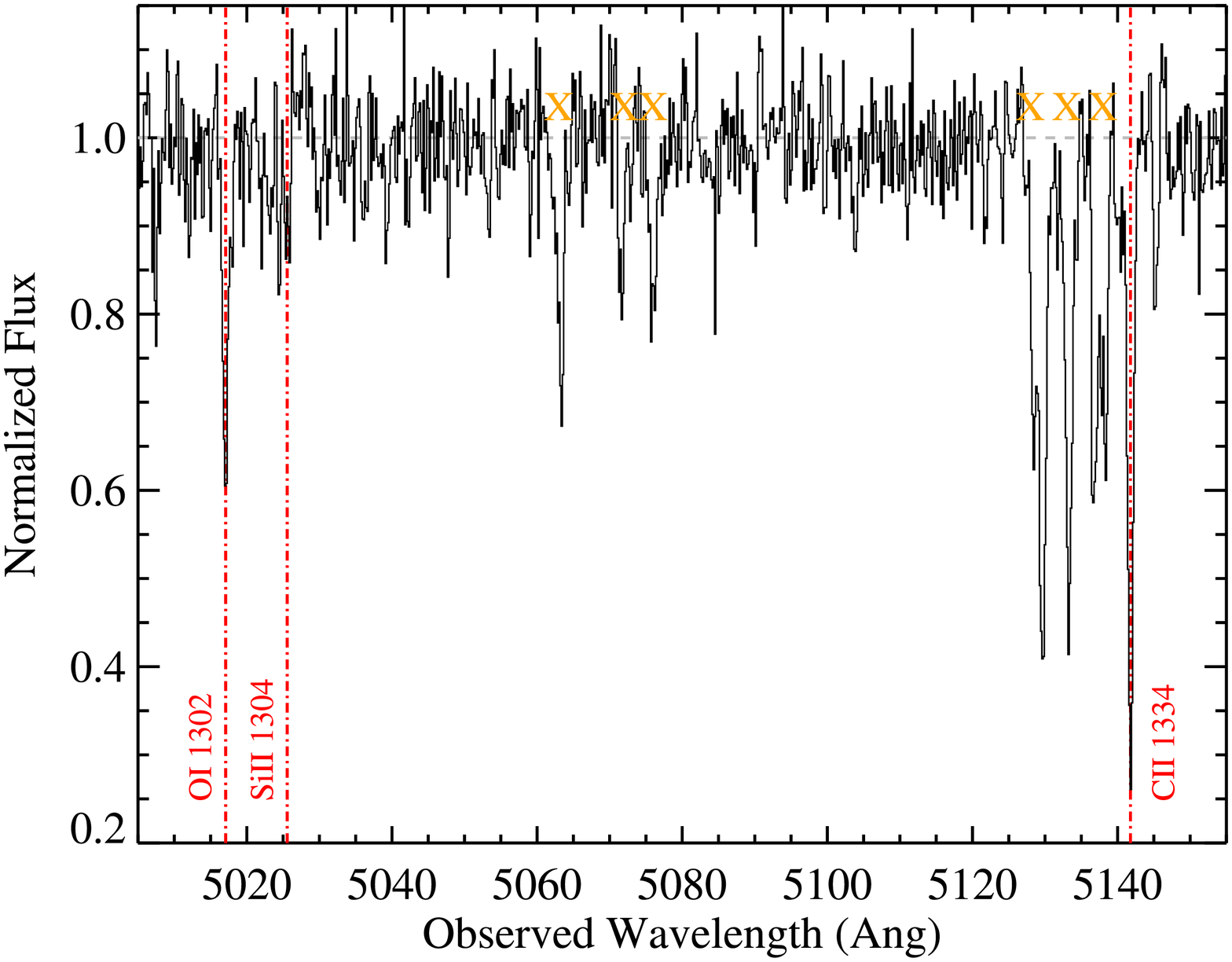}

\includegraphics[width=3.0in,%
angle=0]{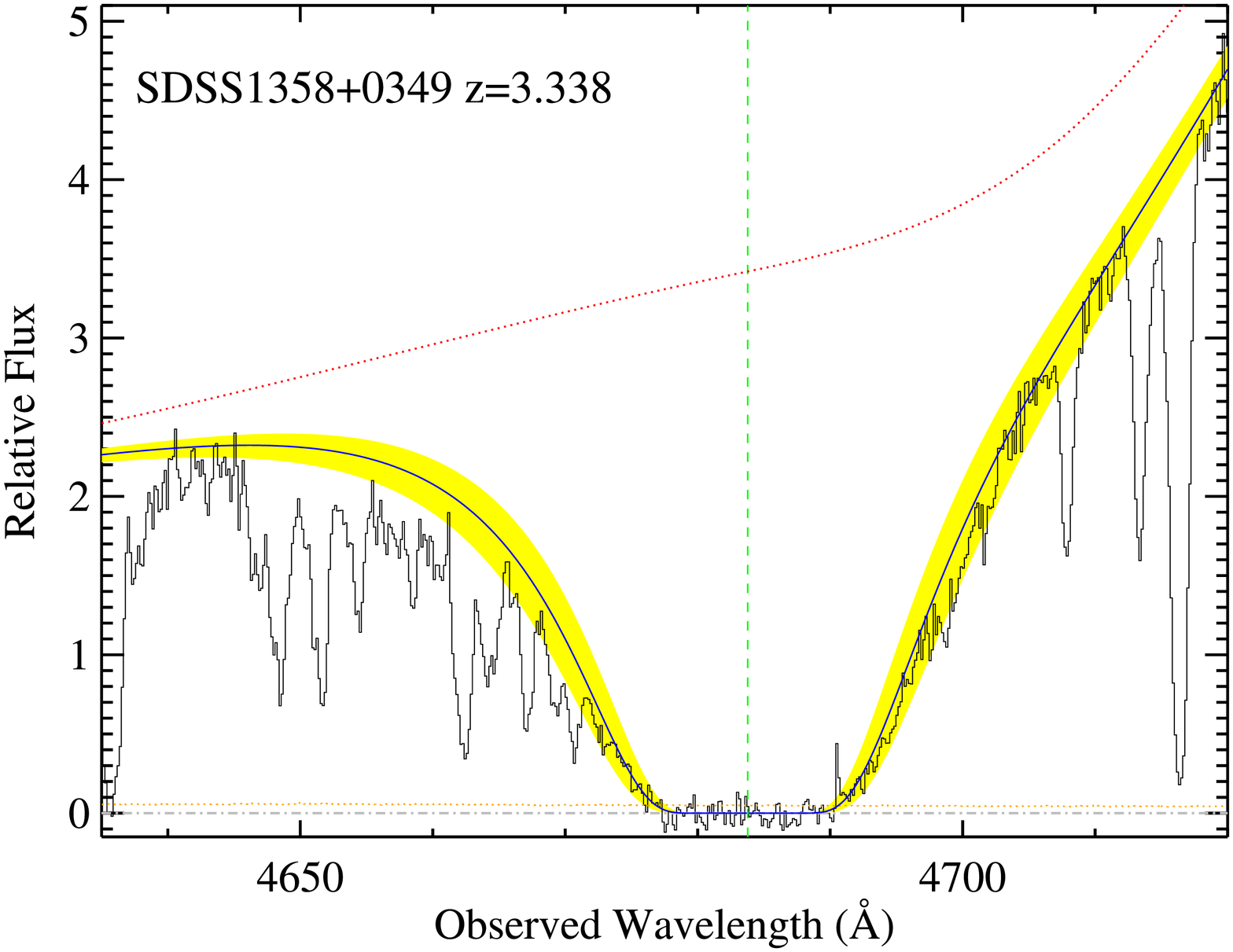}

\end{minipage}
\hspace{0.2cm}
\begin{minipage}[c]{0.45\linewidth}
\centering

\includegraphics[width=2.8in,%
angle=0]{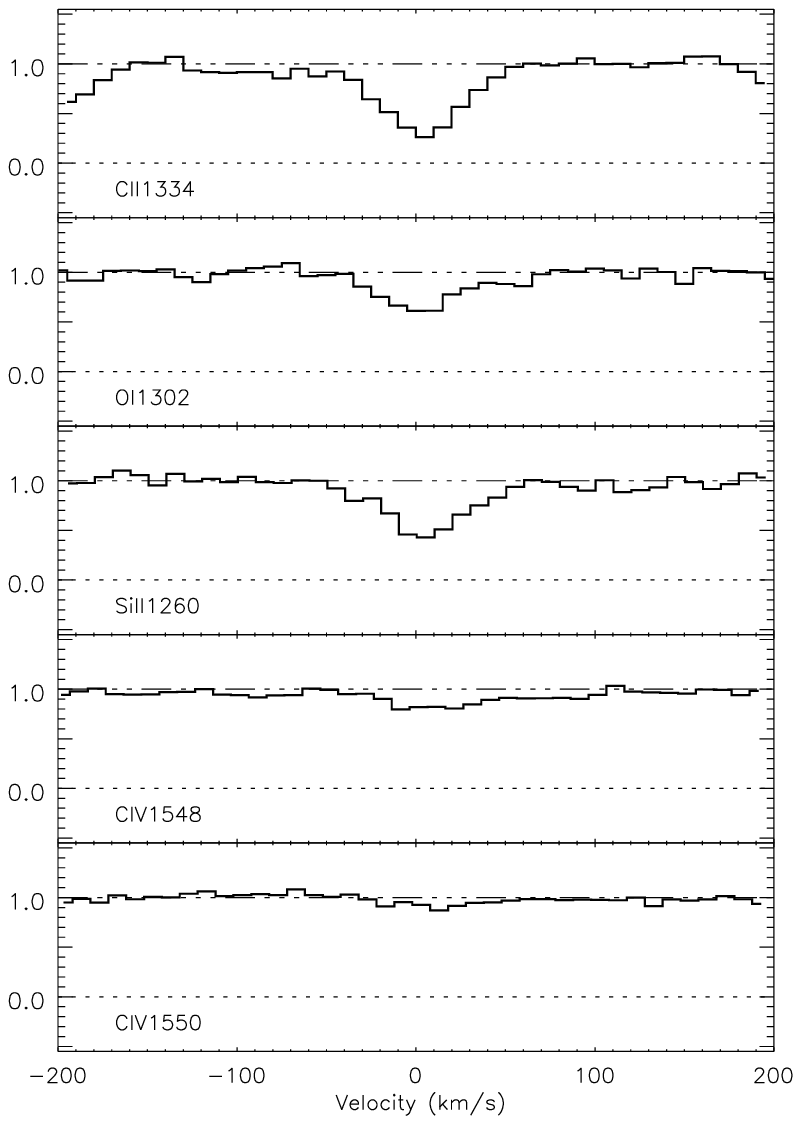}

\end{minipage}
\caption{
Sectio/n of our spectrum SDSS1358+0349 showing the lines of \ion{C}{2}~$\lambda$1334, \ion{Si}{2}~$\lambda$1304, and \ion{O}{1}~$\lambda$1302 for the DLA system at z=2.8528 (left), and the \ion{H}{1} damped Lyman alpha profile at the same redshift (right), along with our best fit model (solid line) for log(N(\ion{H}{1})) = 20.50, and one sigma limits to the HI fit (yellow/shaded region).
}
\label{1358+0349fig}
\end{figure}

Figure \ref{1358+0349fig}  shows details of the spectrum of the DLA system of SDSS 1358+0349, with two panels showing the spectrum near key transitions of \ion{C}{2}, \ion{Si}{2}, and \ion{O}{1} (left), and the damped \ion{H}{1} profile (right)
as in $\S$~\ref{sec:firstind}. Our best fit model for N(\ion{H}{1})=20.50, placing this system well above the DLA threshold of log(N(\ion{H}{1})) = 20.3. A curve of growth fit to the transitions for the SDSS 1358+0349 DLA suggests an underlying b-value of approximately 12.5 \kms, and we have made some allowance for saturation within the regime of 50-100 m\AA, which includes the transitions for \ion{Si}{2} and \ion{O}{1}.  
The column densities of the stronger transitions \ion{Si}{2} and \ion{O}{1} have been corrected to account for the effects of saturation at the level of 0.17 and 0.25 dex, respectively. Sufficient transitions exist with weaker absorption at equivalent widths $<$ 70 m\AA\ for the species of \ion{Fe}{2}, \ion{Si}{2}, and \ion{Al}{2}, and therefore these species have small estimated saturation, which are close to the experimental errors. The results of our adopted column densities for the various species are presented in Table \ref{tab:J1358+0349}. In Table \ref{tab:abunds} we summarize the abundances for this DLA, for which we derive values of [O/H]=-2.88 $\pm$ 0.25, [Si/H]=-2.81 $\pm$ 0.17, and [Al/H]=-2.87 $\pm$ 0.08, and [Fe/H] = -3.03 $\pm$ 0.05. We report only a lower limit for [C/H] $>$ -2.58, due to the very strong absorption of \ion{C}{2} which has an equivalent width of 194 m\AA, The observed value of [O/H]=-2.88 for the SDSS 1358+0349 DLA, even allowing for possible saturation, is the fifth lowest of our sample.

\subsubsection{SDSS 1358+6522}

\begin{figure}[ht]
\begin{minipage}[c]{0.45\linewidth}
\centering

\includegraphics[angle=0,%
width=3.0in]{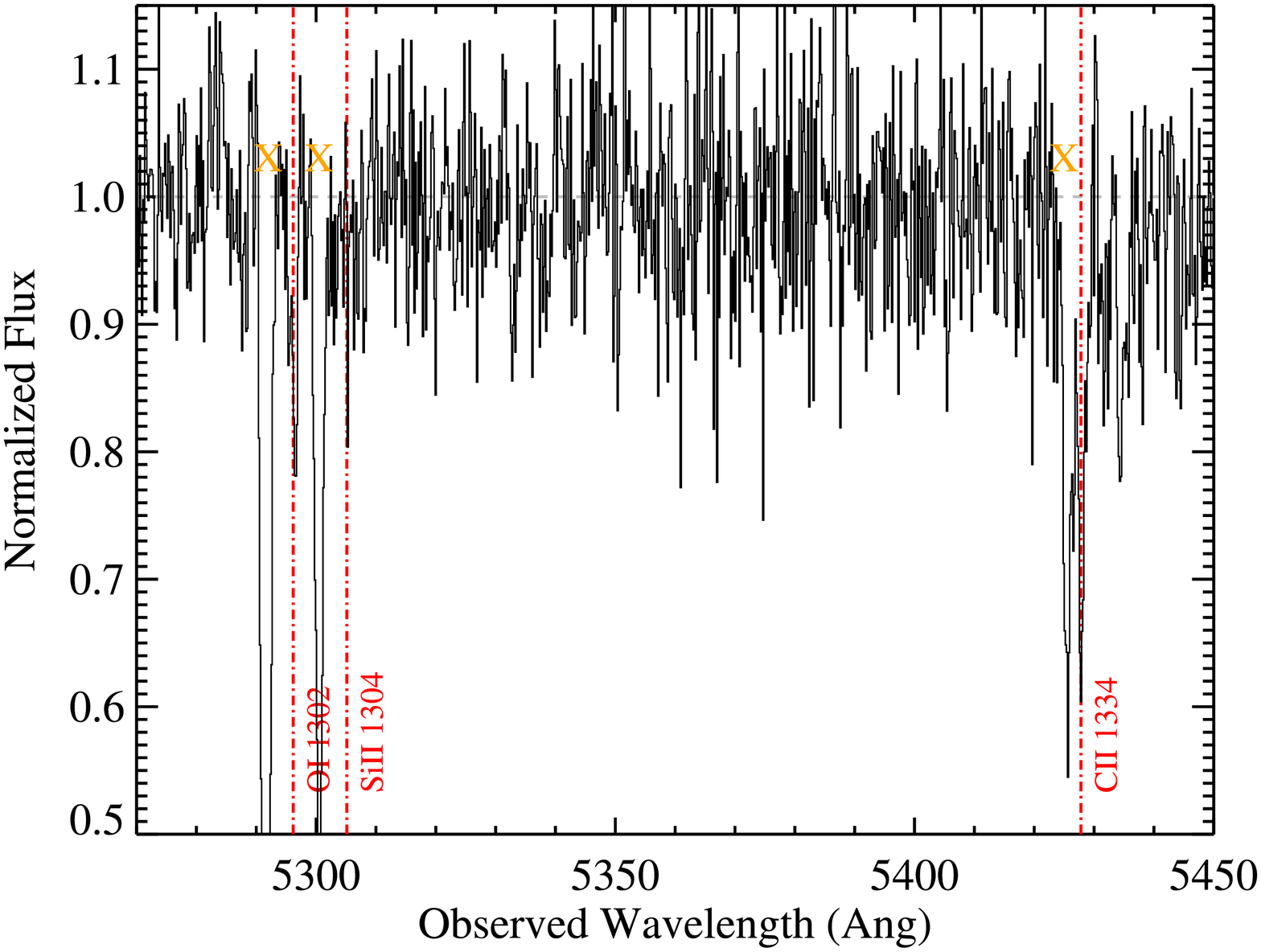}

\includegraphics[width=3.0in,%
angle=0]{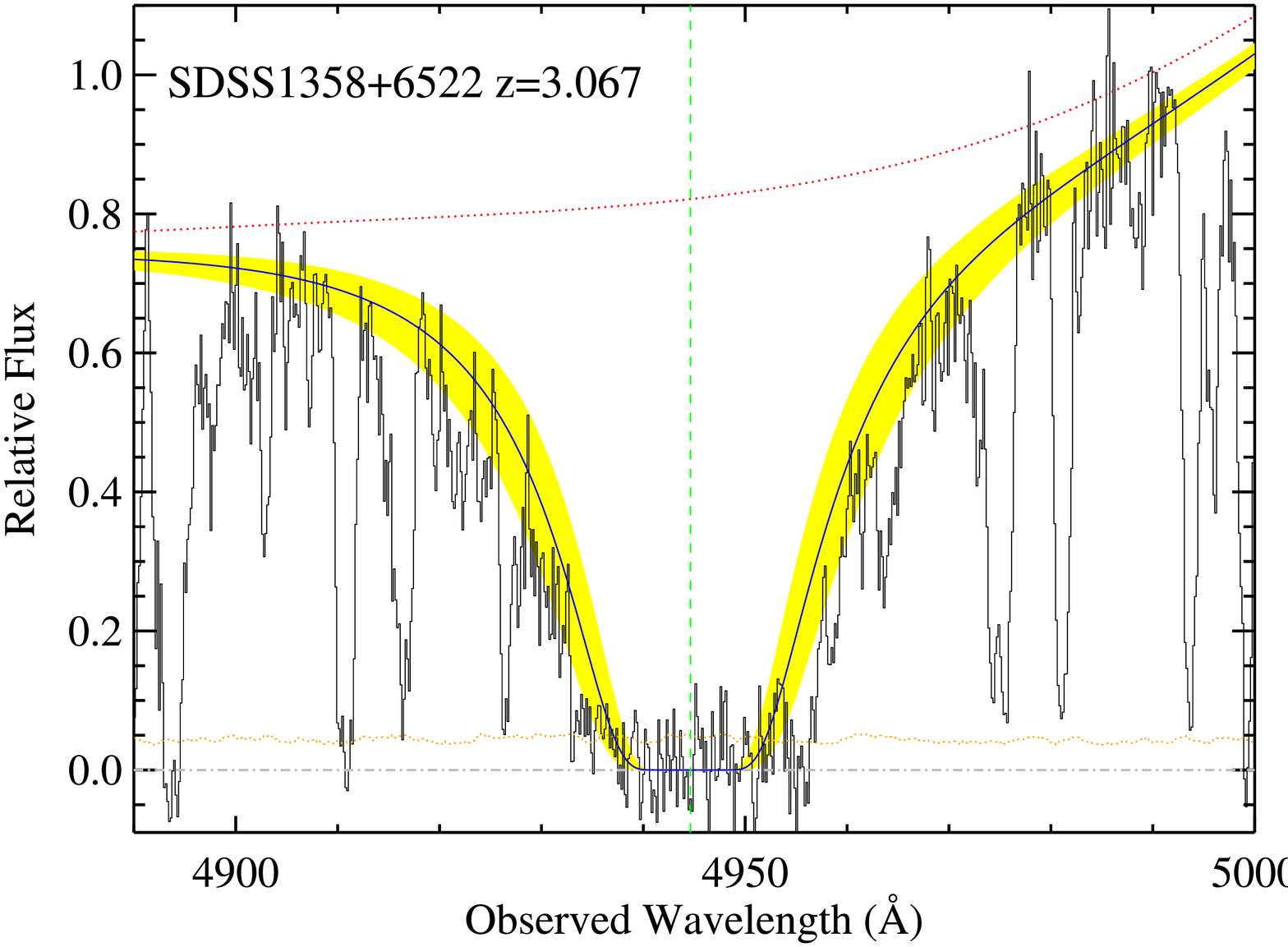}

\end{minipage}
\hspace{0.2cm}
\begin{minipage}[c]{0.45\linewidth}
\centering

\includegraphics[width=2.8in,%
angle=0]{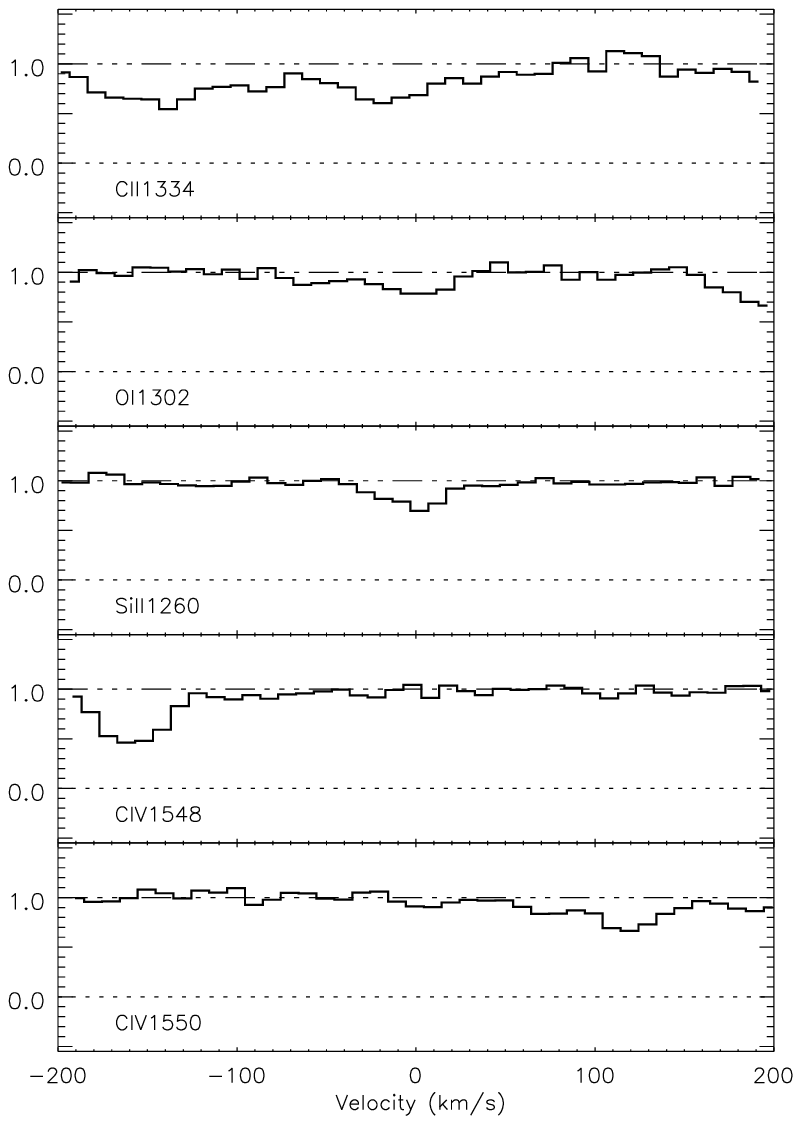}

\end{minipage}
\caption{
Section of our spectrum SDSS1358+6522 showing the lines of \ion{C}{2}~$\lambda$1334, \ion{Si}{2}~$\lambda$1304, and \ion{O}{1}~$\lambda$1302 for the DLA system at z=3.0674 (left), and the \ion{H}{1} damped Lyman alpha profile at the same redshift (right), along with our best fit model (solid line) for log(N(\ion{H}{1})) = 20.35, and one sigma limits to the HI fit (yellow/shaded region).
}
\label{1358+6522fig}
\end{figure}

Figure \ref{1358+6522fig}  shows details of the spectrum of the DLA system of SDSS 1358+6522, with two panels showing the spectrum near key transitions of \ion{C}{2}, \ion{Si}{2}, and \ion{O}{1} (left), and the damped \ion{H}{1} profile (right)
as in $\S$~\ref{sec:firstind}. Our best fit model for N(\ion{H}{1})=20.35, and a curve of growth fit to the absorption lines of the SDSS 1358+6522 DLA indicates small b-values in the range of 5-10 \kms. We have adopted saturation corrections for the absorption lines with equivalent widths between 50 and 130 m\AA\ using the rubric described in $\S$~\ref{sec:saturate}. The adjustment for our \ion{C}{2} transition is significant, at 0.43 dex, and greatly increases the uncertainty of the 122 m\AA\ \ion{C}{2} line, but we have included this transition to enable comparison with the \ion{O}{1} absorption, which is weaker at only 60 m\AA\ equivalent width. The absorption from \ion{Si}{2} is detected in the transitions at 1526 and 1260 \AA, 
but with different resulting AOD column densities, suggesting that some saturation is occurring within the stronger line of the DLA. We have adopted the column density from the weaker of the two lines, and increased the uncertainty estimate for N(\ion{Si}{2}) to encompass the possible effects of saturation.

The absorption from the species of \ion{Fe}{2} is not convincingly detected in any of several transitions, 
and so we present  only upper limits for this ion. 
The results of our adopted column densities for the various species are presented in Table \ref{tab:J1358+6522}. In Table \ref{tab:abunds} we summarize the abundances for this DLA, for which we derive values 
of [C/H] = -2.59 $\pm$ 0.43, [O/H]=-3.08 $\pm$ 0.18, [Si/H]=-2.96 $\pm$ 0.20, and [Al/H]=2.78 $\pm$ 0.16, with an upper limit to [Fe/H] $<$  -3.09.  This DLA also is observed to have measurable column densities for both \ion{C}{2} and \ion{O}{1}, and is one of five DLA systems from our survey where both equivalent widths of these species are less than 130 m\AA, which forms the weak line subsample used in Figure \ref{coratplot}. The measured value of [O/H] is the second lowest of our sample, while the measured value of [C/H] is the fourth lowest of the sample.  Further observations of this system at higher resolution would improve our estimates, since the uncertainty from possible saturation effects dominate our reported values of [C/H] and [O/H], and especially limit our ability to provide more than a very approximate estimate of [C/H].

\subsubsection{SDSS 1456+0407}

\begin{figure}[ht]
\begin{minipage}[c]{0.45\linewidth}
\centering

\includegraphics[angle=0,%
width=3.0in]{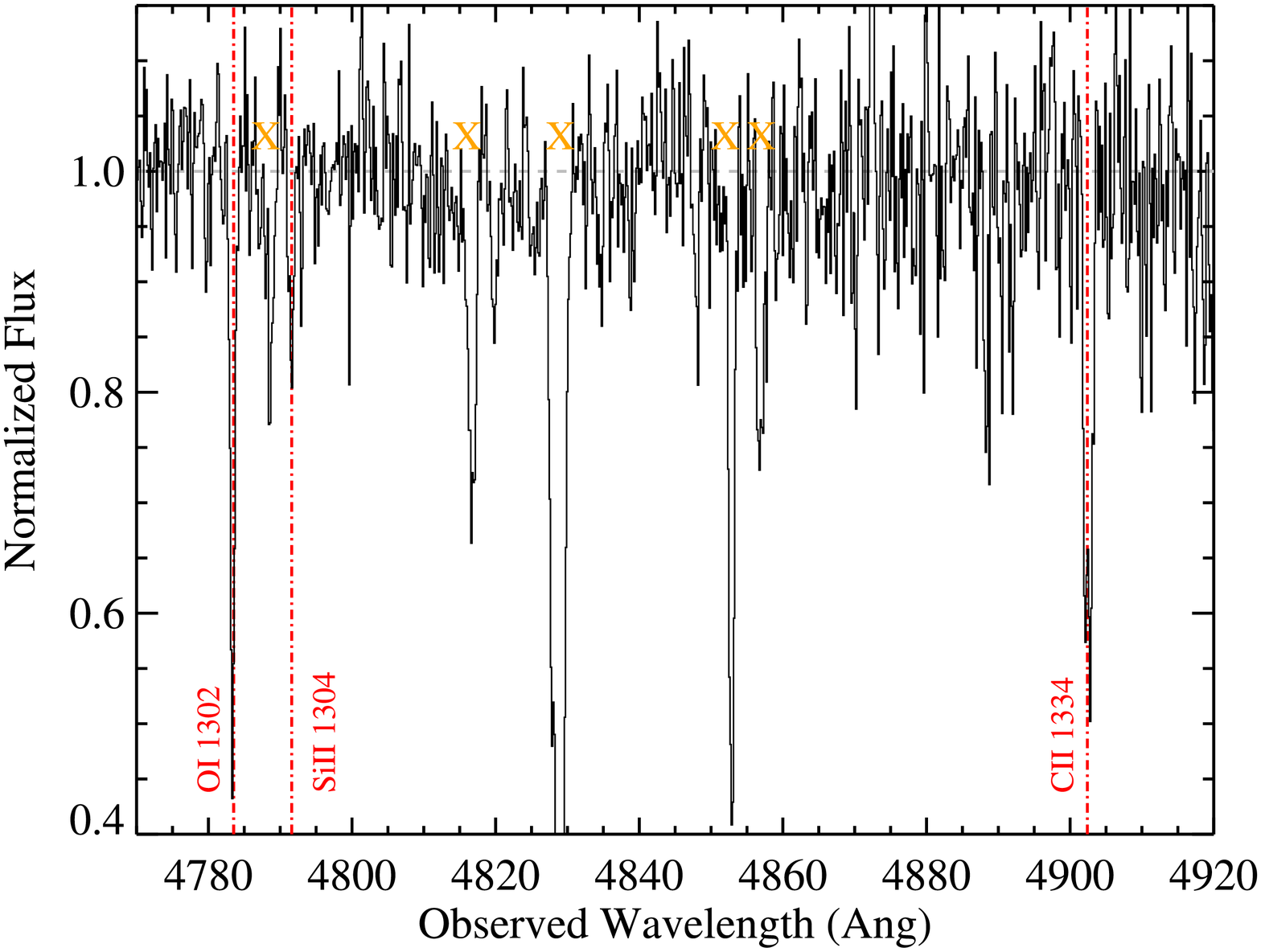}

\includegraphics[width=3.0in,%
angle=0]{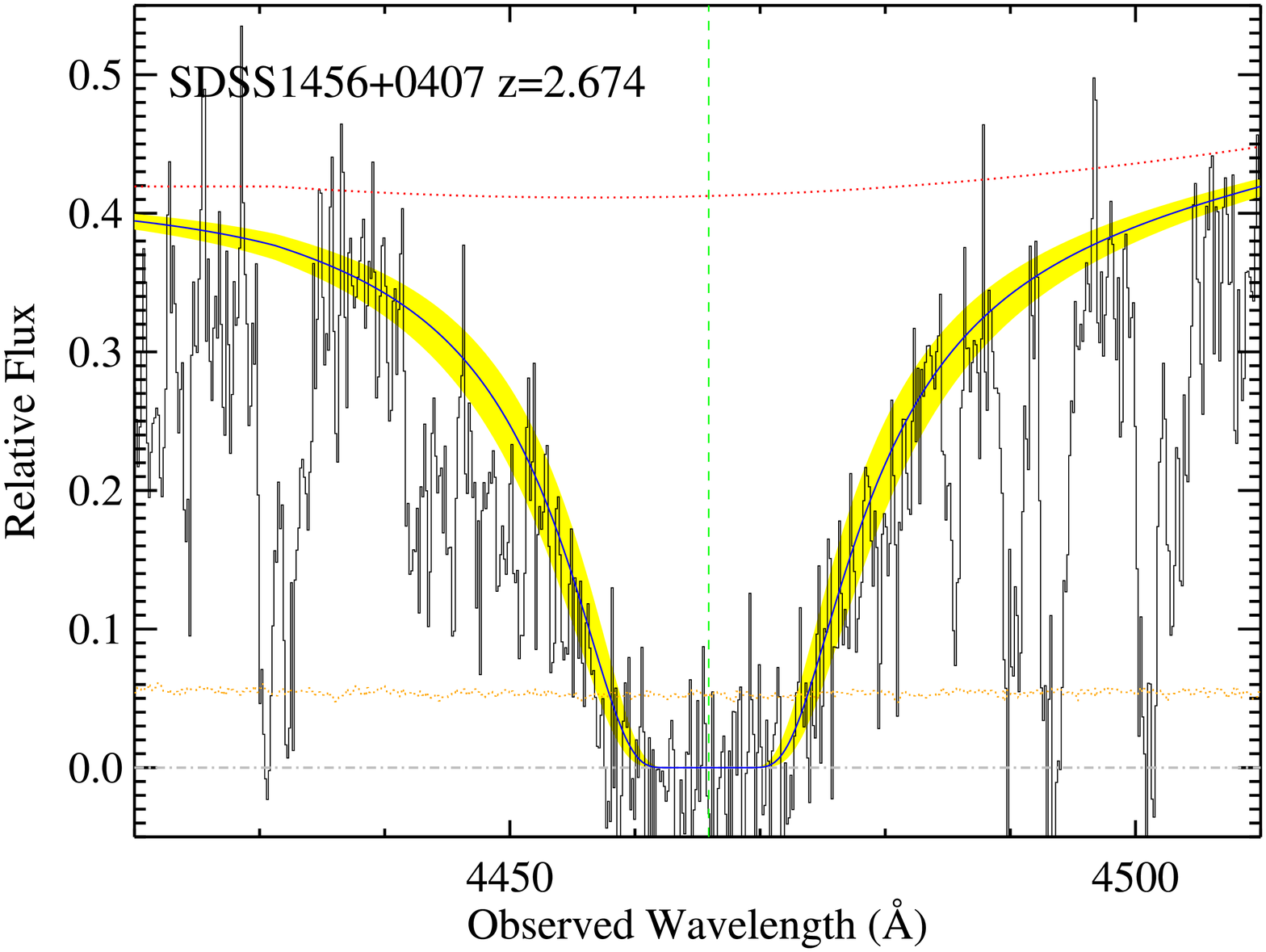}

\end{minipage}
\hspace{0.2cm}
\begin{minipage}[c]{0.45\linewidth}
\centering

\includegraphics[width=2.8in,%
angle=0]{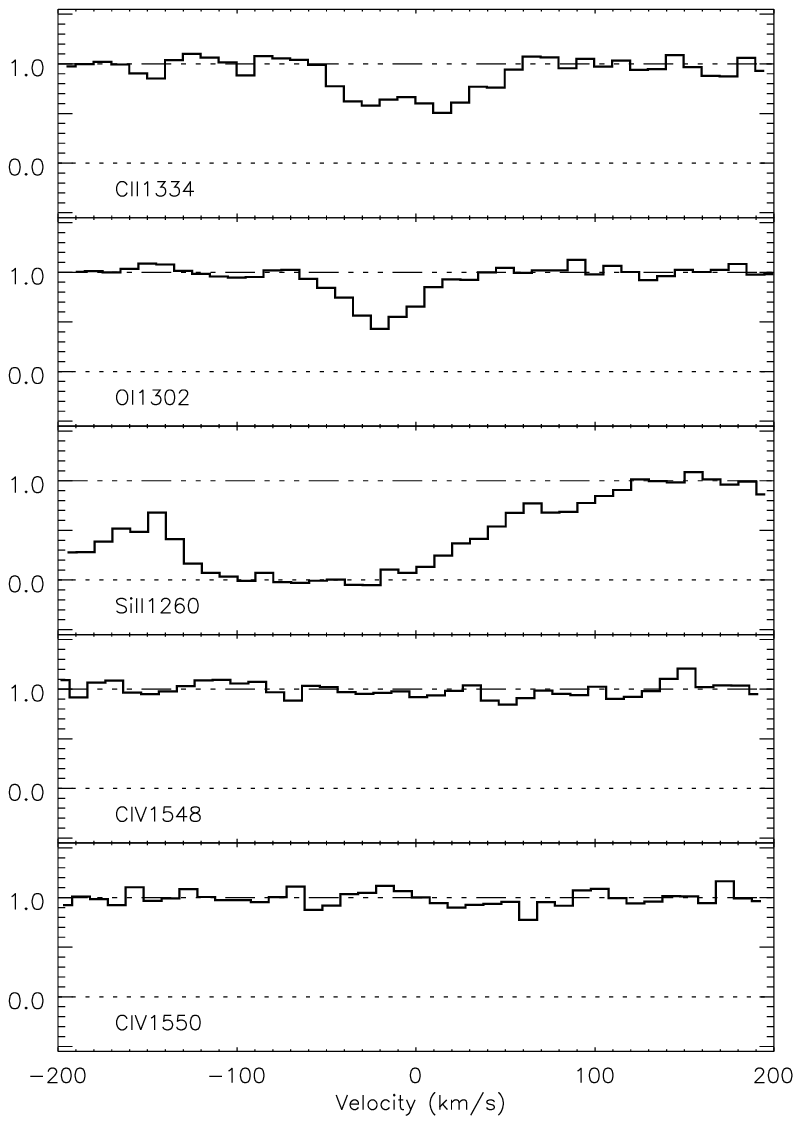}

\end{minipage}
\caption{
Section of our spectrum SDSS1456+0407 showing the lines of \ion{C}{2}~$\lambda$1334, \ion{Si}{2}~$\lambda$1304, and \ion{O}{1}~$\lambda$1302 for the DLA system at z=2.6736 (left), and the \ion{H}{1} damped Lyman alpha profile at the same redshift (right), along with our best fit model (solid line) for log(N(\ion{H}{1})) = 20.35, and one sigma limits to the HI fit (yellow/shaded region).
}
\label{1456+0407fig}
\end{figure}

Figure \ref{1456+0407fig}  shows details of the spectrum of the DLA system of SDSS 1456+0407,  with two panels showing the spectrum near key transitions of \ion{C}{2}, \ion{Si}{2}, and \ion{O}{1} (left), and the damped \ion{H}{1} profile (right)
as in $\S$~\ref{sec:firstind}. Our best fit model for N(\ion{H}{1})=20.35, and the SDSS 1456+0407 DLA includes good detections of \ion{O}{1},\ion{Al}{2},  \ion{Si}{2}, and  \ion{Fe}{2}, which have equivalent widths in the range of 30 to 100 m\AA, and a strong absorption line from \ion{C}{2} with equivalent width in excess of 190 m\AA, which only provides a lower limit to N(\ion{C}{2}).  A set of weak \ion{Fe}{2} lines gives additional data points for our curve of growth, which has a best fit for b $>$ 15 \kms, but with significant uncertainty in the fit.  To be conservative in our estimates of metallicity, we have assumed a b-value of 7.5 \kms, and have adopted the corrections for saturation described in $\S$~\ref{sec:saturate} for the species \ion{O}{1},\ion{Al}{2},\ion{Si}{2}, and \ion{Fe}{2}. The results of our adopted column densities for the various species are presented in Table \ref{tab:J1456+0407}. In Table \ref{tab:abunds} we summarize the abundances for this DLA, for which we derive values of [C/H] $>$ -2.48, [O/H]=-2.56 $\pm$ 0.28, [Si/H]=-2.47 $\pm$ 0.07, [Al/H]=-1.96 $\pm$ 0.45, [Fe/H]=-2.89 $\pm$ 0.1, and [Mg/H]=-2.37 $\pm$ 0.1.

\subsubsection{SDSS 1637+2901}

\begin{figure}[ht]
\begin{minipage}[c]{0.45\linewidth}
\centering

\includegraphics[angle=0,%
width=3.0in]{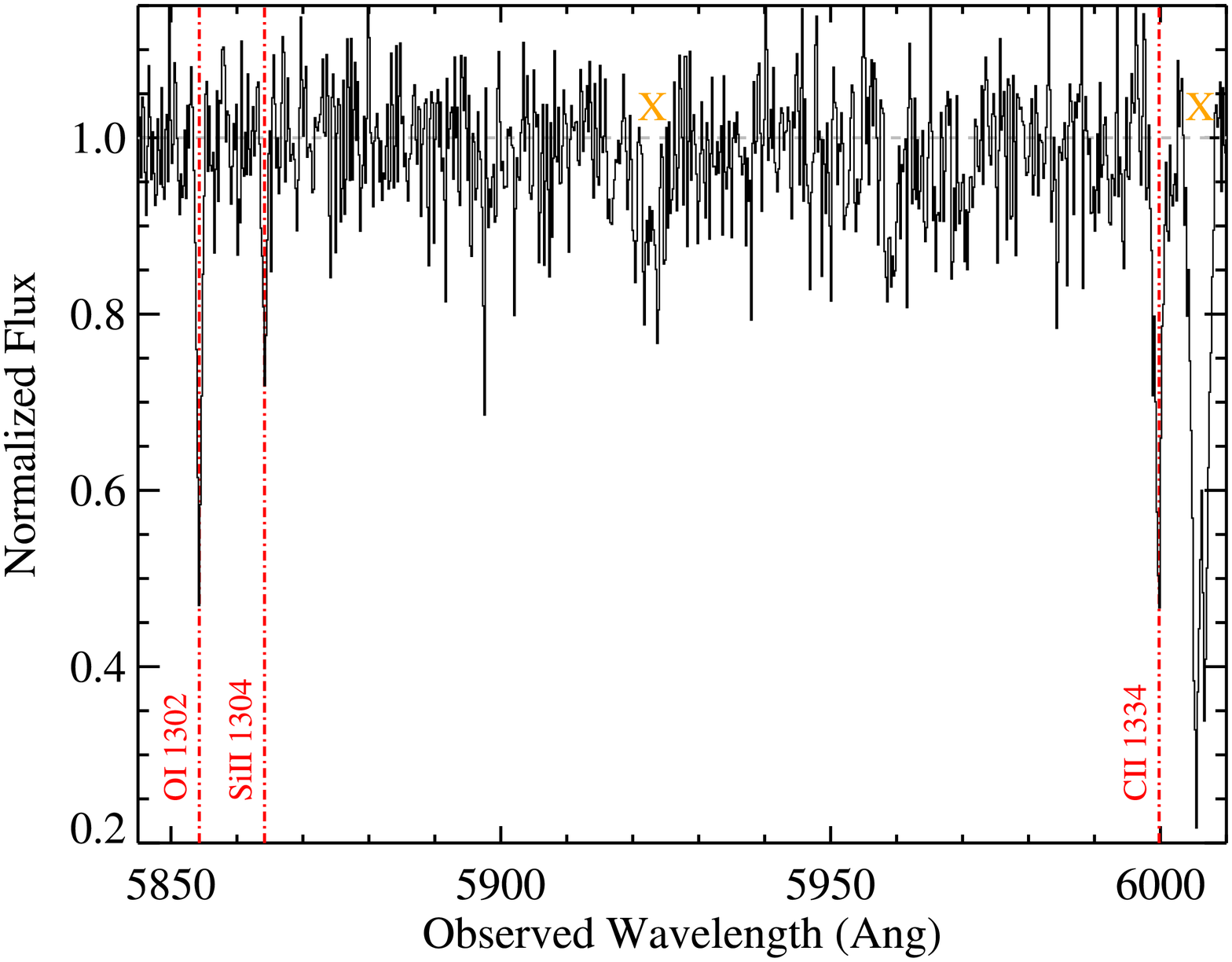}

\includegraphics[width=3.0in,%
angle=0]{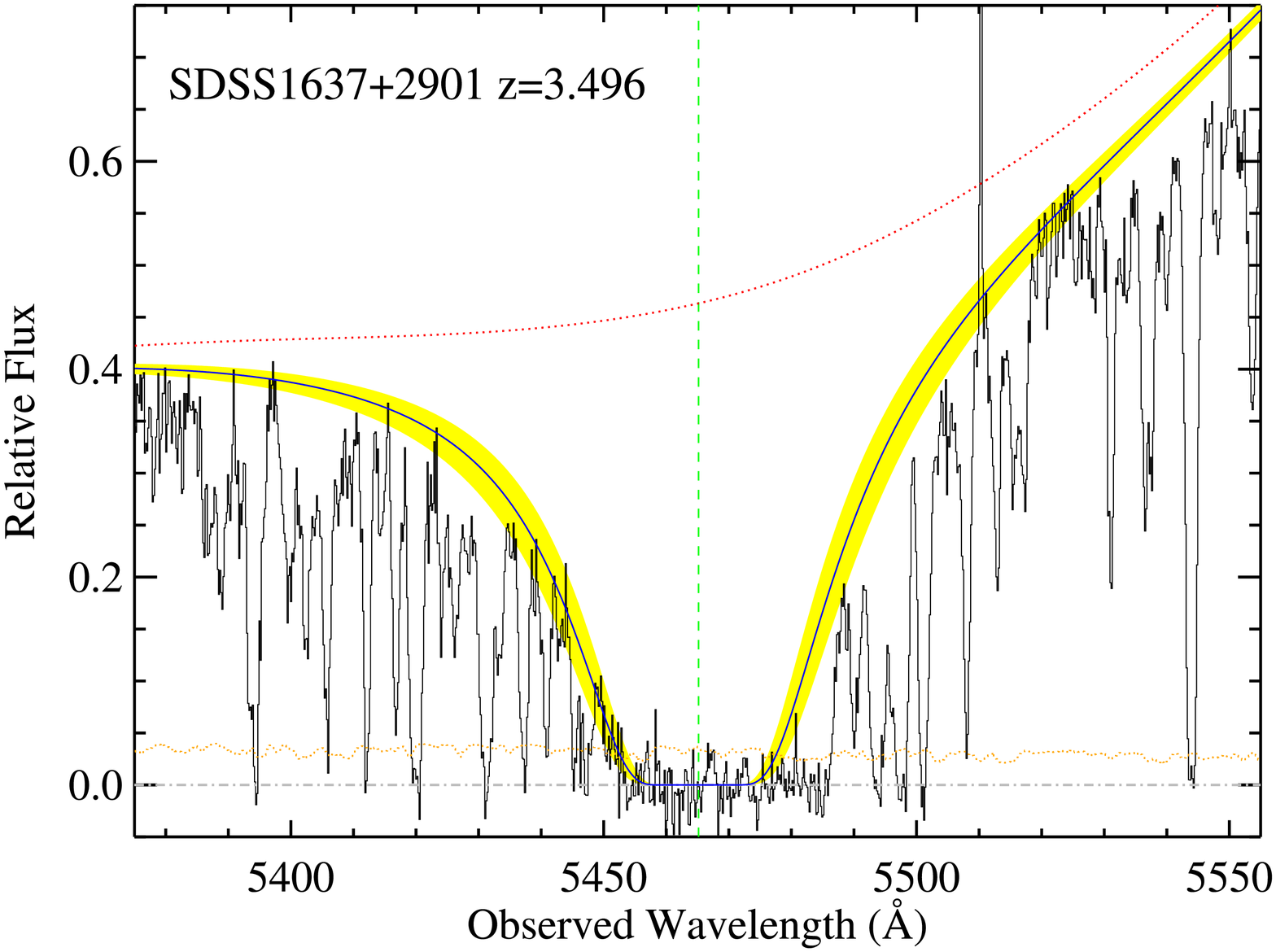}

\end{minipage}
\hspace{0.2cm}
\begin{minipage}[c]{0.45\linewidth}
\centering

\includegraphics[width=2.8in,%
angle=0]{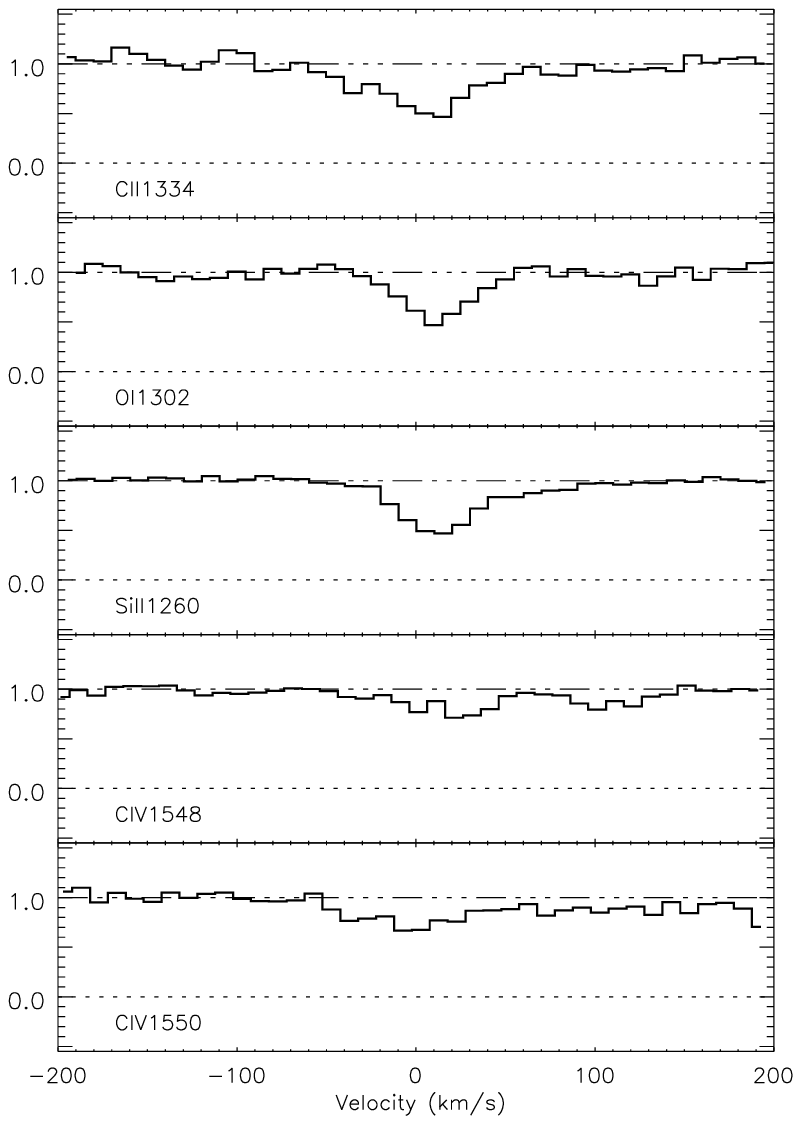}

\end{minipage}
\caption{
Section of our spectrum SDSS1637+2901 showing the lines of \ion{C}{2}~$\lambda$1334, \ion{Si}{2}~$\lambda$1304, and \ion{O}{1}~$\lambda$1302 for the DLA system at z=3.4956 (left), and the \ion{H}{1} damped Lyman alpha profile at the same redshift (right), along with our best fit model (solid line) for log(N(\ion{H}{1})) = 20.70, and one sigma limits to the HI fit (yellow/shaded region).
}
\label{1637+2901fig}
\end{figure}

The SDSS 1637+2901 DLA is one of our lowest metallicity quasar DLA systems, and Figure \ref{1637+2901fig}  shows details of the spectrum of the DLA system of SDSS 1456+0407,  with two panels showing the spectrum near key transitions of \ion{C}{2}, \ion{Si}{2}, and \ion{O}{1} (left), and the damped \ion{H}{1} profile (right)
as in $\S$~\ref{sec:firstind}. Our best fit model for N(\ion{H}{1})=20.70, well above the threshold of log(N(\ion{H}{1})) = 20.3 for DLAs. The SDSS1637+2901 DLA shows evidence of narrow lines, and the lowest b-value of our sample, with b=5 \kms from curve of growth fitting. For this reason we have adopted the saturation corrections described in $\S$~\ref{sec:saturate}. The strong absorption from \ion{C}{2} is accompanied by an even stronger absorption line at the position of the \ion{C}{2}$*$ absorption at 1335.7\AA for z=3.496, 
which is evidence for either extremely strong \ion{C}{2}$*$ excitation or blending from other species, perhaps arising from a \ion{C}{2} absorption at slightly higher redshift.  There also appears to be a pair of absorption lines to the left of the \ion{O}{1} and \ion{Si}{2} lines from Figure \ref{1637+2901fig} arising from an unrelated \ion{C}{4} doublet. The blending allows only a determination of the upper limit for the \ion{C}{2}$*$ column density. The likelihood of saturation in the \ion{C}{2} makes our derived column density of \ion{C}{2} a lower limit. However the absorption from \ion{O}{1} has an equivalent width of 91 m\AA, which we have corrected for saturation effects with a correction of 0.18 dex, according to the saturation correction for b=7.5 \kms described in $\S$~\ref{sec:saturate}. Smaller corrections for saturation of 0.14 and 0.11 dex were applied to the column densities of \ion{Si}{2} and \ion{Al}{2}, respectively. The results of our adopted column densities for the various species are presented in Table \ref{tab:J1637+2901}.In Table \ref{tab:abunds} we summarize the abundances for this DLA, for which we derive values of  [O/H]=-3.17 $\pm$ 0.20, (the lowest of our sample), [Si/H] = -2.90 $\pm$ 0.14, [Al/H] =-2.95 $\pm$ 0.11, [Fe/H]=-2.40 $\pm$ 0.10, and a lower limit of [C/H] $>$ -3.20, due to the very strong absorption at equivalent width of more than 140 m\AA. 

\subsubsection{SDSS 2114-0632}

\begin{figure}[ht]
\begin{minipage}[c]{0.45\linewidth}
\centering

\includegraphics[angle=0,%
width=3.0in]{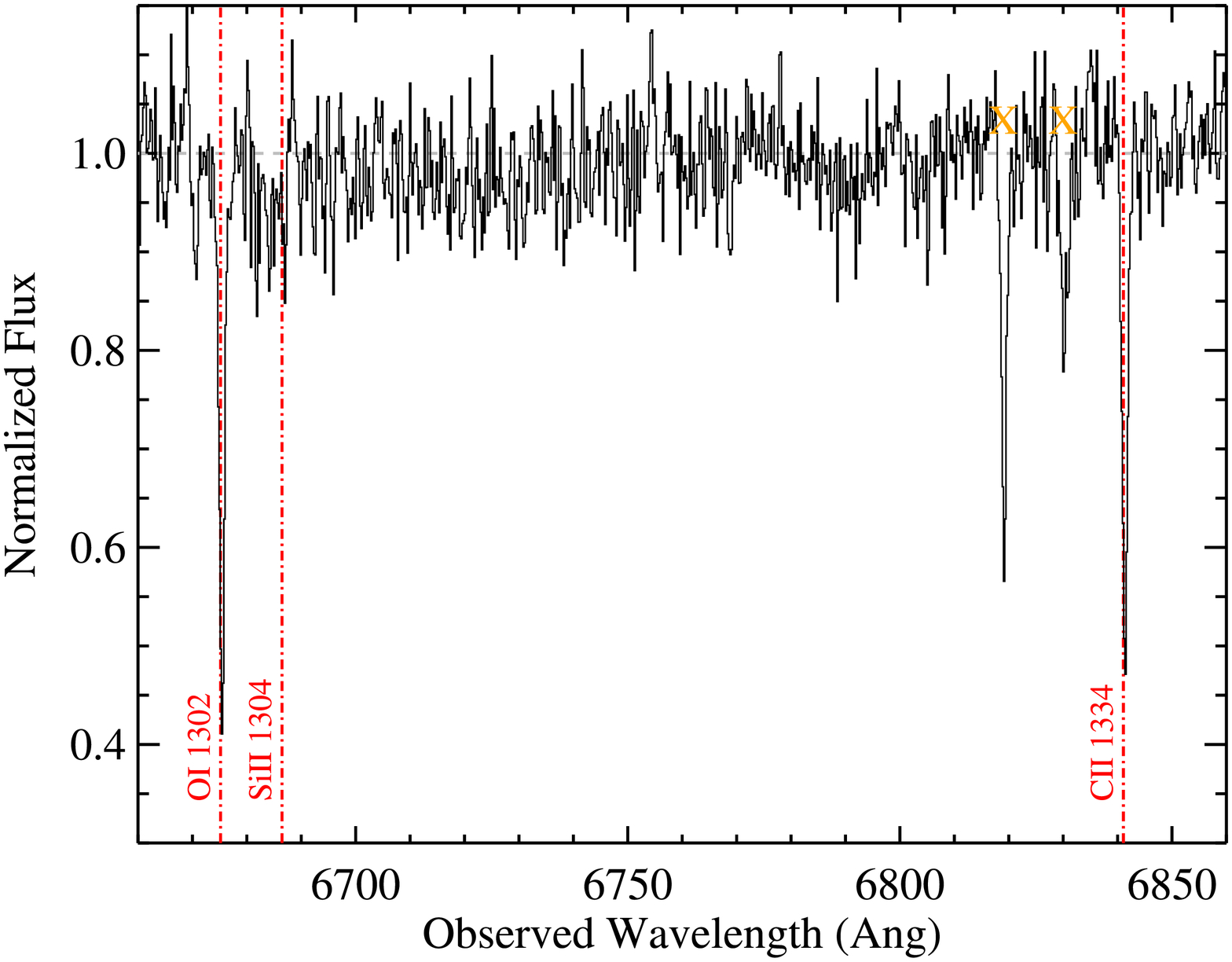}

\includegraphics[width=3.0in,%
angle=0]{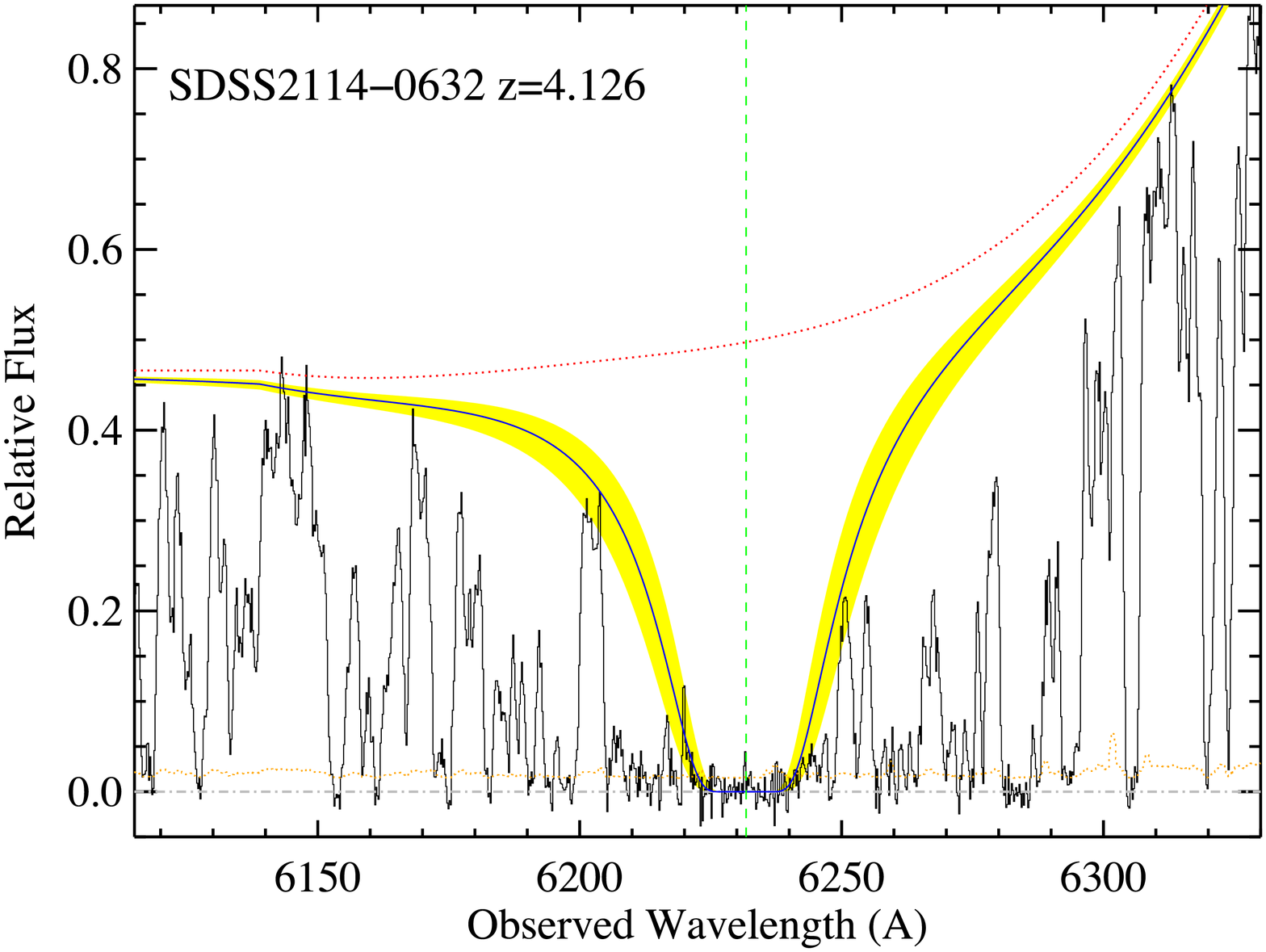}

\end{minipage}
\hspace{0.2cm}
\begin{minipage}[c]{0.45\linewidth}
\centering

\includegraphics[width=2.8in,%
angle=0]{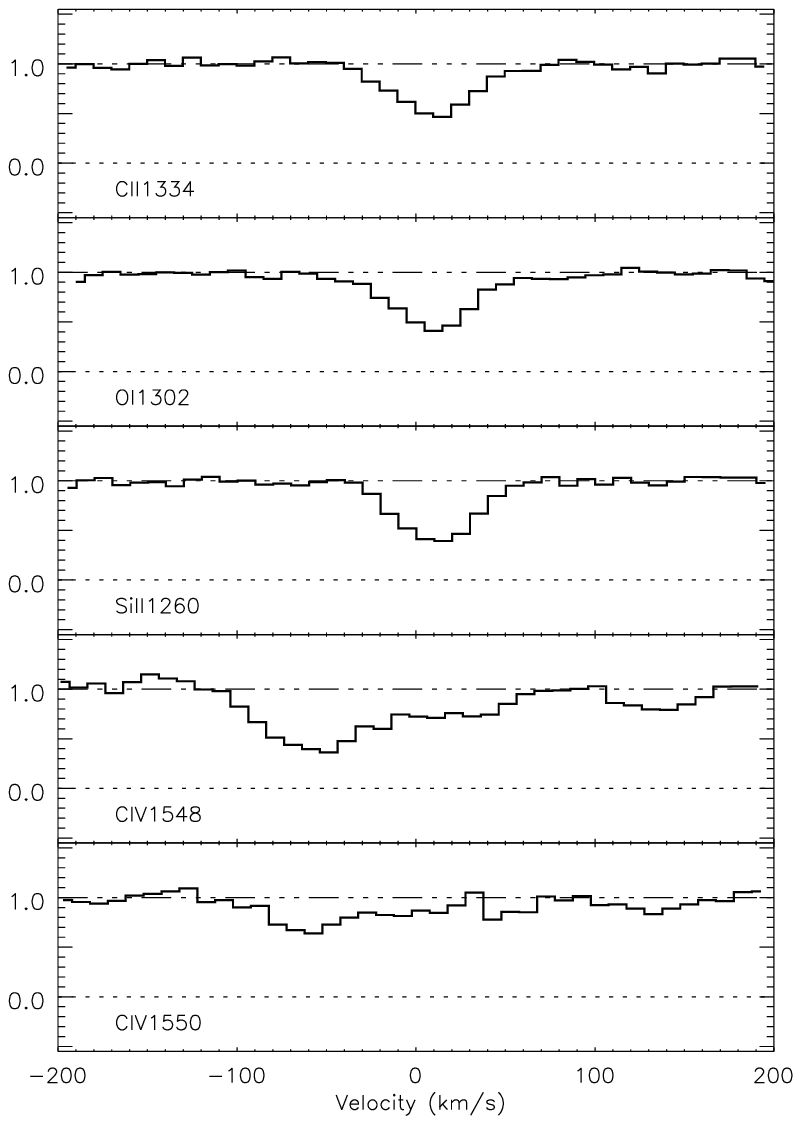}

\end{minipage}
\caption{
Section of our spectrum SDSS2114-0632 showing the lines of \ion{C}{2}~$\lambda$1334, \ion{Si}{2}~$\lambda$1304, and \ion{O}{1}~$\lambda$1302 for the DLA system at z=4.1262 (left), and the \ion{H}{1} damped Lyman alpha profile at the same redshift (right), along with our best fit model (solid line) for log(N(\ion{H}{1})) = 20.40, and one sigma limits to the HI fit (yellow/shaded region).
}
\label{2114-0632fig}
\end{figure}

Figure \ref{2114-0632fig}  shows details of the spectrum of the DLA system of SDSS 2114-0632, with two panels showing the spectrum near key transitions of \ion{C}{2}, \ion{Si}{2}, and \ion{O}{1} (left), and the damped \ion{H}{1} profile (right)
as in $\S$~\ref{sec:firstind}. Our best fit model for N(\ion{H}{1})=20.40, just above the DLA threshold of log(N(\ion{H}{1}))=20.3. The SDSS 2114-0632 DLA is at the highest redshift (z=4.126) of our sample, and includes extremely weak detected absorption lines of the species \ion{C}{2}, \ion{O}{1}, and  \ion{Si}{2}. 
The absorption from the species \ion{Al}{2} and \ion{Fe}{2} is extremely weak, and we are not able to detect these species, and therefore provide only upper limits for both of them. A curve of growth analysis of the detected species suggests a large b-value in the range of 20 \kms, which would greatly reduce the required correction for saturation. Despite the possibility of large b values for the DLA, we account for possible saturation in our reported column densities, as our curve of growth fitting can't rule out the possibility of multiple narrow components within the DLA. We have included the \ion{O}{1} transition in our compilation since the lower bound of the measured equivalent width is within our threshold of 130 m\AA, to enable comparison with \ion{C}{2} and other species.  The results of our adopted column densities for the various species are presented in Table \ref{tab:J2114-0632}. In Table \ref{tab:abunds} we summarize the abundances for this DLA, for which we derive values of [C/H]=-2.63 $\pm$ 0.45 ,[O/H]=-2.44 $\pm$ 0.45, and [Si/H] = -2.78 $\pm$ 0.13,  with upper limits for the metallicities of [Al/H] $<$ -3.17,  and [Fe/H] $<$ -2.43.
This DLA also is observed to have relatively weak absorption lines for both \ion{C}{2} and \ion{O}{1}, and is one of five DLA systems from our survey where both equivalent widths of these species are less than 130 m\AA, which forms the weak line subsample used in Figure \ref{coratplot}. The observed value of [C/H]=-2.63 is the third lowest of our sample.

\subsection{Comparision of ESI-derived HI Column Densities with SDSS Spectra}

\begin{figure}[t!]
\epsscale{0.96}
\plotone{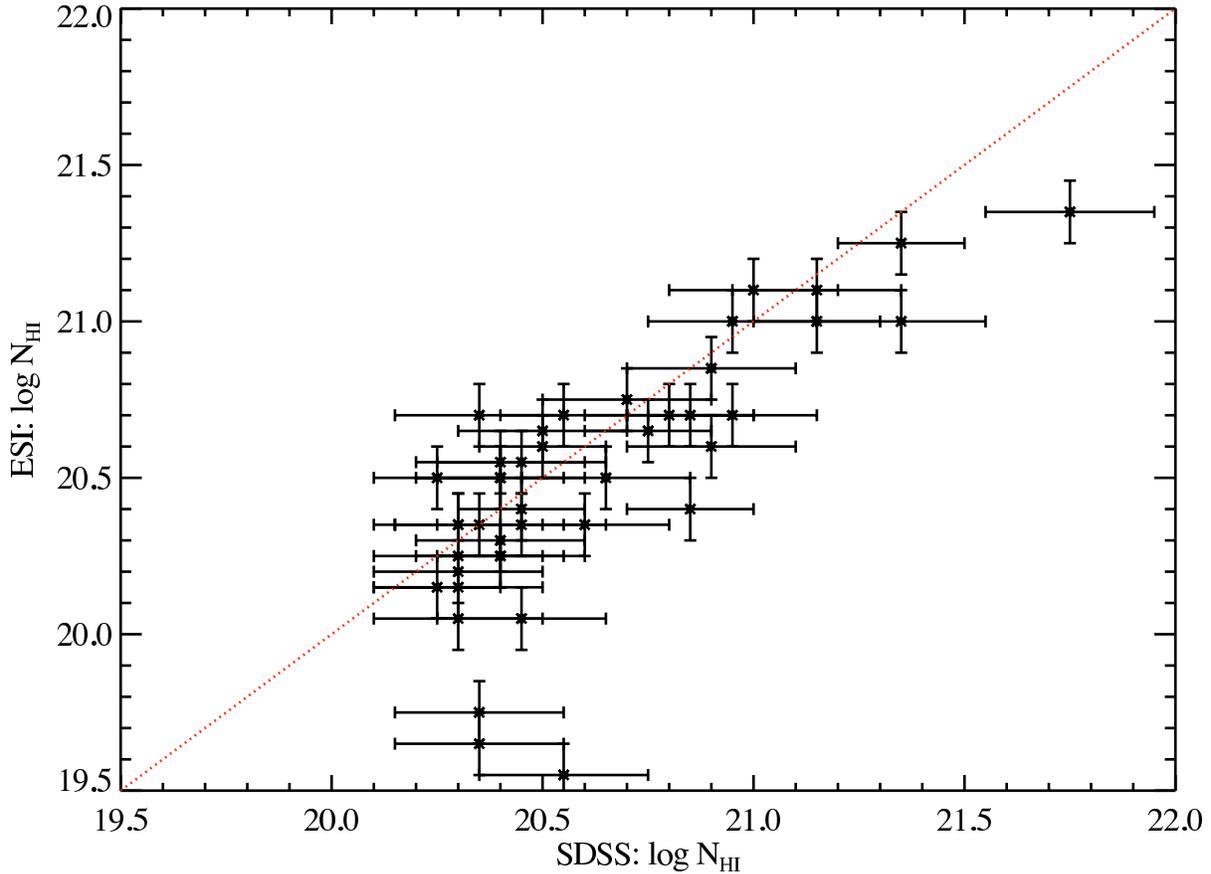}
\caption{ Derived column density of \ion{H}{1} for the ESI spectra, plotted against those for the SDSS spectra. The lower resolution of the SDSS spectra caused some discrepancies due to blending, and continuum placement, 
but the N(HI) values are generally within the $1\sigma$ error estimate.
}
\label{HIcompfig}
\end{figure}

Our sample was selected based on fits to the HI profiles within the
SDSS spectra where the derived values of log(N(\ion{H}{1})) $\ge$
20.3, but we have refitted the \lya\ profiles in the ESI spectra.
Indeed, the SDSS estimates of log(N(\ion{H}{1})) are uncertain due to
the limited resolution of the SDSS spectra, and the possibility of
blending of the DLA system with either other DLA absorbers or strong
Lyman $\alpha$ forest components. Figure \ref{HIcompfig} shows a
comparison of the two values of log(N(\ion{H}{1})), which in most
cases shows that our selection procedure identified DLA systems from
the SDSS spectra.  As a population, we find that the N(HI) values from
SDSS spectra are systematically lower than those derived from 
the ESI data.  We calculate a mean (median) offset of 0.07 (0.1)\,dex
and expect one of two reasons for the differences.  First, if there was no
metal-line transition identiried then one tends to center the \lya\ 
profile on the absorption line.  This will yield the highest value
possible for the data and therefore gives a 
systematic over-estimate for the N(HI) value.
Second, we suspect there is a subtle, statsitical effect related to our
pre-selection of DLAs with weak metal-lines. 
In any case, this offset has no significant implication for this
paper.

Two of the systems drawn from the SDSS database 
(SDSS1156+5513 at $z\approx 2.49$; SDSS1327+4845 at $z \approx 2.61$) 
are now observed to be a pair of strong \lymana\ lines that were analyzed
as a single absorption system by Prochaska et al.\ (2005).
Each of these pair have $N(HI) < 10^{20}$ cm$^{-2}$ and are plotted as a 
single point in the figure.
It is likely that by focusing on systems with very weak metal-line
absorption, we have preferntially identified systems that suffer from
this (rare) systematic effect.

The uncertainty in the SDSS \ion{H}{1} determinations also caused a few of our DLA systems to have column densities of log(N(\ion{H}{1})) $<$ 20.3, due to the uncertainty of the SDSS \ion{H}{1} determination. For this reason a few of our DLA systems could require small corrections for ionization, but these corrections in most cases were below our systematic uncertainty from the limited resolution of the ESI spectrograph. We discuss some of the implications of the sub-DLA systems in $\S$~\ref{sec:nucsynth}.

\section{Elemental Abundances and Comparisions with Literature and Theoretical Models}

\subsection{Distribution of DLA Metallicities}

We present the results of our AOD column density analysis in Table \ref{tab:abunds} for the entire sample, which shows either measurements or limits for the abundances of the available elements.  For each element, we measured the column density of the species, and then adjusted for solar abundances using recently recalibrated values (Lodders, 2003), after combining with the measured \ion{H}{1} column density. The results in Table \ref{tab:abunds}  then are [X/H], or abundance of element X relative to solar on a logarithmic scale. 
In some cases, transitions appear to be blended and provide only an upper limit to the column densities. Due to the limited spectral resolution of the ESI spectrograph, we also considered profiles with normalized flux values  F$_{\lambda}$ $<$ 0.3 in any pixel to be potentially saturated, and many of these column densities are presented in Table \ref{tab:abunds}  as lower limits. 

In most cases, where we have detections of absorption, we can determine values of metallicity to within $\pm$ 0.15 dex, while some of the lower S/N spectra only enable a weaker constraint to within $\pm$ 0.25 dex, and the degree of precision of the abundance is provided with labels within Table \ref{tab:abunds} . Our estimate of the uncertainty is based on comparisons of column densities from multiple transitions of the same species and allows for the combination of uncertainty from continuum fitting, blending, and photon noise within the data. 

\begin{figure}[t!]
\epsscale{0.96}
\plotone{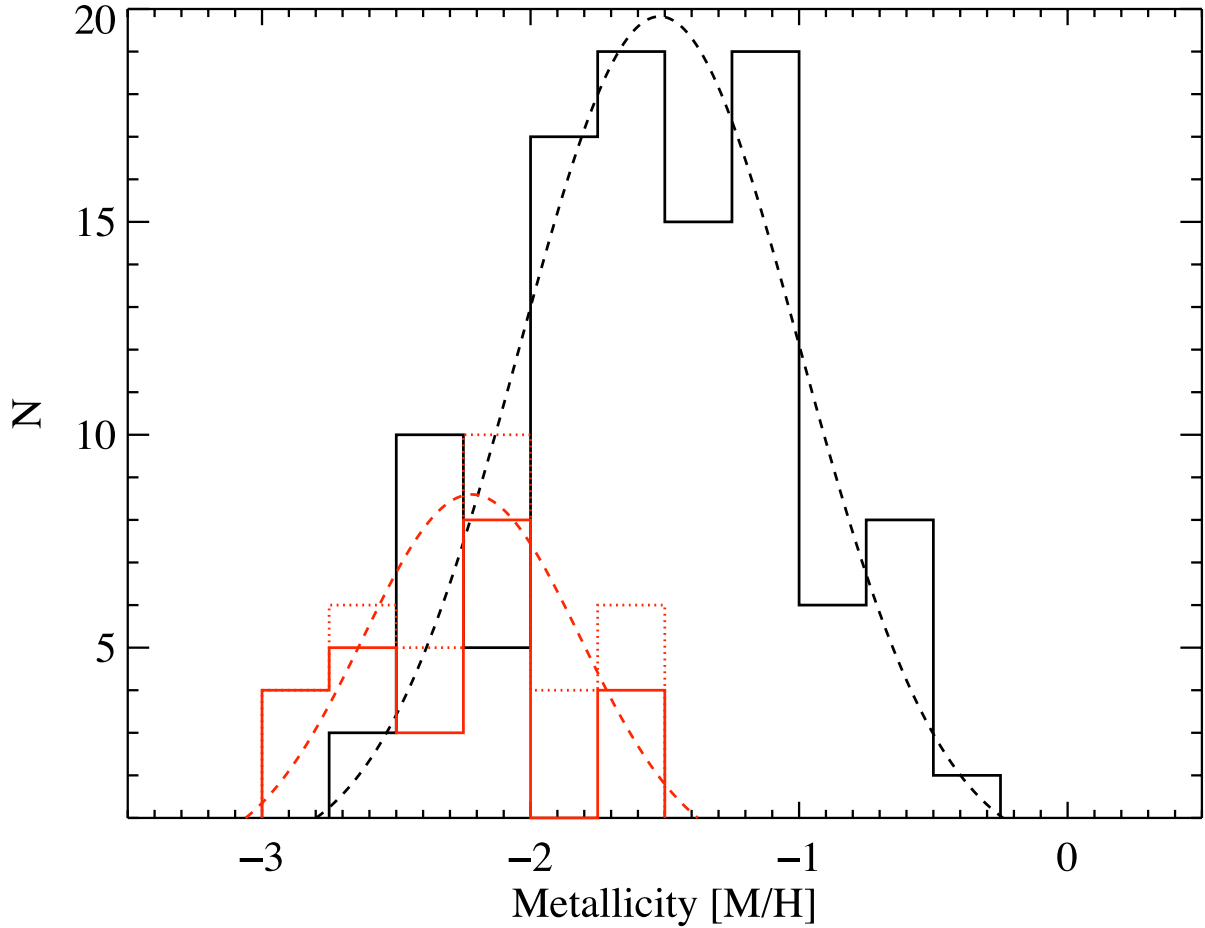}
\caption{
Plot of observed values of [M/H] from the survey of Prochaska et al (2003), and a Gaussian fit (black dashed line). The mean metallicity for this "control" sample in the redshift range of our  sample (2 $<$ z $<$ 4.5) is [M/H] = -1.52, with a width of $\sigma$ = 0.52. We also present a histogram of [M/H] for the low metallicity sample observed in this paper, for which a Gaussian fit (red dashed line) gives a mean metallicity of [M/H] = -2.2.  DLA systems where lower limits of [M/H] have been observed are plotted with the dashed (red) lines on top of the values for the metal poor DLAs. 
}
\label{XHhist}
\end{figure}

Figure \ref{XHhist} shows the histogram of observed values of [M/H] from the survey of Prochaska et al (2003), along with a Gaussian fit to the distribution. The Gaussian fit seems to match the data well, and shows a mean value of metallicity for systems with 2 $<$ z  $<$ 4.5 of [M/H] = -1.52, with a value of $\sigma([X/H])$ = 0.52. Figure~\ref{XHhist} also presents the metallicity distribution for the
metal-poor DLAs analyzed in this paper, which shows a significantly lower mean metallicity of [M/H] = -2.2.  To estimate a [M/H] values, we have adopted the [Si/H] metallicity in those cases where the system has a reported value. For systems with upper/lower limits for [Si/H], we have adopted the [Fe/H] value (when measured) incremented by 0.4\,dex which represents the typical [Si/Fe] offset observed in DLAs.  Systems with only a lower limit to [M/H] are shown separately in the figure with dotted lines. Histograms of our estimated abundances for individual elements are presented in  $\S$~\ref{sec:elhists}.

The present work sampled a large number of DLA systems (458) in the SDSS DR5  to provide the small number of low metallicity DLA systems in this study. If we use the fitted Gaussian distribution we would predict that values of [M/H] $<$ -3.12 would comprise $0.1\%$ of the available DLA systems at this redshift, based on the expectation values for the fitted Gaussian metallicity distribution. Likewise a 3$\sigma$ departure of metallicity [M/H] for the distribution corresponds to a value of [M/H] = -3.08, and would comprise 0.00135 of the available distribution of DLA systems. Given that our SDSS sample consisted of 458 DLA systems, we would then predict 0.6 DLA systems with [M/H] $<$ -3.08, and 0.45 systems with [M/H]. 

Our results are consistent with this distribution, within the small number statistics of our sample, as our lowest metallicity in [C/H] = -2.85, while our lowest metallicity in [O/H] = -3.17, both lie close to the 3$\sigma$ limit of metallicity for DLA systems. We observe 2 systems with [O/H] $\le$ -3.08 (the three $\sigma$ limit for [M/H]), and no systems with [C/H] $\le$ -3.08. It is also interesting to note that extremely low metallicity systems with [X/H] $<$ -4.0 are expected to be extremely rare, as they represent a 4.8 $\sigma$ deviation in metallicity, and therefore would comprise only 8$\times10^{-7}$ of the DLA systems. Put another way, one would need observations of 1.25 million quasar DLA systems to expect to find 1 system with [M/H] $<$ -4.0, making the discovery of such low metallicity systems highly unlikely, under the assumption that  the metallicity distribution obeys a Gaussian distribution.

\subsection{Distribution of Estimated Metallicity for Individual Elements}
\label{sec:elhists}

When we plot the observed elemental abundances for our sample against redshift, we can see that many of the target quasars feature DLA systems of significantly lower metallicity than previously published surveys.  Our sample was chosen to have very weak lines of \ion{C}{2}~$\lambda$1334 and \ion{Si}{2}~$\lambda$1260 in the SDSS spectra, and the derived abundances from the ESI spectra bear out that the selection process was successful in isolating very low metallicity DLAs. 

\begin{figure}[t!]
\epsscale{1.8}
\plottwo{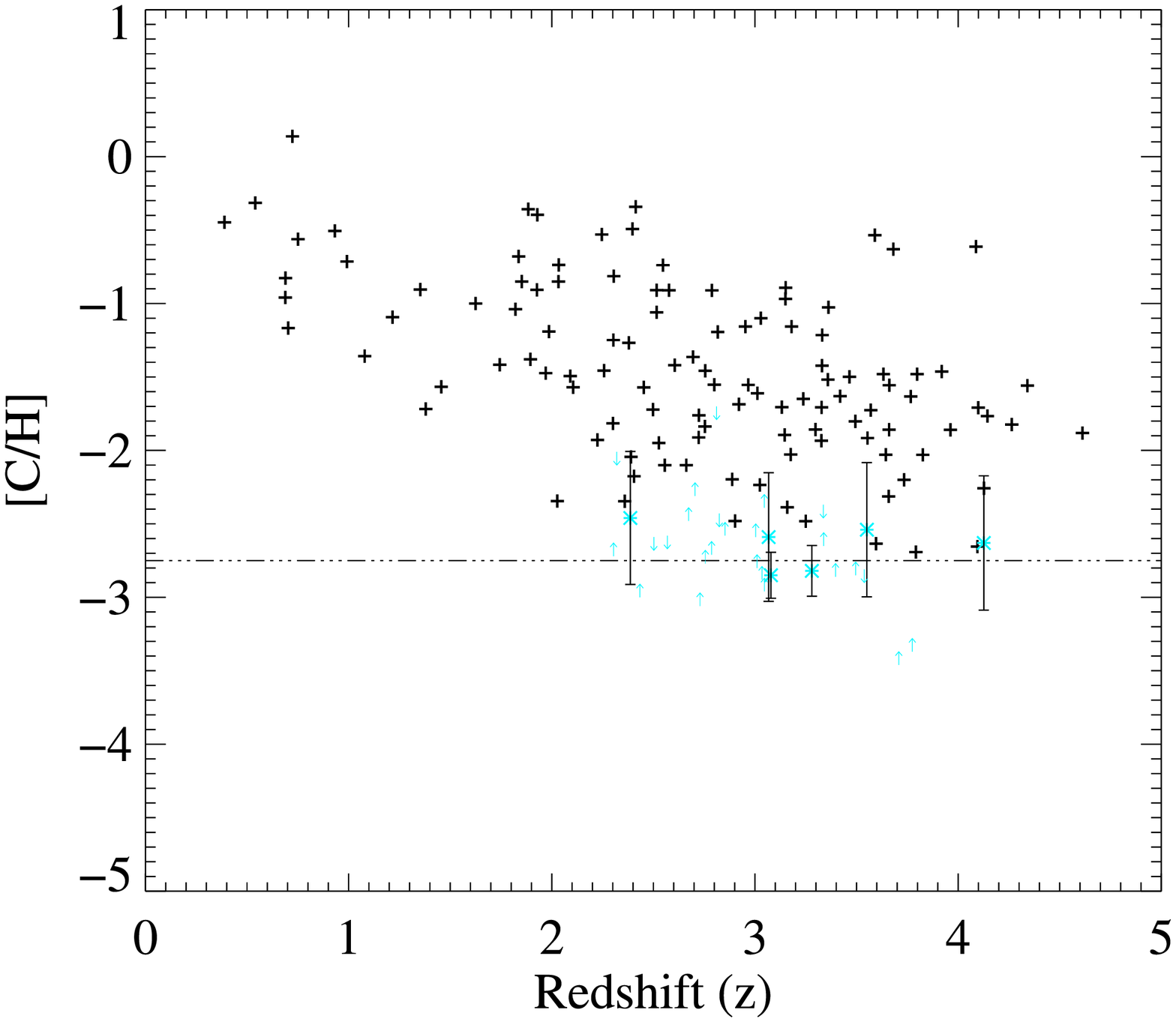}{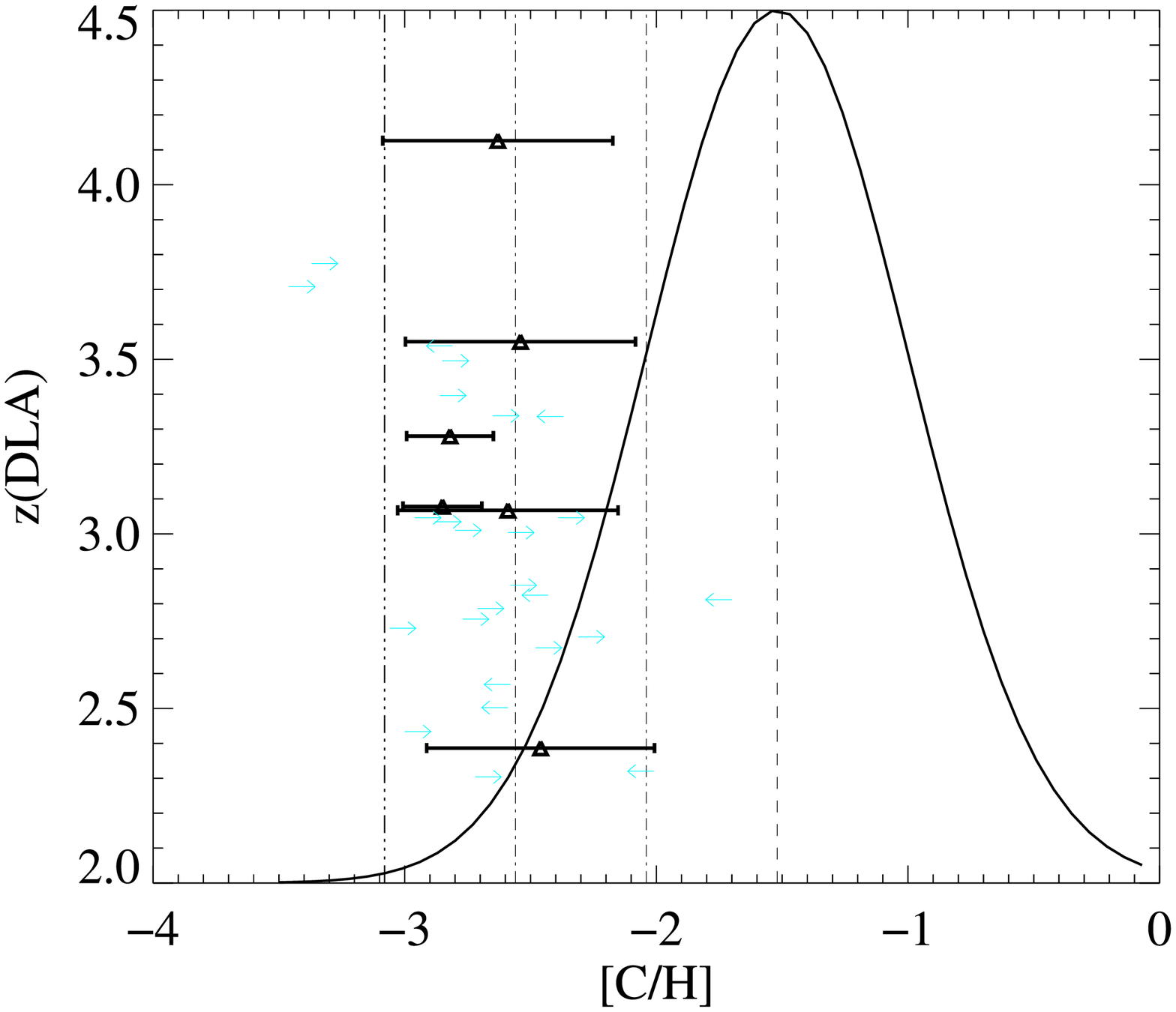}

\caption{
Plot of  DLA element abundance relative to solar ([C/H]) vs redshift for our sample (symbols with error bars), compared with that of Prochaska et al.\ (2003) (top), and compared to a histogram of metallicity for the same sample (bottom).
Vertical lines on the right panel show the location of 1,2 and 3 $\sigma$ departures from the mean DLA metallicity of -1.52 from the Prochaska et al. (2003) survey.}

\label{MHzfigC}
\end{figure}

The top panel of Figure \ref{MHzfigC} shows the values of [C/H] for our sample, including those sightlines which appear to have only limits to the C abundance, either due to saturation effects (lower limits) or due to a non-detection of the line (upper limits) due to Lyman-$\alpha$ blending. These limits are shown in the figure as arrows. The lower panel of Figure \ref{MHzfigC} shows the same data points superimposed on the metallicity distribution from Prochaska et al. (2003), with vertical lines indicating the mean metallicity ([M/H] = -1.52) and 1,2 and 3 $\sigma$ departures from the mean metallicity.

\begin{figure}[t!]
\epsscale{1.8}
\plottwo{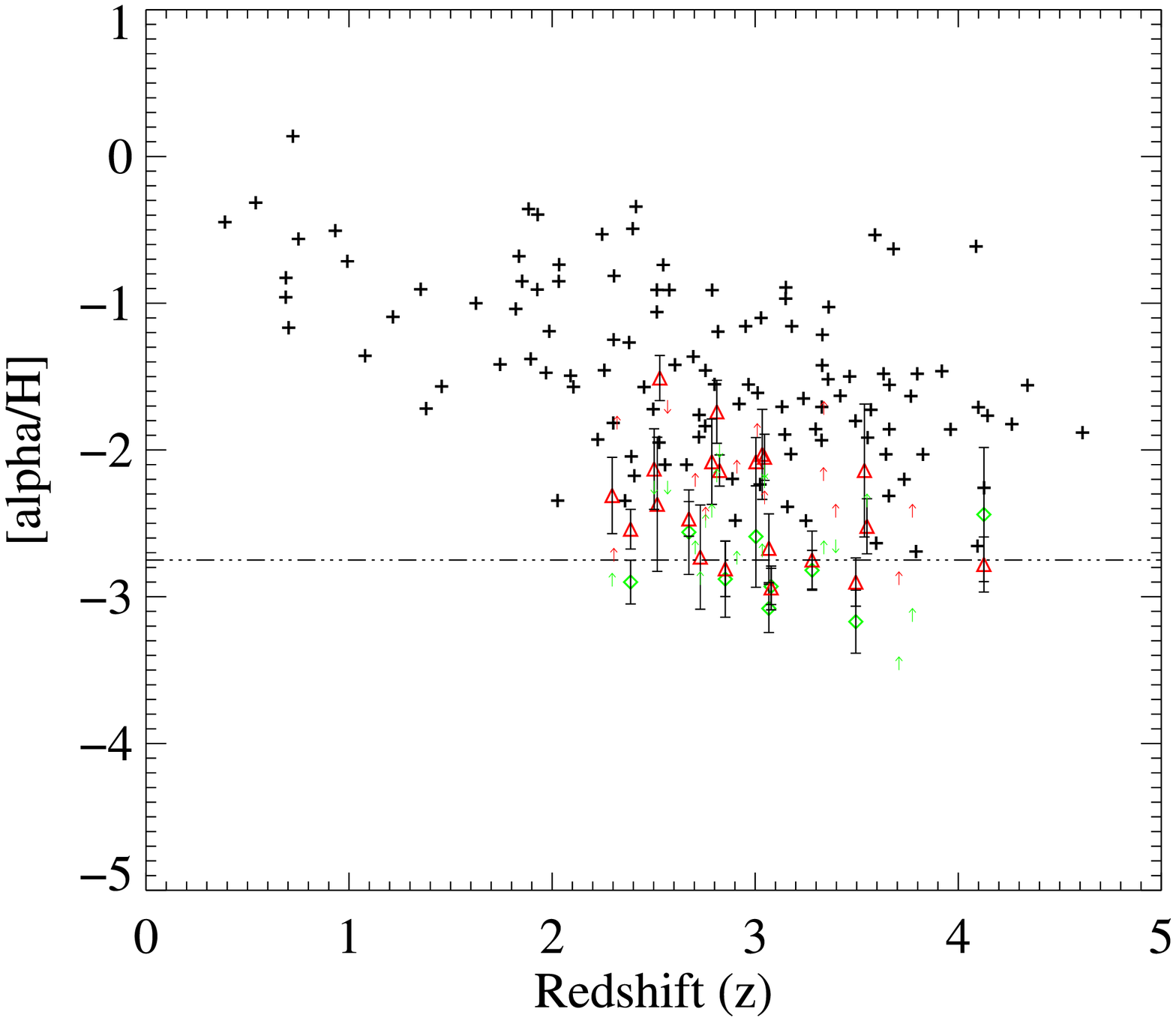}{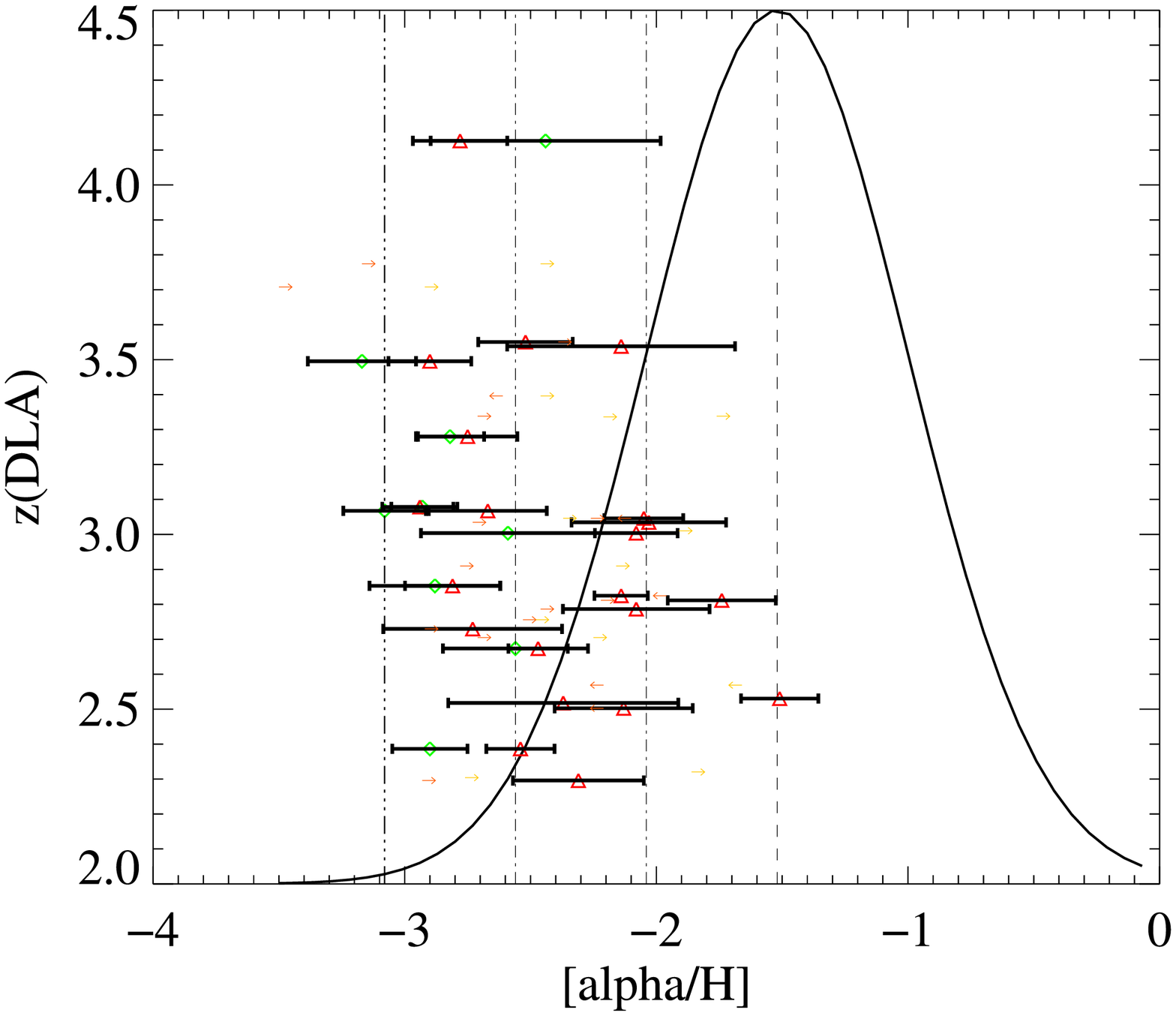}
\caption{
Plot of  DLA element abundance relative to solar ([$\alpha$/H]) vs redshift for our sample (symbols with error bars), compared with that of Prochaska et al.\ (2003) 
(plus signs) for the $\alpha$-process elements (top), and compared to a histogram of metallicity for the same sample (bottom).  The $\alpha$-process elements presented include [O/H] (triangles) and [Si/H] (diamonds). }
\label{MHzfigalpha}
\end{figure}

The top panel of Figure \ref{MHzfigalpha} compares the results from our survey with the metallicity sample of Prochaska et al.\ (2003) for the $\alpha$-process elements of O and Si, for those DLA systems with measurements (i.e. excluding upper and lower limits) of the elements O and Si. Our values of [O/H], and [Si/H] are some of the lowest yet observed, with several values of [$\alpha$/H] $<$ -2.75.  The bottom panel of Figure \ref{MHzfigalpha} compares these same measurements with the distribution of metallicity within Prochaska et al (2003), as in Figure \ref{MHzfigC}.

A similar plot of the metallicity dependence on redshift is show below in Figure \ref{MHzfigothers}, where we present the metallicities of our sample for the elements Fe and Al, against redshift in the top panel, and the comparison of these metallicities against the survey of DLA systems from Prochaska et al (2003) in the bottom panel. We discuss the observed abundances of these elements compared with stellar abundance surveys, and in the contex of  nucleosynthesis models in the following section.

\begin{figure}[t!]
\epsscale{1.8}


\plottwo{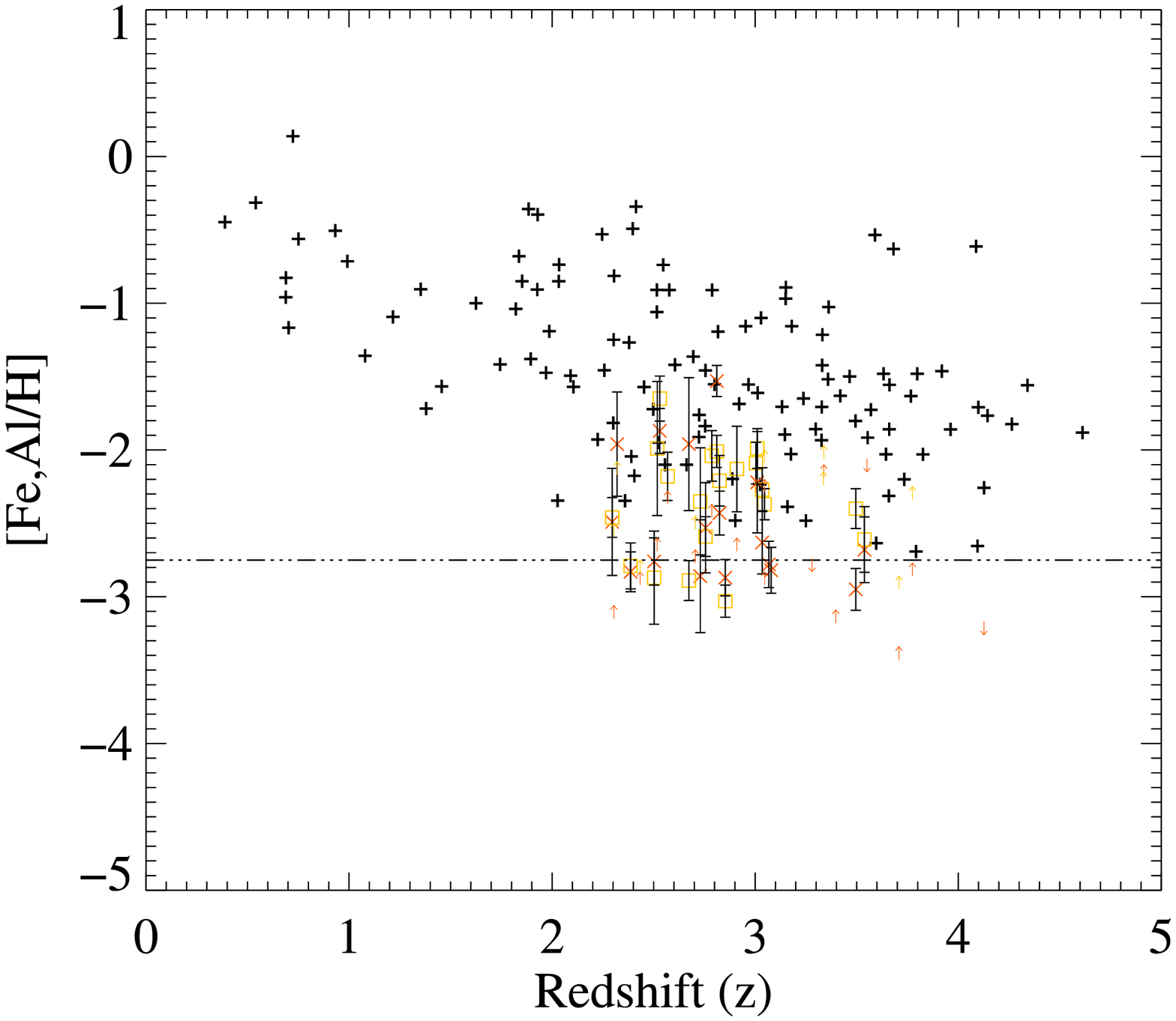} {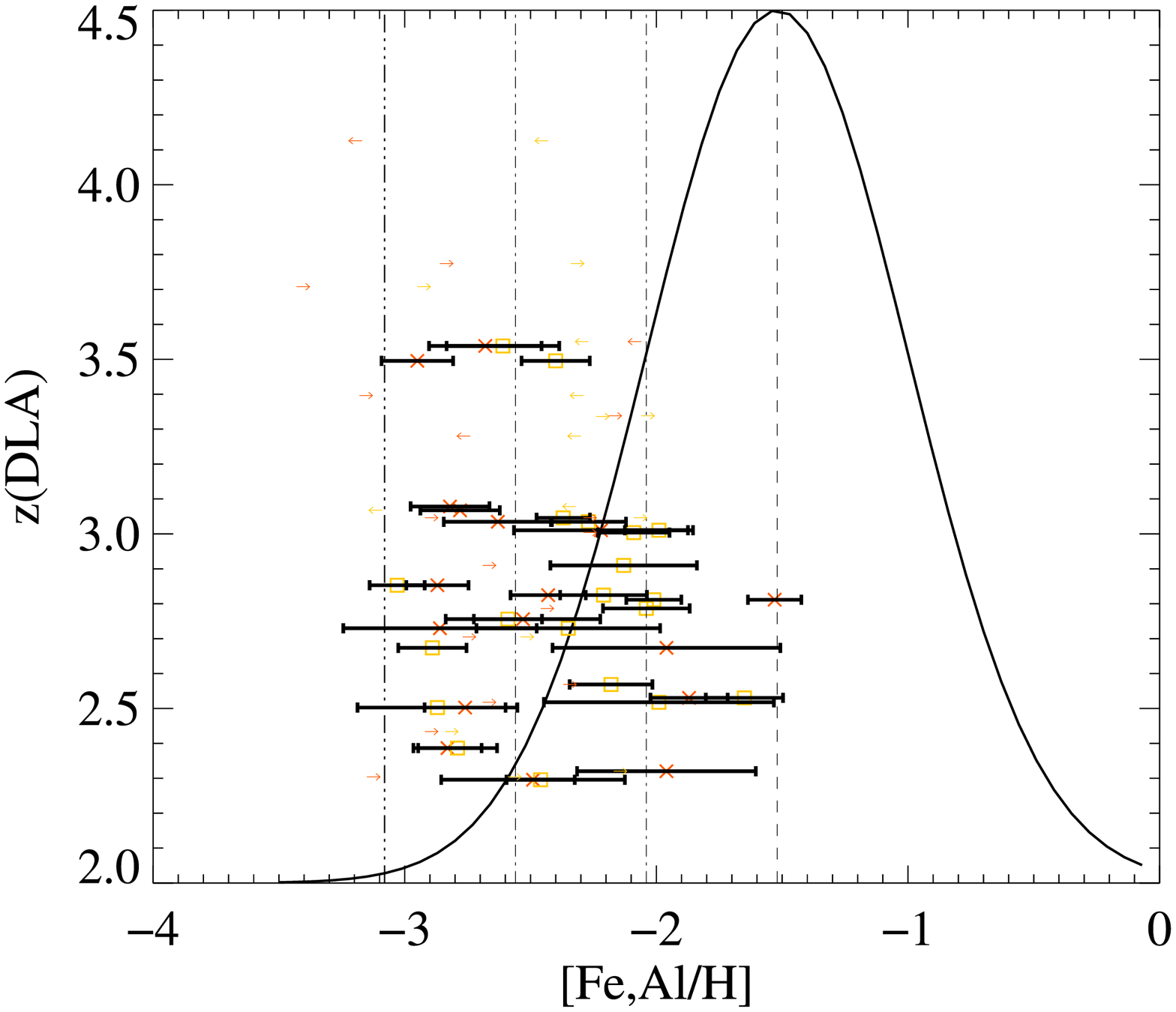}
\caption{Plot of  DLA element abundance relative to solar ([X/H]) vs redshift for our sample (symbols with error bars), compared with that of Prochaska et al.\ (2003) 
(plus signs) for the Fe-peak elements (top), and compared to a histogram of metallicity for the same sample (bottom). The Fe-peak elements presented include [Fe/H] (squares) and [Al/H] (x's). }
\label{MHzfigothers}
\end{figure}

\subsection{C and O Abundances and Comparison with Stellar Sample}
\label{sec:coabund}

The lowest abundances of our sample enable a comparison with metal-poor stellar samples, and as such the DLA abundances provide a complementary probe of important nucleosynthetic quantities such as O/C and Si/O yields within the first stars that enrich the DLA with heavy elements. Models of high mass stellar nucleosynthesis predict small values of [C/O], due to the prediction of a large Oxygen rich core in the final stages of the Pop III star (with 140 \msun $<$ \mstar\ $<$ 260 \msun), and a much larger yield of O compared to C in massive star explosion, typically with a factor of 10 more O than C produced (Heger and Woosley, 2002).  However, a great deal of uncertainty exists in reaction rates for the alpha processing of C into O, and other factors such as rotation, mass loss, and convection will have a significant impact on the observed nuclear abundances \citep[e.g.][]{chie04}.

\begin{figure}[t!]
\epsscale{0.96}
\plotone{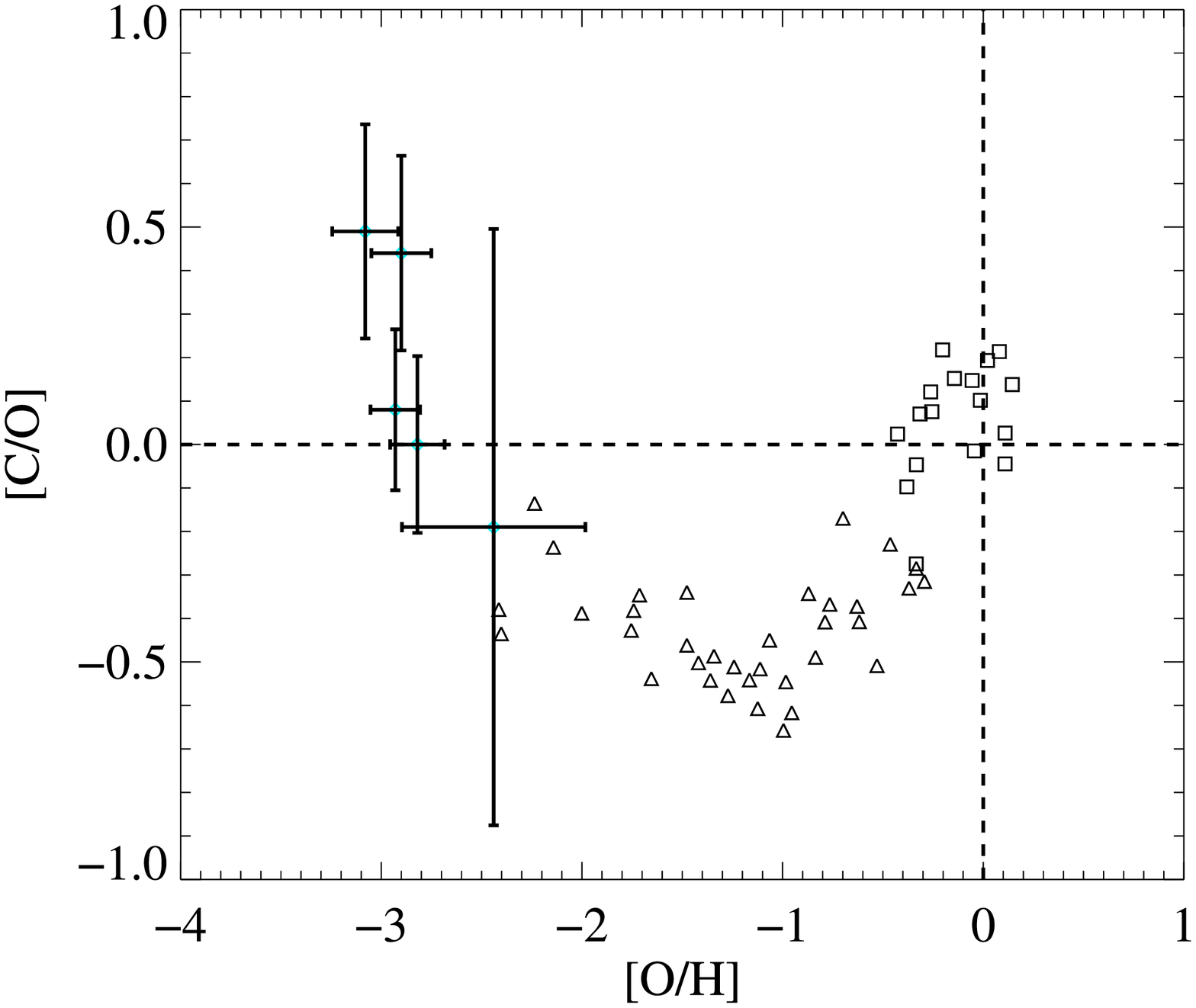}
\caption{ [C/O]  vs [O/H] as in Figure \ref{coratplot} observed only for those points with small equivalent widths with equivalent width less than 130 m\AA  for both \ion{C}{2} and \ion{O}{1}, and corrected for saturation using the conservative procedure described in $\S$~\ref{sec:saturate}, which assumes a b-value of  b = 7.5 \kms. The upward trend in [C/O] persists for this smaller sample, and suggests that the effect is independent of saturation effects. Stellar measurements of [C/O](squares and triangles) from Akerman, et al (2004) are also plotted for comparison.
}
\label{coratplot}
\end{figure}

Figure \ref{coratplot} presents the results for [C/O] of our newly measured low-metallicity DLA systems (diamonds), compared to a the data of Akerman et al (2004) (squares are Galactic disk stars; triangles are Galactic halo stars).  
Our results confirm the upward trend of [C/O] for low metallicity and low [O/H].  We have provided corrections for saturation where relevant to the column densities within Figure \ref{coratplot}, as described in the above sections. We also caution that one or more of these values could be compromised if a system has a very low Doppler parameter. We plan to confirm the results presented here with followup echelle observations.

In some cases it is possible that saturation corrections may be a source of the upward trend shown in Figure \ref{coratplot}, especially since some of our saturation corrections were adjusted to account for b-values determined from curve of growth fitting. 
Recent studies have shown that curve of growth fits of limited resolution spectra for GRB spectra can underestimate column densities, as they occasionally derive larger b-values (and hence lower column densities) if those spectra contain numerous unresolved components of low b-value (Prochaska 2006). 
For those DLA systems with significant corrections for saturation, a lower b-value would cause us to underestimate the columns of our absorbing species, especially in the strong lines of \ion{C}{2} and \ion{O}{1}. 

To limit  this possibility, we have included in Figure \ref{coratplot} only those DLA systems which have weak absorption in both \ion{C}{2} and \ion{O}{1}, with equivalent widths less than 130 m\AA\ in both species. Within this sub-sample, we have estimated the saturation corrections, using corrections to the weak line AOD column densities, appropriate to b-values of 7.5 \kms, as described in $\S$~\ref{sec:saturate}. These corrections are conservative, and in most cases, the weak lines result in small corrections of only 0.1 dex or less.  We also note that corrections for saturation would likely increase the C/O ratio because the C column density derived from the \ion{C}{2} ~$\lambda$1334 transition is more likely to be underestimated than the O column density from the \ion{O}{1} ~$\lambda$1302 transition, due to its larger oscillator strength.  The enhancement of [C/O] is also strongest within the DLA systems with the weakest \ion{C}{2} and \ion{O}{1} lines, suggesting that the trend is not a result of saturation effects in our spectra.

\subsection{Fe Abundances Compared with C, O, and Si abundances}

\begin{figure}[t!]
\epsscale{0.96}
\plotone{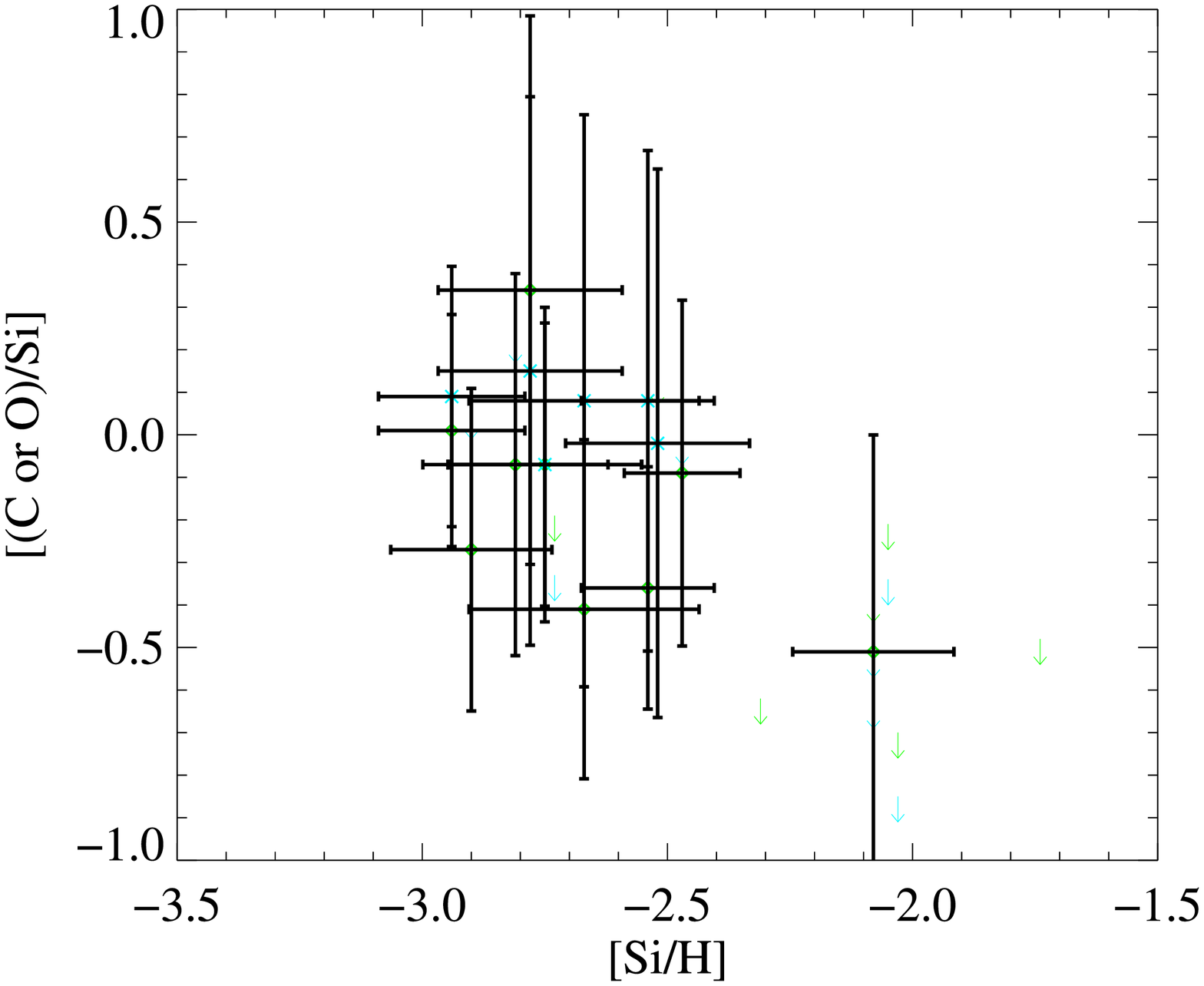}
\caption{ Trend in metal abundances of [C/Si] (turquoise X) and [O/Si] (green diamonds) against [Si/H]. The data shows a systematic dependence of [C/Si] and [O/Si] against Si/H.}
\label{sihratplot}
\end{figure}

\begin{figure}[t!]
\epsscale{0.96}
\plotone{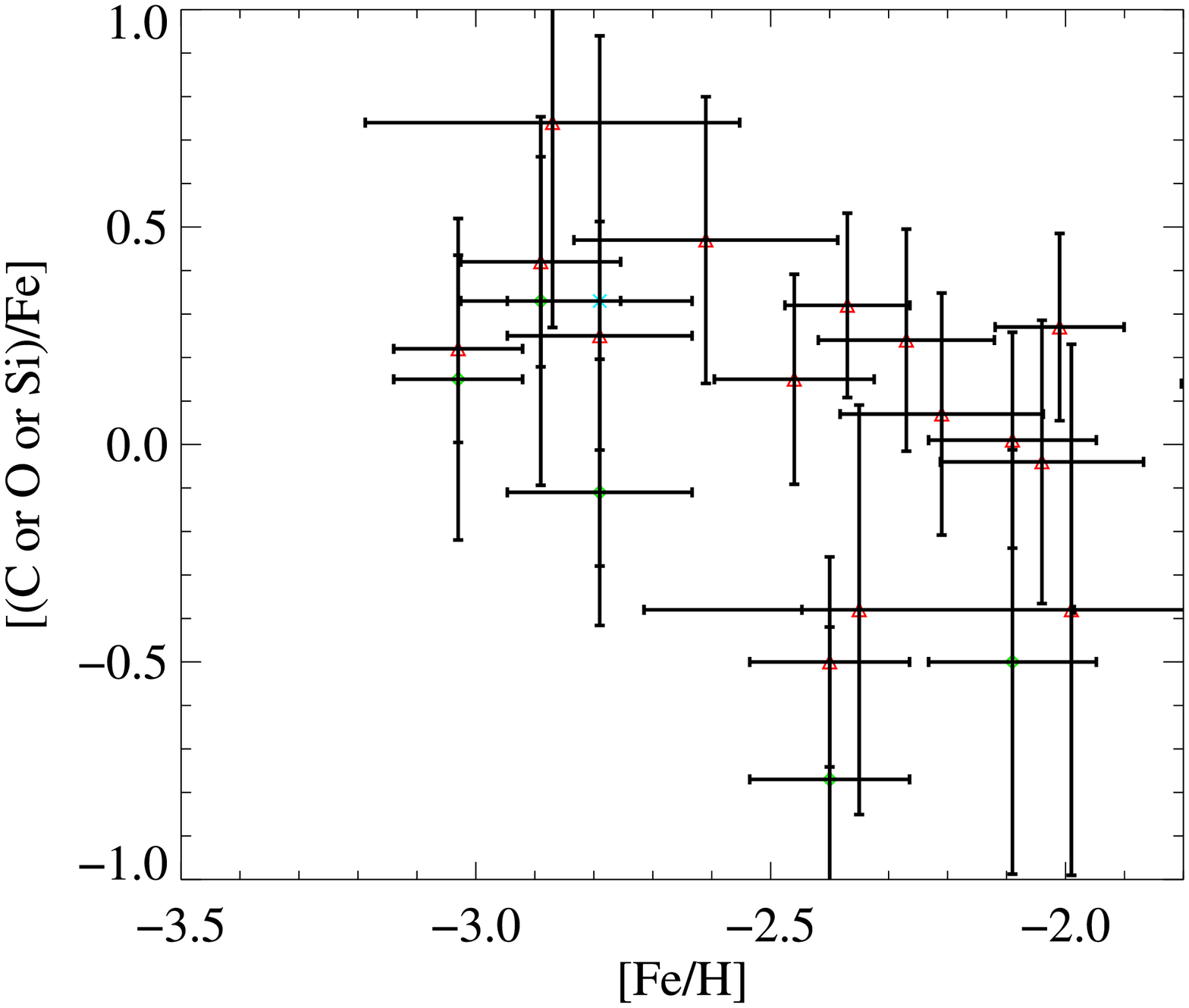}
\caption{ Trend in metal abundances of [C/Fe] ( turquoise X), [O/Fe] (green diamonds) and [Si/Fe] (red triangles) against [Fe/H]. The data shows a systematic increase of lighter element abundances against Si/H.}
\label{fehratplot}
\end{figure}

We have also examined trends in the ratio of metal abundances aginst the [Si/H] and [Fe/H] metallicity, to see if systematic trends in elemental abundances exist at lower metallicities. Figure \ref{sihratplot} shows the trend of [C/Si] and [O/Si], measured against the [Si/H] metallicity, while Figure \ref{fehratplot} shows the element ratios [C/Fe], [O/Fe] and [Si/Fe]. In both cases we see trends of increasing [C/X] and [O/X] for the sample with decreased metallicity. This suggests an upward trend of $\alpha$/Fe, which has implications for the nucleosynthesis which produced the low metal abundances within our DLA sample, perhaps due to a mix of more high-mass stars within DLA systems which would be expected to produce large amounts of $\alpha$-process elements with the lowest values of [X/H].  One might argue that some of the reduced values of $\alpha$/Fe within the sample at higher metallicities could arise from depletion effects, and higher spectral resolution observations which include highly-depleted species such as Cr are needed to help elucidate this effect. However since the depletion of Fe relative to Si increases with dust content, with 0.5 $<$ [Si/Fe] $<$ 1.0 depending on whether the IGM is "halo" or "disk" type ISM (\cite{sav96}), one would expect that the values of [Si/Fe] and $\alpha$/Fe to increase with increasing metallicity, which is the opposite of what is observed.  It therefore seems more likely that our trend of increasing values of $\alpha$/Fe for lower metallicity DLA systems is a result of nucleosynthetic effects.

\subsection{Comparison of Elemental Abundances with Nucleosynthesis Models}
\label{sec:nucsynth}

Our results help provide constraints on the nucleosynthetic yields of high-mass stars, since the low values of [O/H] observed in our sample suggest that they were enriched by only one to a few supernovae, 
depending on the value of [O/H]. The higher yield of [C/O] is not predicted by nucleosynthesis models of Heger and Woosley (2002), and Umeda and Nomoto (2002), but is consistent with the predictions of Chieffi and Limongi (2002, 2004) for stars with masses ranging from 20 \msun $<$ \mstar\ $<$ 80 \msun\ . The higher yield of [C/O] is also consistent with those seen in emission line studies of HII regions (Akerman et al 2004), and is within the range of values of [C/O] estimated for the Lyman-$\alpha$ forest, where near solar values of  [C/O] are predicted using the most realistic ionizing radiation models with the overdensity $\delta$ $<$ 10. (\cite{agu08}).

\begin{figure}[t!]
\epsscale{0.91}
\plotone{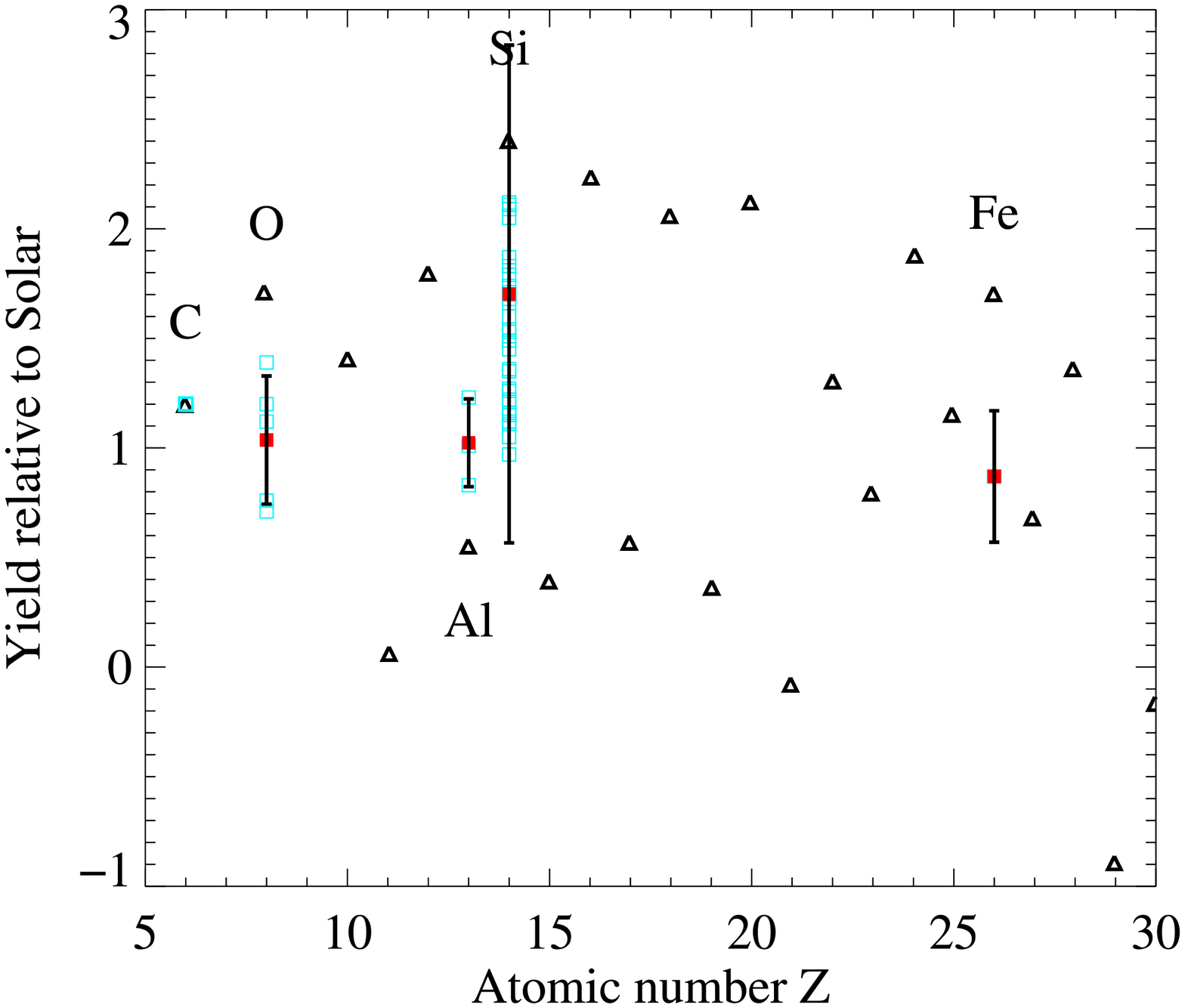}
\caption{ Plot of element abundances for the DLA systems shown in Figure \ref {coratplot} against element number, with open squares representing individual DLA systems, and the mean and standard deviation of the sample of four DLAs shown by the filled squares and error bars. The nucleosynthetic yields of Heger and Woosley (2002) for the various elements are shown with triangles. We observe reduced abundances of O, Si, and Fe relative to C compared to the nucleosynthetic model, and a slight enhancement of Al relative to C.}
\label{abundplot}
\end{figure}

We also present a plot in Figure \ref{abundplot} of our elemental abundances against atomic number, for a subset of 4 of our highest quality low-metallicity DLA systems. We observe that compared to the models of Heger and Woosley, we observe a reduced abundance of O, Si, and Fe compared to the model, and an overabundance of Al.  
This result also suggests that the ratio of [$\alpha$/Fe] in low-metallicity DLA systems appears closer to a value of [$\alpha$/Fe] = 0, instead of higher values  at lower metallicity, as some models have suggested (e.g. Tinsley, 1979). The mean underabundance of the elements O, Si and Fe for our sample is -0.72 dex relative to C when compared to the models of Heger and Woosley (2002), while the Al abundance we observe is +0.52 dex above the model. 

\begin{figure}[t!]
\epsscale{0.91}
\plotone{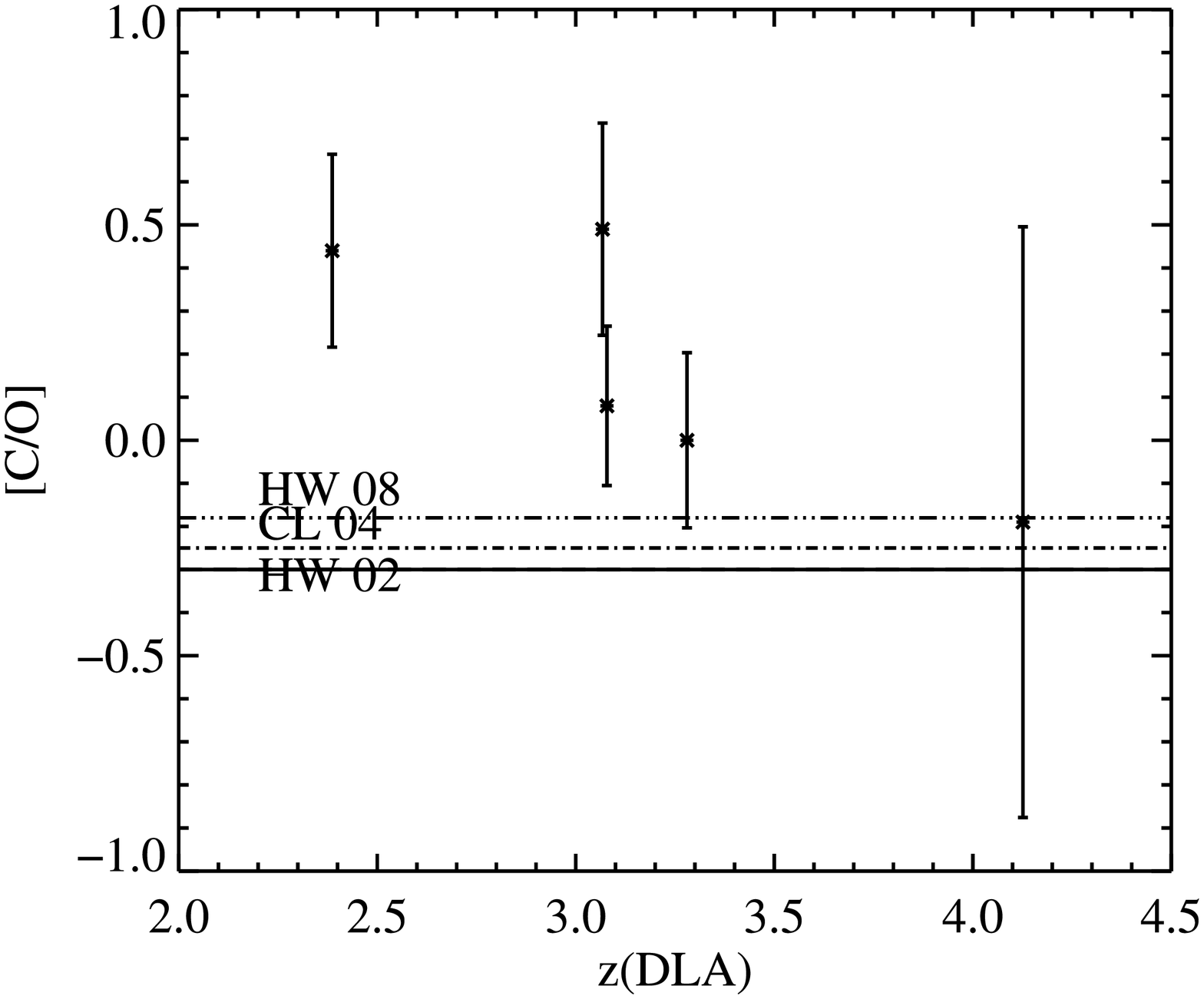}
\caption{ Plot of our observed values of [C/O] against redshift, for our DLA sample. The predicted yields for nucleosynthesis models of massive stars are indicated by the horizontal lines for the models of Heger and Woosley (2008) (top), Chieffi and Limongi (2004) (middle), and Heger and Woosley (2002) (bottom). Also indicated is the mean value of [C/O] = -0.30 for the high redshift DLA systems observed by Becker et al (2006) (solid line)}
\label{abundplot2}
\end{figure}

Figure \ref{abundplot2} summarizes the observed ratios of [C/O] compared with various nucleosynthesis models as a function of redshift.  The horizontal lines indicate the predicted yields from the models of Heger and Woosley (2008), Chieffi and Limongi (2004) and Heger and Woosley (2002). Results from a survey of high redshift DLA systems by Becker et al (2006), which may in some ways be similar to our low metallicity sample, are also indicated on the plot. The results show that the largest values of [C/O] occur for the redshift 2.4 DLA system, and that the higher redshift DLA systems within our sample appear to converge with the value predicted by most nucleosynthesis models of -0.18 $<$ [C/O] $<$ -0.30, which also is close to the mean value of [C/O] = -0.30 for the high redshift DLA systems observed by Becker et al (2006). The data are consistent with a trend of increasing [C/O] with 
decreasing redshift, 
which could indicate the presence of intermediate-mass stars 
30 \msun $<$ \mstar $<$ 60 \msun\ providing additional enrichment of Carbon at later times compared to the more massive stars for which these nucleosynthesis models were calculated.

\begin{figure}[t!]
\epsscale{0.91}
\plotone{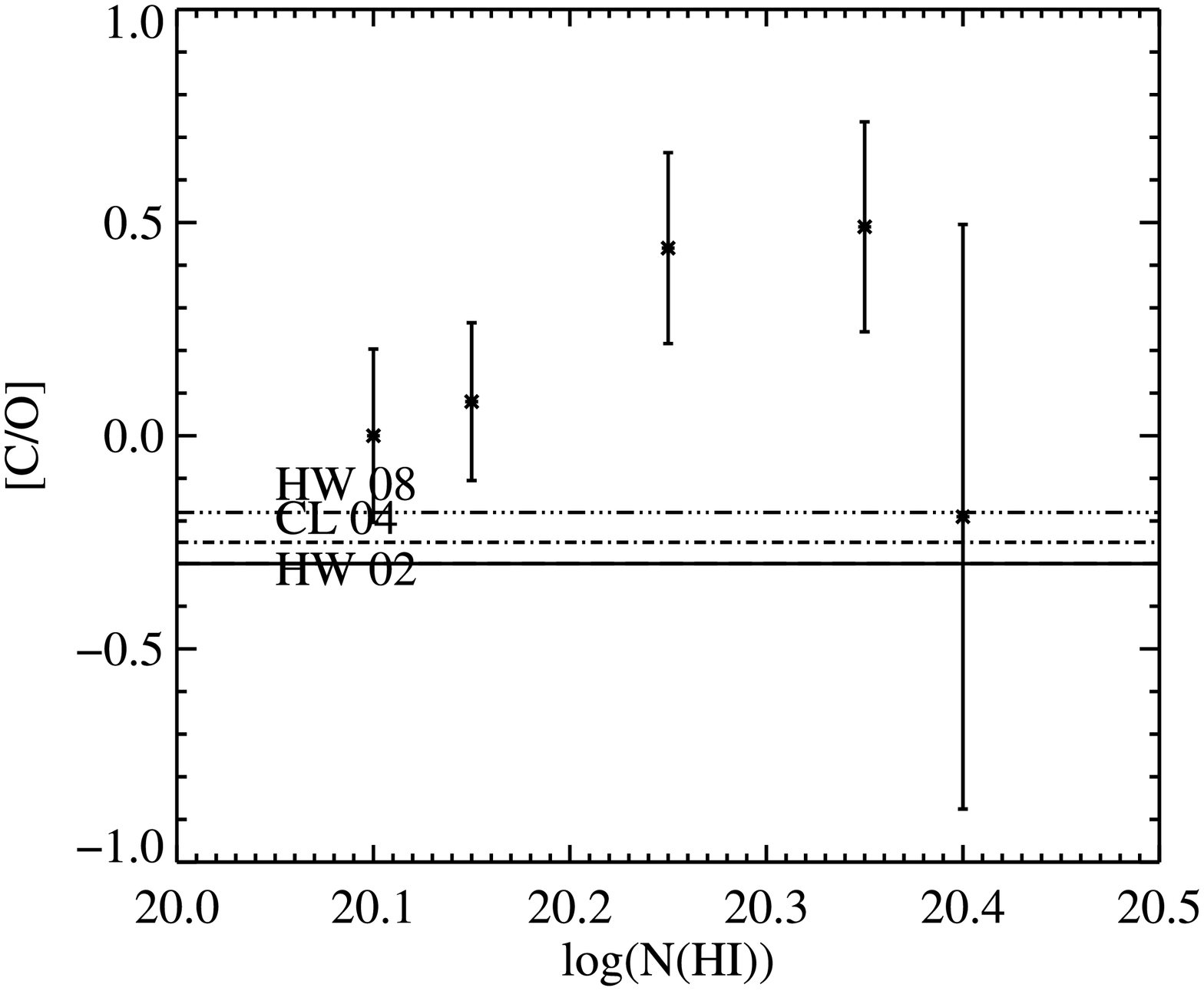}
\caption{ Plot of the observed values of [C/O] as in Figure \ref{abundplot2}, but against the measured column density of \ion{H}{1}. The lowest column densities of both \ion{C}{2} and \ion{O}{1} are observed toward systems which are technically "sub-DLAs", since  N(\ion{H}{1}) $<$ 20.3 for these systems. The observed enhancement of [C/O] for our observed systems is larger than the predicted ionization corrections, which are in the range of 0.1-0.2 dex. }
\label{abundplot3}
\end{figure}

To help determine whether the observed enhancement of [C/O] could result from ionization effects, we have plotted the observed ratio of [C/O] against \ion{H}{1} column in Figure \ref{abundplot3}. 
The lowest column densities of \ion{C}{2} and \ion{O}{1} were observed toward systems which include sub-DLA's, specifically systems with 
values of \ion{HI}{1} ranging from 20.1 $<$ N(\ion{H}{1}) $<$ 20.4. 
For these systems, it is possible that a non-zero ionization correction is
necessary for these species.  
Figure \ref{abundplot3} does not indicate a trend in [C/O] with decreasing column of \ion{H}{1}. Ionization corrections for sub-DLA's for  20.1 $<$ log(N(\ion{H}{1})) $<$ 20.4 based on CLOUDY models are expected to be in the range of 0.1 $< \delta [C/O] < $ 0.3 and produce an enchancement of \ion{C}{2} relative to \ion{O}{1} due to the lower ionization potential of \ion{O}{1} compared to \ion{C}{2}. This effect could help explain some of the super-solar values of [C/O] in our sample, but cannot explain all of 
the observed [C/O] enhancement at low [O/H]. 

\begin{figure}[t!]
\epsscale{0.91}
\plotone{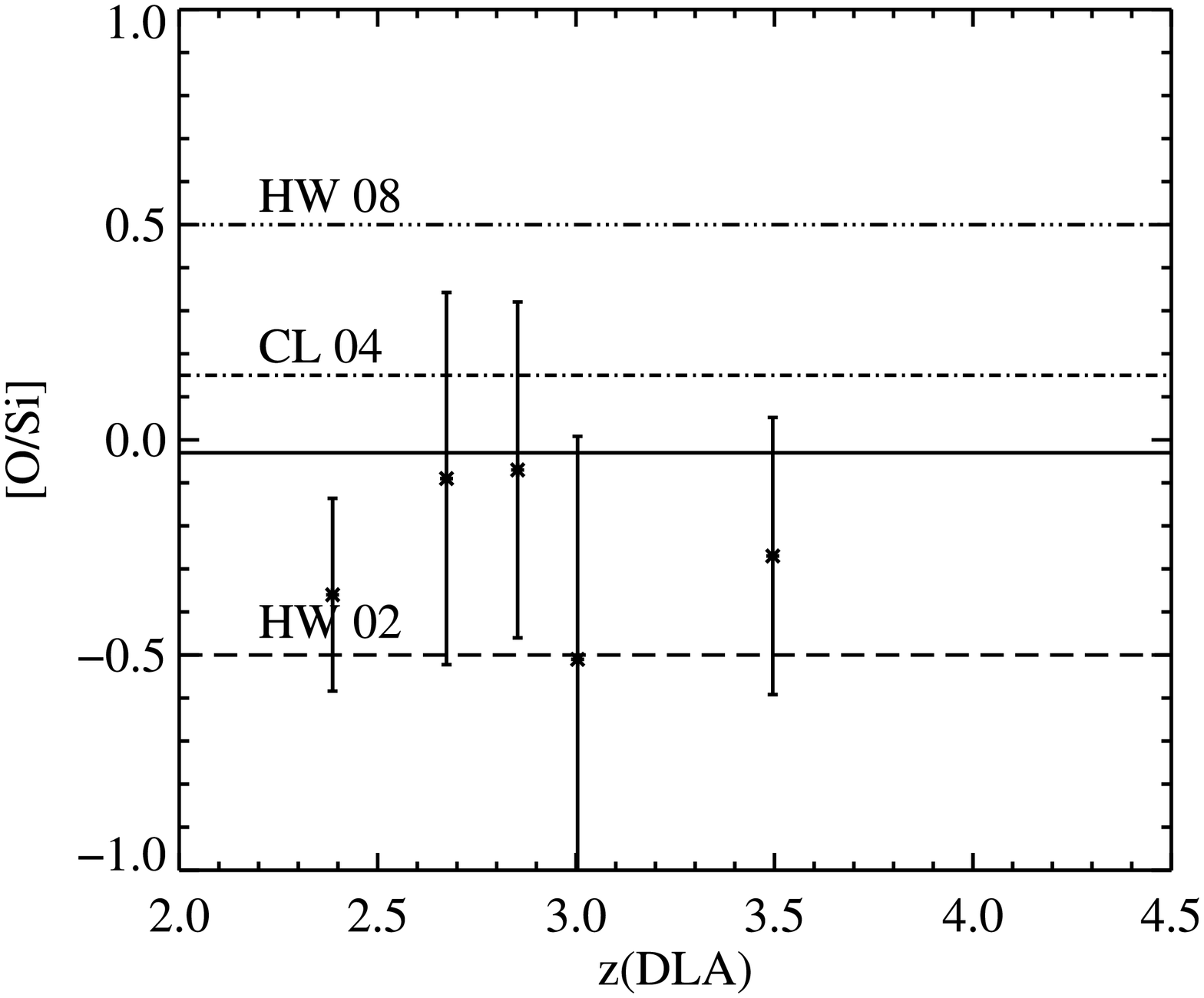}
\caption{ Plot of the observed ratios of the elements [O/Si] the same systems as in Figure \ref{abundplot2}, compared with nucleosynthesis models and the high redshift DLA systems observed by Becker et al (2006) (solid line). The values of [O/Si] for our sample seem to agree with the high redshift DLA sample, and are intermediate between two of the nucleosynthesis models.}
\label{abundplot4}
\end{figure}

We also compare the observed ratios of [O/Si] with nucleosynthesis models as a function of redshift in Figure \ref{abundplot4}. The observed ratios of [O/Si] appear to not be correlated with redshift, and agree well with nucleosynthesis models, and with the high redshift DLA systems observed by Becker et al (2006). 

\subsection{\ion{C}{4} and \ion{Si}{4} absorption within the DLA systems}

\begin{figure}[t!]
\epsscale{0.91}
\plotone{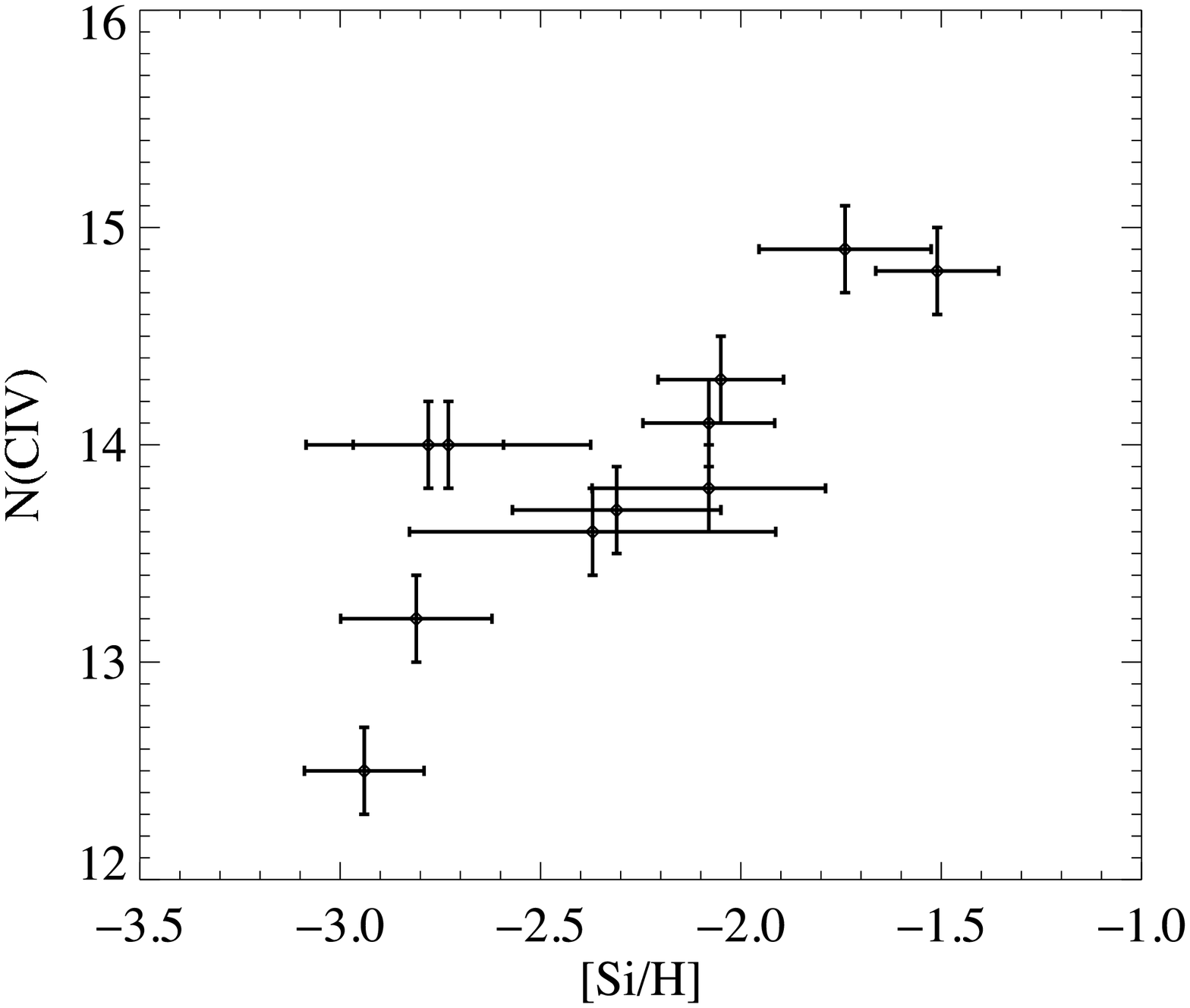}
\caption{ Plot of the observed metallicity (in this case [Si/H]) compared with the CIV column density for our sample, showing an obvious correlation between the N(\ion{C}{4}) and [Si/H], which could reflect an increase in velocity dispersion for higher metallicity DLA systems. }
\label{abundcivplot}
\end{figure}

The presence of \ion{C}{4} and \ion{Si}{4} absorption was detected in most of our DLA systems, but not all of them. 
These latter examples are among the very few DLAs with undetected high-ion
absorption (Lu et al.\ 1996; Wolfe \& Prochaska 2000; Fox et al.\ 2007).
Table \ref{tab:civ} shows the derived column densities of these two
species, for the entire DLA sample. The fact that several of the DLA
systems have no detected \ion{C}{4} or \ion{Si}{4} suggests that the
low metallicity DLA systems could be sampling a more neutral region or
regions with lower ionizing flux than a typical DLA, where these
species are nearly always detected. We also observe within our sample
an increasing trend in both N(\ion{C}{4}) and N(\ion{Si}{4}) with
metallicity (as determined by [Si/H]), which is consistent with the
observations of larger DLA samples (Fox, et al, 2007). The equivalent
widths of \ion{C}{4} and \ion{Si}{4} could be indicative of a larger
velocity dispersion as well as the column densities of these species,
due to saturation effects. If so, then the trend in these species with
[X/H] is consistent with the observation of larger velocity dispersion
and velocity range with increasing metallicity, suggestive of an
increasing mass of the DLA system with metallicity (Wolfe \& Prochaska
1998; Ledoux et al.\ 2006). 
Figure \ref{abundcivplot} shows the trend of N(\ion{C}{4}) with increasing metallicity, and further observations at higher resolution can help constrain whether this correlation is caused by increasing velocity dispersion within the higher metallicity DLA systems.

We also have derived estimates of N(\ion{H}{2})/N(\ion{H}{1}) for the low metallicity DLA sample, using the column densities of \ion{C}{2} and \ion{C}{4} and the prescription of Fox, et al 2007, which assumes f(\ion{C}{4}) = 0.3. For the two systems which have detected \ion{C}{4} and unsaturated \ion{C}{2}, we derive values of N(\ion{H}{2})/N(\ion{H}{1}) in the range -0.5 $<$ log(N(\ion{H}{2})/N(\ion{H}{1})) $<$ -0.1, which is slightly higher than the mean value of N(\ion{H}{2})/N(\ion{H}{1}) for the larger DLA sample of Fox et al (2007).

\section{Summary}
\label{discussion}

Our survey of DLAs pre-selected to have weak metal-line absorption
has established a "floor" of metallicity in the ISM of high $z$ galaxies
at $\approx 1/1000$ solar abundance.
Because these systems represent the most likely systems to exhibit very
low abundance in the available SDSS sample,  
it is unlikely that metallicities much lower than these observed values 
will be found in future observations of DLA systems. The observed floor 
for the DLA systems with detected C, O, Si, Fe, and Al are found to be [C/H] $>$ -2.9, [O/H] $>$ -3.2, [Si/H] $>$ -2.9, [Al/H] $>$ -2.9, and [Fe/H] $>$ -3.2. The existence of a "floor" of metallicity is demonstrated from our observations, regardless of the effects of saturation on our spectra, which could transform some of our reported elemental abundances into lower limits for very low values of Doppler parameter (b $<$ 5 \kms).  In a very few cases, DLA systems without detected Si or Fe lines could have abundances lower than these "floor" values, but higher resolution spectra or higher signal to noise spectra are needed to test this.

From this study, we observe multiple low metallicity systems with multiple transitions which have [X/H] $\approx -2.8$, enabling a comparison with nucleosynthesis models, and providing the first observations of an "ultra-low metallicity" DLA sample. A comparison of our results with stellar abundance studies is consistent, and extends the [C/O] observations to new low values of [O/H], in which an upward trend of [C/O] is observed, consistent with nucleosynthetic yields of high mass stars with masses between 20-80 \msun\ . We observe in the DLA abundances a nucleosynthesis pattern consistent with a Pop III population with  IMF that includes stars from 20 \mstar $<$ \mstar\ $<$ 80 \msun\ , and further studies are ongoing to provide higher resolution spectra of these DLA systems to include more elements, and to also extend our sample to provide a larger statistical sample of low metallicity quasars. 
The low number of DLA systems with [$\alpha$/Fe] $<$ -3.0 even in our carefully selected sample suggests that the fraction of low-metallicity DLA systems and galaxies with [$\alpha$/Fe] $<$ -3.0 may be as low as $\sim$ 0.135 percent, based on a consideration of the statistics of DLA metallicity, and the results from our DLA sample.  

\acknowledgements{The observations from the Keck telescope were made 
possible by Pomona College travel grants to B.E.P., I.T.M., and D.J.B and a Pomona SURP grant
to I.T.M. The authors also wish to recognize the very significant cultural role and reverence that
  the summit of Mauna Kea has always had within the indigenous
  Hawaiian community.  J.X.P. wishes to acknowledge
funding through an NSF CAREER grant (AST--0548180) and an additional NSF grant (AST--0709235). 
  W.W.S and B.E.P. wish to acknowledge support from an NSF grant (AST-0606868).}


\begin{deluxetable}{ccc}
\tablewidth{0pc}
\tablecaption{SDSS-DR5 Metal-Poor DLA Candidates\label{tab:candidates}}
\tabletypesize{\footnotesize}
\tablehead{\colhead{QSO} &
\colhead{$z_{\rm abs}$} &
\colhead{log $N_{\rm HI}$} }
\startdata
J001115.23$+14$4601& 3.612&20.65\\
J001134.52$+15$5137& 4.317&20.50\\
J001328.21$+13$5828.0& 2.612&20.10\\
J003749.19$+15$5208& 3.479&20.35\\
J012211.11$+15$0914.3& 3.519&20.45\\
J012403.77$+00$4432& 3.077&20.30\\
J014609.33$-09$2918& 3.680&20.25\\
J032226.09$-05$5824& 3.763&20.30\\
J033119.66$-07$4143& 4.192&20.65\\
J034300.88$-06$2229.9& 3.511&20.15\\
J073146.99$+36$4346& 3.591&20.80\\
J073149.51$+28$5448& 2.686&20.55\\
J073823.94$+34$0303.8& 3.817&20.15\\
J073938.85$+30$5951& 3.355&20.10\\
J074145.00$+21$5932.9& 2.578&20.80\\
\enddata
\tablecomments{{} List of all candidate metal-poor DLAs flagged by Prochaska and collaborators from their analysis of the SDSS-DR5.  We caution that many of these will not be bona-fide DLAs.}
\tablecomments{[The complete version of this table is in the electronic edition of the Journal.  The printed edition contains only a sample.]}
\end{deluxetable}
 
\clearpage

\clearpage
\begin{deluxetable}{ccccccc}
\tablewidth{0pc}
\tablecaption{Targets Observed, Magnitudes and Exposure Times\label{tab:obstab}}
\tabletypesize{\footnotesize}
\tablehead{\colhead{Quasar} & \colhead{$z_{DLA}$} & \colhead{RA(2000)}
& \colhead{Dec(2000)}
& \colhead{R}
& \colhead{Exp.(2008)}
& \colhead{Exp.(2007)} \\
& & & & &(sec)& (sec) }
\startdata
SDSS0751+4516&3.046&07:51:13.05&+31:20:37&19.90&...&900. \\
SDSS0759+3129&3.035&07:59:51.83&+31:29:41&18.78&600.&...\\
SDSS0811+2838&2.434&08:11:10.32&+28:38:14&18.87&...&1200.\\
SDSS0814+5029&3.708&08:14:35.18&+50:29:46&18.34&...&1800.\\
SDSS0831+3358&2.304&08:31:02.55&+33:58:03&18.76&...&1000.\\
SDSS0844+4624&3.336&08:44:38.63&+46:24:25&19.54&...&1800.\\
SDSS0910+1026&2.398&09:24:59.91&+09:51:03&19.75&...&1200.\\
SDSS0924+0951&3.338&09:24:59.91&+09:51:03&19.75&2400.&...\\
SDSS0928+0939&2.910&09:28:14.93&+09:39:55&19.44&1800.&...\\
SDSS0940+0549&2.577&09:40:54.56&+05:49:03&18.90&1100.&1200.\\
""&2.578&""&""&""&""&""\\
SDSS0955+4116&3.280&09:55:42.12&+41:16:55&19.36&1800&...\\
SDSS1001+0343&3.078&10:01:51.45&+03:43:01&19.02&1222.&1200.\\
SDSS1003+5520&2.502&10:03:21.11&+55:20:59&19.40&2000.&...\\
SDSS1031+4055&2.569&10:31:26.13&+40:55:32&19.43&2000.&...\\
SDSS1037+0139&2.705&10:37:24.40&+01:39:33&19.14&...&1400.\\
SDSS1048+3911&2.296&10:48:26.03&+39:11:10&18.82&1200.&...\\
SDSS1108+1209&3.396&11:08:55.47&+12:09:53&18.63&...&1200.\\
SDSS1156+5513&2.481&11:56:59.59&+55:13:08&18.84&1200.&1200.\\
""&2.481&""&""&""&""&""\\
""&2.498&""&""&""&""&""\\
SDSS1219+1603&3.004&12:19:28.92&+16:03:57&19.47&1800&...\\
SDSS1251+4120&2.730&12:51:25.36&+41:20:00&18.95&1300.&...\\
SDSS1305+2902&2.386&13:05:48.92&+29:02:28&18.77&971.&...\\
SDSS1325+1255&3.550&13:25:54.12&+12:55:46&19.19&1430.&...\\
SDSS1327+4845&2.447&13:27:29.75&+48:45:00&19.30&1800.&...\\
""&2.612&""&""&""&""&...\\
SDSS1331+4838&3.692&13:31:46.21&+48:38:26&19.26&1800.&...\\
SDSS1349+1242&3.743&13:49:39.78&+12:42:30&19.61&2200.&...\\
SDSS1350+5952&2.756&13:50:44.21&+59:52:05&19.26&1800.&...\\
SDSS1358+0349&2.853&13:58:03.97&+03:49:36&18.65&930.&...\\
SDSS1358+6522&3.067&13:58:42.92&+65:22:36&18.71&1000.&...\\
SDSS1402+5909&3.774&14:02:43.97&+59:09:58&19.80&2500.&...\\
SDSS1440+0637&2.518&14:40:51.89&+06:37:09&19.33&1800.&...\\
""&2.825&""&""&""&""&...\\
SDSS1456+0407&2.320&14:56:24.66&+04:07:41&19.23&1600.&...\\
""&2.674&""&""&""&""&...\\
SDSS1557+2320&3.540&15:57:38.39&+23:20:57&19.46&1800.&...\\
SDSS1637+2901&3.496&16:37:47.58&+29:01:35&19.91&2400.&...\\
SDSS1654+3509&2.811&16:54:08.17&+35:09:54&18.92&1200.&...\\
SDSS1709+3417&2.530&17:09:31.01&+34:17:31&19.32&1611.&...\\
""&3.010&""&""&""&""&...\\
SDSS1717+5802&3.046&17:17:19.45&+58:02:18&19.61&2100.&...\\
SDSS2114-0632&4.126&21:14:50.33&-06:32:57&19.75&2400.&...\\
\enddata
\tablenotetext{a}{Velocity interval over which the equivalent width and
column density are measured.}
\tablenotetext{b}{Rest equivalent width.}
\end{deluxetable}

\clearpage
\begin{deluxetable}{c|c c}
\tablewidth{0pc}
\tablecaption{Calculated metallicity limits for different ions, for a typical DLA with N(HI) = 20.5, z=3.0, and with a FWHM of 4 pixels. The resulting calclated limits assume a 4 $\sigma$ detection for spectra with the two values of $S/N$ listed.}
\tablehead{{\em Transition}&{S/N = 10}&{S/N = 20}}

\startdata
OI   1302 	& -3.5  &  -3.8	 \\
SiII 1304       & -2.6  &  -2.9 \\
CII  1334       & -3.8  &  -4.1  \\
SiII 1526       & -2.9  &  -3.2  \\
FeII 1608       & -2.6  &  -2.9  \\
AlII 1670       & -3.1  &  -3.4  \\
\hline
	
\enddata
\label{limitstable}
\end{deluxetable}


\clearpage
\begin{deluxetable}{lcccccc}
\tablewidth{0pc}
\tablecaption{IONIC COLUMN DENSITIES FOR J0955+4116 $z=3.280$\label{tab:J0955+4116}}
\tabletypesize{\footnotesize}
\tablehead{\colhead{Ion} & \colhead{$\lambda$} & \colhead{$\log f$}
& \colhead{$v_{int}^a$} 
& \colhead{$W_\lambda^b$} 
& \colhead{$\log N$} & \colhead{$\log N_{adopt}$} \\
& (\AA) & & (\kms) & (m\AA) & & }
\startdata
\ion{C}{2}\\
&1334.5323 &$ -0.8935$&$[-100,  60]$&$  77.8 \pm  10.4$&$ 13.58$&$ 13.72 \pm 0.14$\\
\ion{C}{2}*\\
&1335.7077 &$ -0.9397$&$[-100,  50]$&$<  21.2$&$< 13.26$& \\
\ion{C}{4}\\
&1548.1950 &$ -0.7194$&$[-100,  60]$&$  81.3 \pm  13.3$&$ 13.33 \pm 0.07$&$ 13.33 \pm 0.07$\\
&1550.7700 &$ -1.0213$&$[-100,  60]$&$<  27.3$&$< 13.32$& \\
\ion{O}{1}\\
&1302.1685 &$ -1.3110$&$[-100,  60]$&$  65.1 \pm   9.0$&$ 13.94 \pm 0.05$&$ 14.04 \pm 0.10$\\
\ion{Al}{2}\\
&1670.7874 &$  0.2742$&$[-100,  60]$&$<  25.2$&$< 11.92$&$< 11.92$\\
\ion{Si}{2}\\
&1260.4221 &$  0.0030$&$[-100,  60]$&$  87.0 \pm   6.1$&$ 12.79 \pm 0.03$&$ 12.96 \pm 0.18$\\
&1304.3702 &$ -1.0269$&$[-100,  60]$&$<  19.2$&$< 13.31$& \\
&1526.7066 &$ -0.8962$&$[-100,  60]$&$<  29.3$&$< 13.24$& \\
&1808.0130 &$ -2.6603$&$[-100,  60]$&$<  27.2$&$< 14.79$& \\
\ion{Si}{4}\\
&1393.7550 &$ -0.2774$&$[-100,  60]$&$  99.4 \pm  11.6$&$ 13.08 \pm 0.05$&$ 13.07 \pm 0.05$\\
&1402.7700 &$ -0.5817$&$[-100,  60]$&$  47.2 \pm  12.9$&$ 13.04 \pm 0.12$& \\
\ion{Fe}{2}\\
&1608.4511 &$ -1.2366$&$[-100,  60]$&$<  20.0$&$< 13.34$&$< 13.34$\\
\enddata
\tablenotetext{a}{Velocity interval over which the equivalent width and
column density are measured.}
\tablenotetext{b}{Rest equivalent width.}
\end{deluxetable}

\clearpage
\begin{deluxetable}{lcccccc}
\tablewidth{0pc}
\tablecaption{IONIC COLUMN DENSITIES FOR J1001+0343 $z=3.078$\label{tab:J1001+0343}}
\tabletypesize{\footnotesize}
\tablehead{\colhead{Ion} & \colhead{$\lambda$} & \colhead{$\log f$}
& \colhead{$v_{int}^a$} 
& \colhead{$W_\lambda^b$} 
& \colhead{$\log N$} & \colhead{$\log N_{adopt}$} \\
& (\AA) & & (\kms) & (m\AA) & & }
\startdata
\ion{C}{2}\\
&1334.5323 &$ -0.8935$&$[ -50,  50]$&$  76.7 \pm   9.5$&$ 13.63 \pm 0.06$&$ 13.76 \pm 0.13$\\
\ion{C}{2}*\\
&1335.7077 &$ -0.9397$&$[ -50,  50]$&$<  19.5$&$< 13.20$& \\
\ion{C}{4}\\
&1548.1950 &$ -0.7194$&$[-200,  50]$&$ 136.6 \pm  16.3$&$ 13.58 \pm 0.05$&$ 13.56 \pm 0.05$\\
&1550.7700 &$ -1.0213$&$[-200,  50]$&$  59.0 \pm  17.0$&$ 13.50 \pm 0.12$& \\
\ion{O}{1}\\
&1302.1685 &$ -1.3110$&$[ -50,  50]$&$  57.6 \pm   9.0$&$ 13.90 \pm 0.07$&$ 13.98 \pm 0.08$\\
\ion{Al}{2}\\
&1670.7874 &$  0.2742$&$[ -50,  50]$&$  32.6 \pm   9.4$&$ 11.87 \pm 0.13$&$ 11.87 \pm 0.13$\\
\ion{Si}{2}\\
&1260.4221 &$  0.0030$&$[ -50,  50]$&$  72.5 \pm   4.3$&$ 12.70 \pm 0.05$&$ 12.82 \pm 0.12$\\
&1304.3702 &$ -1.0269$&$[ -50,  50]$&$<  19.3$&$< 13.32$& \\
&1526.7066 &$ -0.8962$&$[ -50,  50]$&$<  19.0$&$< 13.04$& \\
&1808.0130 &$ -2.6603$&$[ -50,  50]$&$<  26.9$&$< 14.81$& \\
\ion{Si}{4}\\
&1393.7550 &$ -0.2774$&$[ -50,  50]$&$  47.7 \pm   9.8$&$ 12.75 \pm 0.09$&$ 12.75 \pm 0.09$\\
\ion{Fe}{2}\\
&1608.4511 &$ -1.2366$&$[ -50,  50]$&$< 38.7 $&$< 13.49$&$< 13.31$\\
&2344.2140 &$ -0.9431$&$[ -50,  50]$&$<  67.5$&$< 13.31$& \\
\enddata
\tablenotetext{a}{Velocity interval over which the equivalent width and
column density are measured.}
\tablenotetext{b}{Rest equivalent width.}
\end{deluxetable}

\clearpage
\begin{deluxetable}{lcccccc}
\tablewidth{0pc}
\tablecaption{IONIC COLUMN DENSITIES FOR J1219+1603 $z=3.003$\label{tab:J1219+1603}}
\tabletypesize{\footnotesize}
\tablehead{\colhead{Ion} & \colhead{$\lambda$} & \colhead{$\log f$}
& \colhead{$v_{int}^a$} 
& \colhead{$W_\lambda^b$} 
& \colhead{$\log N$} & \colhead{$\log N_{adopt}$} \\
& (\AA) & & (\kms) & (m\AA) & & }
\startdata
\ion{C}{2}\\
&1334.5323 &$ -0.8935$&$[ -70,  70]$&$  208.0 \pm   10.0$&$ 14.02 \pm 0.06$&$ > $14.22 \\
\ion{C}{4}\\
&1548.1950 &$ -0.7194$&$[-150,  100]$&$ 391.0 \pm  16.3$&$ 13.98 \pm 0.05$&$ 14.05 \pm 0.05$\\
&1550.7700 &$ -1.0213$&$[-150,  100]$&$ 231.0 \pm  17.0$&$ 14.05 \pm 0.12$& \\
\ion{O}{1}\\
&1302.1685 &$ -1.3110$&$[ -70,  70]$&$  112.0 \pm   9.0$&$ 14.18 \pm 0.07$&$ 14.52 \pm 0.34$\\
\ion{Al}{2}\\
&1670.7874 &$  0.2742$&$[ -70,  70]$&$  149.0 \pm   9.4$&$ 12.50 \pm 0.10$&$ > 12.60$\\
\ion{Si}{2}\\
&1260.4221 &$  0.0030$&$[ -70,  70]$&$  186.7 \pm   9.3$&$ > 13.11 $&$13.88  \pm 0.15$\\
&1304.3702 &$ -1.0269$&$[ -70,  70]$&$  79.1 \pm 7.0 $&$13.74 \pm 0.05$& \\
&1526.7066 &$ -0.8962$&$[ -70,  70]$&$  135.0 \pm   7.0$&$> 13.70$& \\
\ion{Si}{4}\\
&1393.7550 &$ -0.2774$&$[ -150,  100]$&$  256.0 \pm   9.8$&$ 13.45 \pm 0.10$&$ 13.45 \pm 0.10$\\
\ion{Fe}{2}\\
&1608.4511 &$ -1.2366$&$[ -70,  70]$&$  66.8 \pm   6.0$&$13.70 \pm 0.05$&$ 13.80 \pm 0.10$\\
\enddata
\tablenotetext{a}{Velocity interval over which the equivalent width and
column density are measured.}
\tablenotetext{b}{Rest equivalent width.}
\end{deluxetable}


\clearpage
\begin{deluxetable}{lcccccc}
\tablewidth{0pc}
\tablecaption{IONIC COLUMN DENSITIES FOR J1305+2902 $z=2.386$\label{tab:J1305+2902}}
\tabletypesize{\footnotesize}
\tablehead{\colhead{Ion} & \colhead{$\lambda$} & \colhead{$\log f$}
& \colhead{$v_{int}^a$} 
& \colhead{$W_\lambda^b$} 
& \colhead{$\log N$} & \colhead{$\log N_{adopt}$} \\
& (\AA) & & (\kms) & (m\AA) & & }
\startdata
\ion{C}{2}\\
&1334.5323 &$ -0.8935$&$[ -50,  70]$&$ 130.8 \pm  15.4$&$ 13.80 \pm 0.10$&$ 14.25 \pm 0.45$\\
\ion{C}{2}*\\
&1335.7077 &$ -0.9397$&$[ -90,  70]$&$<  35.0$&$< 13.46$& \\
\ion{C}{4}\\
&1548.1950 &$ -0.7194$&$[ -90,  70]$&$ 109.8 \pm  15.5$&$ < 13.49 \pm 0.06$&$ < 13.52 $\\
&1550.7700 &$ -1.0213$&$[ -90,  70]$&$  77.6 \pm  15.8$&$ < 13.61 \pm 0.09$& \\
\ion{O}{1}\\
&1302.1685 &$ -1.3110$&$[ -90,  70]$&$  69.9 \pm  10.3$&$ 13.99 \pm 0.06$&$ 14.11 \pm 0.12$\\
\ion{Mg}{1}\\
&2852.9642 &$  0.2577$&$[ -90,  70]$&$< 109.6$&$< 12.11$&$< 12.11$\\
\ion{Mg}{2}\\
&2796.3520 &$ -0.2130$&$[ -90,  70]$&$ 254.5 \pm  54.4$&$> 12.88$&$> 12.88$\\
&2803.5310 &$ -0.5151$&$[ -90,  70]$&$ 115.8$&$> 12.46$& \\
\ion{Al}{2}\\
&1670.7874 &$  0.2742$&$[ -90,  70]$&$  26.7 \pm 5.0$&$ 11.82 \pm 0.10$&$ 11.96 \pm 0.10$\\
\ion{Si}{2}\\
&1193.2891 &$ -0.3019$&$[ -90,  70]$&$ 180.8 \pm  20$&$> 13.45 $&$  13.32 \pm 0.1$\\
&1260.4221 &$  0.0030$&$[ -90,  70]$&$ 220.9 \pm  21.4$&$> 13.33$&\\
&1304.3702 &$ -1.0269$&$[ -90,  70]$&$  21.5 \pm  5$&$ 13.37 \pm 0.1$& \\
&1526.7066 &$ -0.8962$&$[ -90,  70]$&$  30.7 \pm 5$&$ 13.26 \pm 0.1$& \\
&1808.0130 &$ -2.6603$&$[ -90,  70]$&$< 56.5 $&$< 14.97$& \\
\ion{Si}{4}\\
&1393.7550 &$ -0.2774$&$[ -90,  70]$&$  59.5 \pm  17.3$&$ 12.86 \pm 0.12$&$ 12.86 \pm 0.12$\\
&1402.7700 &$ -0.5817$&$[ -90,  70]$&$  35.7 \pm  12.0$&$ 12.89 \pm 0.12$& \\
\ion{Fe}{2}\\
&1608.4511 &$ -1.2366$&$[ -90,  70]$&$  25.5 \pm  10.6$&$ 13.24 \pm 0.13$&$ 13.00 \pm 0.13$\\
&2344.2140 &$ -0.9431$&$[ -90,  70]$&$<  42.4$&$> 12.99$& \\
&2374.4612 &$ -1.5045$&$[ -90,  70]$&$<  43.7$&$< 13.63$& \\
&2382.7650 &$ -0.4949$&$[ -90,  70]$&$ 118.3 \pm  21.4$&$> 12.90$& \\
&2600.1729 &$ -0.6216$&$[ -90,  70]$&$ 126.8 \pm  20.0$&$12.98 \pm 0.13$& \\
\enddata
\tablenotetext{a}{Velocity interval over which the equivalent width and
column density are measured.}
\tablenotetext{b}{Rest equivalent width.}
\end{deluxetable}

\clearpage
\begin{deluxetable}{lcccccc}
\tablewidth{0pc}
\tablecaption{IONIC COLUMN DENSITIES FOR J1358+0349 $z=2.853$\label{tab:J1358+0349}}
\tabletypesize{\footnotesize}
\tablehead{\colhead{Ion} & \colhead{$\lambda$} & \colhead{$\log f$}
& \colhead{$v_{int}^a$} 
& \colhead{$W_\lambda^b$} 
& \colhead{$\log N$} & \colhead{$\log N_{adopt}$} \\
& (\AA) & & (\kms) & (m\AA) & & }
\startdata
\ion{C}{2}\\
&1334.5323 &$ -0.8935$&$[ -80,  70]$&$ 174.0 \pm   9.0$&$ 13.93 \pm 0.15$&$ > 14.38 $\\
\ion{C}{2}*\\
&1335.7077 &$ -0.9397$&$[ -80,  70]$&$  19.9 \pm 8.0$&$ 13.24 \pm 0.14$& \\
\ion{C}{4}\\
&1548.1950 &$ -0.7194$&$[ -80,  70]$&$  86.6 \pm   8.9$&$ 13.33 \pm 0.04$&$ 13.25 \pm 0.04$\\
&1550.7700 &$ -1.0213$&$[ -80,  70]$&$  31.5 \pm   8.2$&$ 13.19 \pm 0.10$& \\
\ion{O}{1}\\
&1302.1685 &$ -1.3110$&$[ -80,  70]$&$ 100.5 \pm  10.7$&$ 14.13 \pm 0.05$&$ 14.38 \pm 0.25$\\
\ion{Al}{2}\\
&1670.7874 &$  0.2742$&$[ -80,  70]$&$  57.7 \pm  11.1$&$ 12.09 \pm 0.08$&$ 12.17 \pm 0.08$\\
\ion{Si}{2}\\
&1193.2891 &$ -0.3019$&$[ -80,  70]$&$ 86.0 \pm  8.0$&$ 13.13 \pm 0.05 $&$  13.30 \pm 0.17$\\
&1260.4221 &$  0.0030$&$[ -80,  70]$&$ 141.1 \pm   8.7$&$ 13.11 \pm 0.05$&\\
&1304.3702 &$ -1.0269$&$[ -40,  70]$&$<  19.0$&$< 13.32$& \\
&1526.7066 &$ -0.8962$&$[ -80,  70]$&$  64.0 \pm  10.2$&$ 13.41 \pm 0.05$& \\
&1808.0130 &$ -2.6603$&$[ -80,  70]$&$<  21.4$&$< 14.70$& \\
\ion{Si}{4}\\
&1393.7550 &$ -0.2774$&$[ -80,  70]$&$  76.2 \pm   9.9$&$ 12.96 \pm 0.06$&$ 12.96 \pm 0.06$\\
\ion{Fe}{2}\\
&1608.4511 &$ -1.2366$&$[ -80,  70]$&$<  19.9$&$< 13.35$&$ 13.01 \pm 0.05$\\
&2344.2140 &$ -0.9431$&$[ -80,  70]$&$  51.3 \pm  15.6$&$ 13.00 \pm 0.13$& \\
&2374.4612 &$ -1.5045$&$[ -80,  70]$&$  19.15 \pm 8.0 $&$ 13.08 \pm 0.10$& \\
&2382.7650 &$ -0.4949$&$[ -80,  70]$&$ 153.1 \pm  17.4$&$ 13.02 \pm 0.05$& \\
\enddata
\tablenotetext{a}{Velocity interval over which the equivalent width and
column density are measured.}
\tablenotetext{b}{Rest equivalent width.}
\end{deluxetable}

\clearpage
\begin{deluxetable}{lcccccc}
\tablewidth{0pc}
\tablecaption{IONIC COLUMN DENSITIES FOR J1358+6522 $z=3.067$\label{tab:J1358+6522}}
\tabletypesize{\footnotesize}
\tablehead{\colhead{Ion} & \colhead{$\lambda$} & \colhead{$\log f$}
& \colhead{$v_{int}^a$} 
& \colhead{$W_\lambda^b$} 
& \colhead{$\log N$} & \colhead{$\log N_{adopt}$} \\
& (\AA) & & (\kms) & (m\AA) & & }
\startdata
\ion{C}{2}\\
&1334.5323 &$ -0.8935$&$[ -70,  80]$&$ 122.0 \pm  12.5$&$ 13.78 \pm 0.04$&$ 14.22 \pm 0.43$\\
\ion{C}{2}*\\
&1335.7077 &$ -0.9397$&$[ -50,  30]$&$<  19.6$&$< 13.23$& \\
\ion{C}{4}\\
&1548.1950 &$ -0.7194$&$[ -50,  50]$&$<  20.8$&$< 12.90$&$< 12.90$\\
\ion{O}{1}\\
&1302.1685 &$ -1.3110$&$[ -90,  70]$&$  60.0 \pm  12.2$&$13.95 \pm 0.04$&$ 14.03 \pm  0.08$\\
\ion{Al}{2}\\
&1670.7874 &$  0.2742$&$[ -50,  70]$&$  55.5 \pm  10.3$&$ 12.11 \pm 0.10$&$ 12.11 \pm 0.10$\\
\ion{Si}{2}\\
&1260.4221 &$  0.0030$&$[ -60,  70]$&$  55.0 \pm   7.6$&$ 12.64 \pm 0.10$&$ 13.29 \pm 0.25$\\
&1304.3702 &$ -1.0269$&$[ -60,  70]$&$<  23.0$&$< 13.41$& \\
&1526.7066 &$ -0.8962$&$[ -50,  70]$&$  50.9 \pm 5.0$&$ 13.29 \pm 0.10$& \\
&1808.0130 &$ -2.6603$&$[ -50,  70]$&$<  29.4$&$< 14.84$& \\
\ion{Si}{4}\\
&1393.7550 &$ -0.2774$&$[ -50,  50]$&$<  20.2$&$< 12.52$&$< 12.52$\\
\ion{Fe}{2}\\
&1608.4511 &$ -1.2366$&$[ -50,  70]$&$<  5.3$&$< 12.8$&$< 12.80$\\
\enddata
\tablenotetext{a}{Velocity interval over which the equivalent width and
column density are measured.}
\tablenotetext{b}{Rest equivalent width.}
\end{deluxetable}

\clearpage
\begin{deluxetable}{lcccccc}
\tablewidth{0pc}
\tablecaption{IONIC COLUMN DENSITIES FOR J1456+0407 $z=2.674$\label{tab:J1456+0407}}
\tabletypesize{\footnotesize}
\tablehead{\colhead{Ion} & \colhead{$\lambda$} & \colhead{$\log f$}
& \colhead{$v_{int}^a$} 
& \colhead{$W_\lambda^b$} 
& \colhead{$\log N$} & \colhead{$\log N_{adopt}$} \\
& (\AA) & & (\kms) & (m\AA) & & }
\startdata
\ion{C}{2}\\
&1334.5323 &$ -0.8935$&$[ -80,  60]$&$ 127.0 \pm  14.4$&$ 13.88 \pm 0.04$&$ > 14.33 $\\
\ion{C}{2}*\\
&1335.7077 &$ -0.9397$&$[ -80,  60]$&$<  32.0$&$< 13.42$& \\
\ion{C}{4}\\
&1548.1950 &$ -0.7194$&$[ -80,  60]$&$<  35.5$&$< 13.14$&$< 13.14$\\
&1550.7700 &$ -1.0213$&$[ -80,  60]$&$<  36.4$&$< 13.45$& \\
\ion{O}{1}\\
&1302.1685 &$ -1.3110$&$[ -80,  60]$&$ 100.3 \pm   8.2$&$ 14.13 \pm 0.03$&$ 14.55 \pm 0.28$\\
\ion{Al}{2}\\
&1670.7874 &$  0.2742$&$[ -80,  60]$&$ 87.0 \pm  10.2$&$ 12.27 \pm 0.05$&$ 12.93 \pm 0.45$\\
\ion{Si}{2}\\
&1304.3702 &$ -1.0269$&$[ -80,  60]$&$  32.6 \pm   8.9$&$ 13.40 \pm 0.12$&$ 13.49 \pm 0.07$\\
&1526.7066 &$ -0.8962$&$[ -80,  60]$&$  84.4 \pm  16.7$&$ 13.58 \pm 0.08$& \\
&1808.0130 &$ -2.6603$&$[ -80,  60]$&$<  44.6$&$< 15.02$& \\
\ion{Si}{4}\\
&1393.7550 &$ -0.2774$&$[ -80,  60]$&$  83.9 \pm  15.8$&$ 13.01 \pm 0.09$&$ 13.01 \pm 0.09$\\
\ion{Fe}{2}\\
&1608.4511 &$ -1.2366$&$[ -80,  60]$&$  30.1 \pm 8.00$&$ 13.35 \pm 0.10$&$ 13.00 \pm 0.10$\\
&2344.2140 &$ -0.9431$&$[ -80,  60]$&$  41.2 \pm 10.00$&$12.86 \pm 0.10$& \\
&2374.4612 &$ -1.5045$&$[ -80,  60]$&$  47.4$&$< 13.68$& \\
&2382.7650 &$ -0.4949$&$[ -80,  60]$&$ 167.0 \pm  13.2$&$ 13.01 \pm 0.06$& \\
\enddata
\tablenotetext{a}{Velocity interval over which the equivalent width and
column density are measured.}
\tablenotetext{b}{Rest equivalent width.}
\end{deluxetable}

\clearpage
\begin{deluxetable}{lcccccc}
\tablewidth{0pc}
\tablecaption{IONIC COLUMN DENSITIES FOR J1637+2901 $z=3.496$\label{tab:J1637+2901}}
\tabletypesize{\footnotesize}
\tablehead{\colhead{Ion} & \colhead{$\lambda$} & \colhead{$\log f$}
& \colhead{$v_{int}^a$} 
& \colhead{$W_\lambda^b$} 
& \colhead{$\log N$} & \colhead{$\log N_{adopt}$} \\
& (\AA) & & (\kms) & (m\AA) & & }
\startdata
\ion{C}{2}\\
&1334.5323 &$ -0.8935$&$[ -70,  80]$&$ 147.0 \pm  12.4$&$> 13.86$&$> 14.30$\\
\ion{C}{2}\\
&1335.7077 &$ -0.9397$&$[ -70,  80]$&$ 274.8 \pm  11.7$&$< 14.35$& \\
\ion{C}{4}\\
&1548.1950 &$ -0.7194$&$[ -70, 150]$&$ 130.3 \pm  14.0$&$ 13.55 \pm 0.05$&$ 13.55 \pm 0.05$\\
\ion{O}{1}\\
&1302.1685 &$ -1.3110$&$[ -70,  80]$&$  91.7 \pm  10.6$&$ 14.09 \pm 0.05$&$ 14.29 \pm 0.20$\\
\ion{Al}{2}\\
&1670.7874 &$  0.2742$&$[ -70,  80]$&$  71.0 \pm  12.2$&$ 12.18 \pm 0.07$&$ 12.29 \pm 0.11$\\
\ion{Si}{2}\\
&1260.4221 &$  0.0030$&$[ -70,  80]$&$ 132.7 \pm   5.0$&$> 13.07$&$ 13.41 \pm 0.14$\\
&1304.3702 &$ -1.0269$&$[ -70,  80]$&$  42.3 \pm  11.0$&$ 13.54 \pm 0.11$& \\
&1526.7066 &$ -0.8962$&$[ -70,  80]$&$<  32.3$&$ 13.30 \pm 0.14$& \\
&1808.0130 &$ -2.6603$&$[ -70,  80]$&$<  56.4 $&$< 14.98 $& \\
\ion{Si}{4}\\
&1393.7550 &$ -0.2774$&$[ -70,  80]$&$  40.7 \pm  13.1$&$ 12.69 \pm 0.14$&$ 12.69 \pm 0.14$\\
&1402.7700 &$ -0.5817$&$[ -70,  80]$&$<  26.3$&$< 12.95$& \\
\ion{Fe}{2}\\
&1608.4511 &$ -1.2366$&$[ -70,  80]$&$  67.8 \pm   9.7$&$ 13.74 \pm 0.06$&$ 13.84 \pm 0.10$\\
\enddata
\tablenotetext{a}{Velocity interval over which the equivalent width and
column density are measured.}
\tablenotetext{b}{Rest equivalent width.}
\end{deluxetable}

\clearpage
\begin{deluxetable}{lcccccc}
\tablewidth{0pc}
\tablecaption{IONIC COLUMN DENSITIES FOR J2114-0632 $z=4.126$\label{tab:J2114-0632}}
\tabletypesize{\footnotesize}
\tablehead{\colhead{Ion} & \colhead{$\lambda$} & \colhead{$\log f$}
& \colhead{$v_{int}^a$} 
& \colhead{$W_\lambda^b$} 
& \colhead{$\log N$} & \colhead{$\log N_{adopt}$} \\
& (\AA) & & (\kms) & (m\AA) & & }
\startdata
\ion{C}{2}\\
&1334.5323 &$ -0.8935$&$[ -70,  80]$&$ 118.4 \pm   8.0$&$ 13.80 \pm 0.03$&$ 14.25 \pm 0.45$\\
\ion{C}{2}*\\
&1335.7077 &$ -0.9397$&$[ -70,  80]$&$<  16.1$&$< 13.13$& \\
\ion{C}{4}\\
&1548.1950 &$ -0.7194$&$[-120,  80]$&$ 337.7 \pm  12.8$&$ 14.05 \pm 0.02$&$ 14.04 \pm 0.02$\\
&1550.7700 &$ -1.0213$&$[-120,  80]$&$ 192.3 \pm  14.6$&$ 14.04 \pm 0.04$& \\
\ion{O}{1}\\
&1302.1685 &$ -1.3110$&$[ -70,  80]$&$ 138.5 \pm   8.6$&$ 14.27 \pm 0.03$&$ 14.72 \pm 0.45$\\
\ion{Al}{2}\\
&1670.7874 &$  0.2742$&$[ -70,  80]$&$<  20.2$&$< 11.83$&$< 11.83$\\
\ion{Si}{2}\\
&1260.4221 &$  0.0030$&$[ -70,  80]$&$ 135.9 \pm   5.4$&$> 12.98$&$ 13.23 \pm 0.13$\\
&1304.3702 &$ -1.0269$&$[ -70,  80]$&$<  18.5$&$< 13.31$& \\
&1526.7066 &$ -0.8962$&$[ -70,  80]$&$  39.0 \pm   9.7$&$ 13.17 \pm 0.13$& \\
&1808.0130 &$ -2.6603$&$[ -70,  80]$&$<  36.4$&$< 14.93$& \\
\ion{Si}{4}\\
&1393.7550 &$ -0.2774$&$[-100,  80]$&$  57.5 \pm  10.7$&$ 12.84 \pm 0.08$&$ 12.90 \pm 0.06$\\
&1402.7700 &$ -0.5817$&$[-100,  80]$&$  48.8 \pm  10.4$&$ 13.06 \pm 0.09$& \\
\ion{Fe}{2}\\
&1608.4511 &$ -1.2366$&$[ -40,  80]$&$< 39.6 $&$< 13.51$&$< 13.51$\\
\enddata
\tablenotetext{a}{Velocity interval over which the equivalent width and
column density are measured.}
\tablenotetext{b}{Rest equivalent width.}
\end{deluxetable}

\clearpage
\begin{deluxetable}{cccrrrrrr}
\tablewidth{0pt}
\tabletypesize{\footnotesize}
\tablecaption{Elemental Abundances for DLA Systems \label{tab:abunds}}
\tablehead {
\colhead {Quasar}                      &
\colhead {z$_{DLA}$}                   &
\colhead {log(N(\ion{H}{1}))}          &
\colhead {[C/H]}                     &
\colhead {[O/H]}                     &
\colhead {[Si/H]}                     &
\colhead {[Al/H]}                     &
\colhead {[Fe/H]}                     &
\colhead {[Mg/H]}                     \\
                                  &
                                  &
          (cm$^{-2}$)
}
\startdata
(1)&(2)&(3)&(4)&(5)&(6)&(7)&(8)&(9)\nl
\tableline
SDSS0751+4516 & 3.0462 & 20.35 $\pm$ 0.15 & $>$-2.96$^{b}$&$<$-2.10$^{b}$&$>$-2.37$^{b}$&$>$-2.92$^{b}$&$>$-2.09$^{b}$&\nodata\nl
SDSS0759+3129 & 3.0346 & 20.60 $\pm$ 0.10 & $>$-2.88$^{a}$&$>$-2.73$^{a}$&-2.03$^{c}$&-2.63$^{c}$&-2.27$^{b}$&\nodata\nl
SDSS0811+2838 & 2.4338 & 20.50 $\pm$ 0.10 & $>$-3.00$^{a}$&\nodata&\nodata&$>$-2.92$^{a}$&$>$-2.84$^{a}$&\nodata\nl
SDSS0814+5029 & 3.7079 & 21.25 $\pm$ 0.15 & $>$-3.46$^{b}$&$>$-3.50$^{b}$&$>$-2.92$^{b}$&$>$-3.43$^{b}$&$>$-2.95$^{b}$&\nodata\nl
SDSS0831+3358 & 2.3039 & 20.25 $\pm$ 0.15 & $>$-2.72$^{b}$&\nodata&$>$-2.76$^{b}$&$>$-3.15$^{b}$&$>$-2.59$^{b}$&\nodata\nl
SDSS0844+4624 & 3.3363 & 20.70 $\pm$ 0.15 & $<$-2.37$^{b}$&\nodata&$>$-2.21$^{b}$&\nodata&$>$-2.24$^{b}$&\nodata\nl
SDSS0924+0951 & 3.3382 & 20.85 $\pm$ 0.10 & $>$-2.65$^{c}$&$>$-2.71$^{b}$&$>$-1.76$^{c}$&$>$-2.19$^{c}$&$>$-2.06$^{c}$&\nodata\nl
SDSS0928+0939 & 2.9098 & 20.75 $\pm$ 0.15 & -2.41$^{b}$&$>$-2.78$^{b}$&$>$-2.16$^{b}$&$>$-2.69$^{b}$&-2.13$^{c}$&\nodata\nl
SDSS0955+4116 & 3.2801 & 20.10 $\pm$ 0.10 & -2.82$^{b}$&-2.82$^{b}$&-2.75$^{b}$&$<$-2.74$^{a}$&$<$-2.30$^{a}$&\nodata\nl
SDSS1001+0343 & 3.0785 & 20.15 $\pm$ 0.10 & -2.85$^{b}$&-2.93$^{b}$&-2.94$^{b}$&-2.82$^{b}$&$<$-2.32$^{a}$&\nodata\nl
SDSS1003+5520 & 2.5024 & 20.35 $\pm$ 0.15 & $<$-2.59$^{b}$&$<$-2.21$^{b}$&-2.13$^{c}$&-2.76$^{b}$&-2.87$^{c}$&$>$-2.67$^{b}$\nl
SDSS1031+4055 & 2.5686 & 20.55 $\pm$ 0.10 & $<$-2.58$^{a}$&$<$-2.21$^{a}$&$<$-1.66$^{a}$&$>$-2.37$^{a}$&-2.18$^{b}$&$>$-2.51$^{a}$\nl
SDSS1037+0139 & 2.7050 & 20.40 $\pm$ 0.25 & $>$-2.31$^{c}$&$>$-2.71$^{c}$&$>$-2.25$^{c}$&$>$-2.77$^{c}$&$>$-2.54$^{c}$&\nodata\nl
SDSS1043+6151 & 2.7865 & 20.60 $\pm$ 0.15 & $>$-2.71$^{b}$&$>$-2.46$^{b}$&-2.08$^{c}$&$>$-2.46$^{b}$&-2.04$^{b}$&\nodata\nl
SDSS1048+3911 & 2.2957 & 20.70 $\pm$ 0.10 & -2.79$^{a}$&$>$-2.93$^{c}$&-2.31$^{c}$&-2.49$^{c}$&-2.46$^{b}$&$>$-2.82$^{a}$\nl
SDSS1108+1209 & 3.3964 & 20.55 $\pm$ 0.15 & $>$-2.86$^{b}$&$<$-2.61$^{b}$&$>$-2.46$^{b}$&$>$-3.18$^{b}$&$<$-2.29$^{b}$&\nodata\nl
SDSS1219+1603 & 3.0037 & 20.35 $\pm$ 0.10 & $>$-2.59$^{a}$&-2.59$^{c}$&-2.08$^{b}$&$>$-2.29$^{a}$&-2.09$^{b}$&\nodata\nl
SDSS1251+4120 & 2.7296 & 21.10 $\pm$ 0.10 & $>$-3.06$^{a}$&$>$-2.92$^{a}$&-2.73$^{c}$&-2.86$^{c}$&-2.35$^{c}$&\nodata\nl
SDSS1305+2902 & 2.3865 & 20.25 $\pm$ 0.10 & -2.46$^{c}$&-2.90$^{b}$&-2.54$^{b}$&-2.83$^{b}$&-2.79$^{b}$&$>$-2.99$^{a}$\nl
SDSS1325+1255 & 3.5507 & 20.50 $\pm$ 0.15 & -2.54$^{c}$&$>$-2.39$^{c}$&-2.52$^{b}$&$<$-2.06$^{b}$&$<$-2.27$^{b}$&\nodata\nl
SDSS1350+5952 & 2.7558 & 20.65 $\pm$ 0.10 & $>$-2.77$^{a}$&$>$-2.53$^{a}$&$>$-2.48$^{a}$&-2.53$^{c}$&-2.59$^{b}$&\nodata\nl
SDSS1358+0349 & 2.8528 & 20.50 $\pm$ 0.10 & $>$-2.58$^{c}$&-2.88$^{c}$&-2.81$^{b}$&-2.87$^{b}$&-3.03$^{a}$&\nodata\nl
SDSS1358+6522 & 3.0675 & 20.35 $\pm$ 0.15 & -2.59$^{c}$&-3.08$^{b}$&-2.67$^{c}$&-2.78$^{b}$&$<$-3.09$^{b}$&\nodata\nl
SDSS1402+5909 & 3.7745 & 21.35 $\pm$ 0.10 & $>$-3.37$^{c}$&$>$-3.17$^{c}$&$>$-2.46$^{c}$&$>$-2.86$^{c}$&$>$-2.34$^{c}$&\nodata\nl
SDSS1440+0637 & 2.5177 & 21.00 $\pm$ 0.15 & \nodata&\nodata&-2.37$^{c}$&$>$-2.69$^{c}$&-1.99$^{c}$&$>$-3.22$^{b}$\nl
SDSS1440+0637 & 2.8246 & 20.20 $\pm$ 0.10 & $<$-2.43$^{a}$&$<$-1.96$^{a}$&-2.14$^{a}$&-2.43$^{b}$&-2.21$^{b}$&\nodata\nl
SDSS1456+0407 & 2.3201 & 20.15 $\pm$ 0.10 & $<$-2.01$^{a}$&\nodata&$>$-1.86$^{a}$&-1.96$^{c}$&$>$-2.17$^{b}$&$>$-2.37$^{a}$\nl
SDSS1456+0407 & 2.6736 & 20.35 $\pm$ 0.10 & $>$-2.48$^{c}$&-2.56$^{c}$&-2.47$^{b}$&-1.96$^{c}$&-2.89$^{b}$&\nodata\nl
SDSS1557+2320 & 3.5383 & 20.65 $\pm$ 0.10 & $<$-2.81$^{a}$&-2.21$^{a}$&-2.14$^{c}$&-2.68$^{c}$&-2.61$^{c}$&\nodata\nl
SDSS1637+2901 & 3.4956 & 20.70 $\pm$ 0.10 & $>$-2.85$^{a}$&-3.17$^{c}$&-2.90$^{b}$&-2.95$^{b}$&-2.40$^{b}$&\nodata\nl
SDSS1654+3509 & 2.8113 & 20.10 $\pm$ 0.10 & $<$-1.70$^{a}$&$>$-2.22$^{a}$&-1.74$^{c}$&-1.53$^{a}$&-2.01$^{a}$&\nodata\nl
SDSS1709+3417 & 3.0104 & 20.40 $\pm$ 0.10 & $>$-2.80$^{a}$&\nodata&$>$-1.91$^{a}$&-2.22$^{c}$&-1.99$^{b}$&\nodata\nl
SDSS1709+3417 & 2.5303 & 20.45 $\pm$ 0.15 & -2.01$^{b}$&\nodata&-1.51$^{b}$&-1.87$^{b}$&-1.65$^{b}$&$>$-2.27$^{b}$\nl
SDSS1717+5802 & 3.0461 & 20.25 $\pm$ 0.10 & $>$-2.39$^{b}$&$>$-2.26$^{a}$&-2.05$^{b}$&$>$-2.29$^{a}$&-2.37$^{a}$&\nodata\nl
SDSS2114-0632 & 4.1262 & 20.40 $\pm$ 0.15 & -2.63$^{c}$&-2.44$^{c}$&-2.78$^{b}$&$<$-3.17$^{b}$&$<$-2.43$^{b}$&\nodata\nl
\enddata
\tablenotetext{a} { Uncertainty less than 0.11 dex }
\tablenotetext{b} { Metallicity uncertainty between 0.11-0.20 dex }
\tablenotetext{c} { Metallicity uncertainty greater than 0.20 dex }
\label{mycldtable}
\end{deluxetable}

clearpage
\begin{deluxetable}{lllrlrl}
\tablewidth{0pc}
\tablecaption{Column densities for the species \ion{C}{4} and \ion{Si}{4}\label{tab:civ}}
\tabletypesize{\footnotesize}
\tablehead{\colhead{Quasar} & \colhead{z$_{DLA}$}&\colhead{logN(\ion{H}{1})} & \colhead{logN(\ion{C}{4})}
& \colhead{err}
& \colhead{logN(\ion{Si}{4})}
& \colhead{err} }
\startdata
SDSS0751+4516&3.0462&20.35&13.70&0.10&12.71&0.10 \\
SDSS0759+3129&3.0346&20.6&$<$12.9&\nodata&$<$12.0&\nodata\\
SDSS0811+2838&2.4338&20.50&13.90&0.10&12.40&0.10\\
SDSS0814+5029&3.7079&21.25&$<$12.4&\nodata&$<$12.0&\nodata\\
SDSS0831+3358&2.3039&20.25&$<$12.6&\nodata&$<$12.3&\nodata\\
SDSS0844+4624&3.3363&20.70&13.31&0.10&12.30&0.10\\
SDSS0924+0951&3.3382&20.9&12.95&0.043&$<$12.5&\nodata\\
SDSS0928+0939&2.9098&20.8&13.60&0.043&$<$13.8&\nodata\\
SDSS0955+4116&3.2801&20.1&$<$13.1&\nodata&13.0&0.043\\
SDSS1001+0343&3.0785&20.1&13.24&0.043&$<$12.0&\nodata\\
SDSS1003+5520&2.5024&20.4&$<$12.9&\nodata&$<$13.6&\nodata\\
SDSS1031+4055&2.5686&20.5&13.20&0.038&13.0&0.043\\
SDSS1037+0139&2.7050&20.40&13.38&0.10&$<$12.7&\nodata\\
SDSS1043+6151&2.7865&20.6&13.74&0.043&13.40&0.039\\
SDSS1048+3911&2.2957&20.7&13.70&0.042&13.20&0.031\\
SDSS1108+1209&3.3964&20.55&13.73&0.010&13.50&0.010\\
SDSS1156+5513&2.4808&19.6&14.28&0.04&\nodata&\nodata\\
SDSS1156+5513&2.4975&19.6&$<$13.30&\nodata&\nodata&\nodata\\
SDSS1219+1603&3.0037&20.4&14.05&0.038&13.50&0.042\\
SDSS1251+4120&2.7296&21.1&13.93&0.039&13.20&0.042\\
SDSS1305+2902&2.3865&20.3&$<$13.4&\nodata&$<$12.9&\nodata\\
SDSS1325+1255&3.5507&20.5&$<$12.9&\nodata&$<$12.5&\nodata\\
SDSS1327+4845&2.4468&20.61&13.85&0.040&\nodata&\nodata\\
SDSS1327+4845&2.6116&19.60&$<$12.70&\nodata&\nodata&\nodata\\
SDSS1331+4838&3.6923&21.07&13.90&0.10&\nodata&\nodata\\
SDSS1349+1242&3.7433&20.00&$<$13.00&\nodata&\nodata&\nodata\\
SDSS1350+5952&2.7558&20.6&$<$13.1&\nodata&$<$12.5&\nodata\\
SDSS1358+0349&2.8528&20.5&13.20&0.043&12.80&0.030\\
SDSS1358+6522&3.0675&20.4&$<$12.8&\nodata&$<$12.5&\nodata\\
SDSS1402+5909&3.7745&21.4&$<$13.4&\nodata&13.1&0.044\\
SDSS1440+0637&2.5177&21.0&13.40&0.041&13.60&0.044\\
SDSS1440+0637&2.8246&20.2&$<$13.0&\nodata&13.90&\nodata\\
SDSS1456+0407&2.3201&20.1&$<$13.2&\nodata&$<$13.7&\nodata\\
SDSS1456+0407&2.6736&20.3&$<$12.6&\nodata&13.10&0.035\\
SDSS1557+2320&3.5398&20.6&$<$12.9&\nodata&12.90&0.041\\
SDSS1637+2901&3.4956&20.7&$<$13.8&\nodata&11.80&0.043\\
SDSS1654+3509&2.8113&20.1&14.80&0.041&14.10&0.037\\
SDSS1709+3417&3.0104&20.4&13.40&0.037&12.80&0.043\\
SDSS1709+3417&2.5303&20.5&14.80&0.057&14.50&0.041\\
SDSS1717+5802&3.0461&20.3&14.30&0.041&13.50&0.039\\
SDSS2114-0632&4.1262&20.4&14.00&0.040&12.90&0.040\\
\enddata
\end{deluxetable}

\end{document}